\documentclass[preprint,10pt,3p]{elsarticle}

\RequirePackage[OT1]{fontenc}
\RequirePackage{amsthm,amsmath}
\RequirePackage[numbers]{natbib}
\usepackage[table]{xcolor}
\RequirePackage[colorlinks=true,backref=page]{hyperref}
\hypersetup{
    linkcolor = {blue},
    citecolor = {blue},
    urlcolor = {blue}
    }
\usepackage[utf8]{inputenc}
\usepackage{times}
\usepackage{fullpage}   
\usepackage[british,UKenglish,english]{babel}

\usepackage{amsfonts,amssymb}	
\usepackage{mathrsfs}	
\usepackage{dsfont}
\usepackage[english]{babel}
\usepackage{bbm}
\usepackage{euscript,graphicx}
\usepackage{float}
\usepackage{subfigure}
\usepackage{xspace}
\usepackage{booktabs}
\usepackage[font=small,labelfont=bf]{caption}
\usepackage{appendix}
\usepackage{longtable}
\usepackage{siunitx} 
\usepackage[framemethod=TikZ]{mdframed}
\usepackage{pgfplots,pgfplotstable}  
\usepackage{microtype}
\usepackage{bm} 

\newtheorem{theorem}{Theorem}[section]

\newtheorem{lem}[theorem]{Lemma}
\newtheorem{prop}[theorem]{Proposition}
\newtheorem{rem}[theorem]{Remark}

\numberwithin{equation}{section}

\newcommand{\pdissip}{{\varepsilon}}
\newcommand{\productke}{\mathcal{P}}
\newcommand{\av}[1]{{{\displaystyle{\langle} {#1}\displaystyle{\rangle}}}}

\newcommand{\tke}{{\rm k}}
\newcommand{\ttke}{{\widetilde{{\rm k}}}}
\newcommand{\ustar}{u^{\ast}}
\newcommand{\lm}{{\ell_{\textrm{m}}}}
\newcommand{\zlm}{z_{\ell_{\textrm{m}}}}
\newcommand{\nut}{\nu_{\textrm{turb}}}
\newcommand{\xobs}{x_{\textrm{obs}}}

\newcommand{\argmax}[1]{\underset{#1}{\operatorname{arg}\,\operatorname{max}}\;}

\def\RR{\mathbb{R}}

\def\EE{\mathbb{E}}
\def\PP{\mathbb{P}}
\def\qnn{\widehat{q}_{t_n}}
\def\qbn{q^{\text{obs}}_{t_n}}
\def\qn1{\widehat{q}_{t_{n+1}}}
\def\qb1{q^{\text{obs}}_{t_{n+1}}}

\newcommand{\Ee}{{\mathbb E}}

\newcommand{\x}{{\rm x}}

\usepackage{dsfont}
\newcommand{\ind}{\mathds{1}}

\allowdisplaybreaks

\newcommand{\tabhead}[1]{\textbf{#1}}

\journal{Journal of Computational Physics}

\begin{document}

\begin{frontmatter}

\title{Instantaneous turbulent kinetic energy modelling based on Lagrangian stochastic approach in CFD and application to wind energy}

\author[inria-sop]{Mireille Bossy}
\ead{mireille.bossy@inria.fr}

\author[HSE]{Jean-Fran\c{c}ois Jabir}
\ead{jjabir@hse.ru}
\author[inria-sop, UV]{Kerlyns Mart\'inez Rodr\'iguez}
\ead{kerlyns.martinez-rodriguez@inria.fr}

\address[inria-sop]{Universit\'e C\^ote d'Azur, Inria, CNRS,  Sophia-Antipolis, France}
\address[HSE]{HSE University, Moscow, Russian Federation}
\address[UV]{Institute of Statistics, University of Valpara\'iso, Valpara\'iso, Chile}

\begin{abstract}

We present the construction of an original stochastic model for the instantaneous turbulent kinetic energy  at a given   point of a flow, and we validate estimator methods on this model with observational data examples.
Motivated by the need for wind energy industry of acquiring relevant statistical information  of air motion at a local place, we adopt the Lagrangian description of fluid flows to derive, from the $3$D+time equations of the physics, a $0$D+time-stochastic model for the time series of the instantaneous turbulent kinetic energy at a given position.  
Specifically, based on the Lagrangian stochastic description of a generic fluid-particles, we derive a family of mean-field dynamics featuring the square norm  of the turbulent velocity.  By approximating at equilibrium the characteristic nonlinear terms of the dynamics, we recover the so called Cox-Ingersoll-Ross process, which was previously suggested in the literature for modelling wind speed. We then propose a calibration procedure for the parameters employing both direct methods and Bayesian inference. In particular, we show the consistency of the estimators and validate the model through the quantification of uncertainty, with respect to the range of values given in the literature for some physical constants of turbulence modelling. 
\end{abstract}


\end{frontmatter}

\tableofcontents

\section{Introduction}

The need of statistical information on the wind, at a given location and on large time period, is of major importance in many applications  such as the  structural safety of large construction projects or the economy of a wind farm, whether it concerns an investment project, a wind farm operation or its repowering.   
The evaluation of the local wind  is expressed on  different time scales: monthly, annually or over several decades for resource assessment, daily, hourly or even less  for dynamical forecasting (these scales being addressed with an increasing panel of  methodologies, see e.g. \cite{chang2014literature}). 
In the literature, wind forecasting models are generally classified into physical models (numerical weather prediction models), statistical approaches (time-series models, machine learning models, and more recently deep learning methods), and hybrid physical and statistical models, see e.g. \cite{soman2010review, chang2014literature, hanifi2020critical}.
At  a given site and height in the atmospheric boundary layer,  measuring instruments (anemometer mast  or nacelle anemometer) record time series of characteristics of the wind. For instance, wind speed characterizing load conditions,  wind direction, kinetic energy and possibly power production. Such observation should feed into forecasting, but also uncertainty modelling. These observations present various time scales, some large ones  such  as  diurnal,  weekly  and seasonal changes and variations, and some  small scales often referred to as atmospheric turbulence \cite{lauren2001analysis}. 
In this context, probabilistic or statistical approaches are widely used, helping to characterize uncertainty through quantile indicators.  

Recently, dynamical diffusion models have been proposed in the literature, featuring a continuous time description of wind dynamics.  In \cite{bensoussan2016cox}, the Cox, Ingersoll and Ross (CIR) stochastic process  --originally introduced in mathematical finance to model short term  interest rates -- is proposed to describe the dynamics of the squared wind speed. 
In \cite{arenas2020ornstein,arenas2020stochastic},  an Ornstein-Uhlenbeck (OU) process is proposed and combined with a statistical measure of atmospheric turbulence called turbulent intensity.  
 
In this paper,  we derive a statistical model for the local wind speed, obtaining  a reduced $0$D+time equation from the $3$D+time  averaged  Navier-Stokes equation with subgrid turbulence model. More precisely, we derive a continuous-time stochastic diffusion model for the  instantaneous wind speed fluctuation called instantaneous turbulent kinetic energy   measured at a given point. For that purpose, we start from the physical description of the fluid flow in a Lagrangian formulation that represents the fluid-particle dynamics with a stochastic diffusion process. 

\paragraph{\bf Turbulent flow models}
Airflow is described by the incompressible Navier-Stokes equations. 
In the atmospheric boundary layer (ABL), the vicinity of the earth's surface, many vortex scales are present and interact. The flow is then described as turbulent, a situation that   complexes any prediction by numerical approach. In particular, this  makes the direct numerical approach (DNS) impossible,  requiring to refine the spatial discretisation scales below the Kolmogorov scale (the scale from which viscosity allows to dissipate the kinetic energy, of a few millimetres in the atmospheric boundary layer).  
Common computational models are  based on statistical approximations that replace unsolved subgrid scales, and on the idea that numerical estimations on averaged quantities of the flow (rather  than in the details of the fluctuations) concentrated the major interest in many applications. Among the most well-known averaging methods, 
the Reynolds averaging of the Navier-Stokes equations --leading to the RANS models-- use a statistical averaging decomposing  each instantaneous variable of a flow into the sum of a mean part and a fluctuating part:
 \[u '= U- \av{U} \]
where $U$ is the velocity field, $\av{U}$ is the mean velocity (or average velocity) and $u'$ represents the fluctuating or turbulent velocity. 
Averaging the Navier-Stokes equations has the effect of removing the fast fluctuation terms, but introduce unclosed term  such as turbulent velocity covariance known as the Reynolds stress tensor $\av{(u^{\prime(i)} u^{\prime(j)})}$. 
 The question of {\it what might be  a good model for this tensor ?} has been subject to a great  interest and abundant literature (see e.g. in \cite{P00}), in the context of computational fluid dynamics (CFD). Before introducing classical closure models used  in the ABL,  we  first present the Lagrangian point of view of the  averaged Navier-Stokes equation, corresponding to the starting point of the stochastic reduced model analysed in this paper. 

\paragraph{\bf Fluid-particle based models}
Stochastic Lagrangian approaches  for turbulent flow were introduced at first for laden turbulent flow, 
in order to  represent the  turbulent subgrid-scale fluctuations of particle velocities that cannot 
be adequately resolved by mesh computation alone \cite{P94}.
 These approaches are used in the case of disperse two-phase flows where one phase is a set of discrete 
elements or ‘particles’ \cite{Pope85,MINIER20161} and they have been  implemented in various complex industrial applications  \cite{MinierPeirano-01}. In the context of atmospheric flow, the so-called Lagrangian Particle Dispersion Models  (LPDMs)  are widely used  for the analysis of  air pollutants dispersion  (see \cite{STOHL1998947} and the references therein). Such methods adopts the perspective of an 'air parcel'  by tracking a number of fictitious (or statistical) particles with position $X_t$ released into a flow field:  
\begin{align}\label{eq:lpdm}
{d X_t} = \av{U}(t,X_t)dt   + u'(t) dt,
\end{align}
where $u'(t)$ is a random fluctuation of the particle velocity, $\av{U}$ is the mean velocity computed on a mesh. 
The velocity fluctuation is modelled using a stochastic differential equation (SDE) 
the complexity of which varies with the number of physical variables to be represented. 
But restricted to the turbulent velocity, it is generally declined with the simple Langevin model:
\begin{align}
d  u'(t) =  - \frac{u'(t)}{T_L} dt + \sqrt{C_0 \pdissip(t,X_t)}  dB_t,  
\end{align}
where the stochastic (or fast) part of the  motion is described by the 3-dimensional Brownian motion  $B$,  amplified with the turbulent pseudo dissipation of the flow $\pdissip$. 

From a modelling view point, a great interest of the Lagrangian approach is its ability 
to represent the mean velocity and the Reynolds stress tensor as the first 
and second moments of the probability density function (PDF) of the Lagrangian model. 
A PDF method for turbulent flow\footnote{We consider here only the case of constant 
mass density flow, for the sake of clarity.} 
 (firstly introduced by Pope~\cite{Pope85}) considers  stochastic processes 
$((X_t,U_t);0\leq t\leq T)$,  describing the instantaneous position $X_t\in\mathbb{R}^3$ of a fluid 
particle and its instantaneous velocity $U_t\in\mathbb{R}^3$. 
Therefore, denoting the joint probability density of the process $((X_t,U_t);0\leq t\leq T)$ by $\varrho$, 
for all suitable map $g:v\in\mathbb{R}^3\mapsto g(v)\in\mathbb{R}$,  
the Reynolds operator $\av{\cdot}$ applied to $g(U)$ is interpreted as the probabilistic  
conditional expectation of the particle velocity $U_t$, knowing that  its position $X_t$ is  $x$, 
under the probability $\PP$ of the model provided with expectation symbol $\EE$: 
\begin{equation}\label{eq:Reynolds_average_interpretation}
\langle{g(U)}\rangle(t,x)=\mathbb{E}\left[g(U_t)|\;X_t=x\right]=\frac{\int_{\mathbb{R}^3}g(u)\varrho(t,x,u)du}{\int_{\mathbb{R}^3}\varrho(t,x,u)du}.
\end{equation}
Macroscopic quantities of interest can be then identified with this rule, as long as the chosen Reynolds 
stress closure order allows to represent them in the model (we refer to Section \ref{sec:Models_and_Theory} for more details). 

For the use of the fluid-particle method in meteorological context,  we particularly refer the interested reader to \cite{Baehr-10, SuBaDa-11} for application to  filtering of wind data, \cite{BeBoChDrRoSa-09, BeBoChJaRo10} for application to  refined wind computation, and more recently \cite{B16, B18} for wind farm simulation.

\paragraph{\bf Aim and layout of the paper}
In this paper, starting from the 3D+time physical description of the flow given by the stochastic Lagrangian model framework, we construct  a simplified 0D+time model featuring the evolution of the turbulent kinetic energy of the flow 
at a fixed location. In addition we  construct and analyse a calibration procedure that includes uncertainty 
as a part of the calibration methodology. We test the model and its calibration  on wind measurements 
obtained from a typical measure mast used in wind energy potential assessment (a 30\xspace\si{\meter}  
height mast, with a frequency of 10\xspace\si{\hertz})   on a large  interval of time (a year) to capture 
a larger part of the variability of the wind during a seasonal cycle.   
We validate the method, comparing  the posterior distribution of the model-parameter issued from the    
$G_{ij}$ tensor  coefficient in \eqref{eq:SDM} below against the  interval of the values proposed in 
the CFD literature. 
Next, we evaluate the ability of  the calibrated model to replicate  the observed time series. 
Finally, we come back on the calibration procedure, considering the 10-minutes-averaged  turbulence intensity as a good statistical prediction value for  the turbulent source term in the proposed stochastic model  \eqref{eq:tke_equilibrated_CIR}. In particular this method allows to well approximate the Weibull distribution form of  the turbulent wind speed $\|u'\|$, showing the consistency of this simpler variant of the calibration method  and the reliability of the model to predict efficiently the wind distribution.

In Section~\ref{sec:Models_and_Theory}, we set the theoretical basis of the  reduced model 
by shortly introducing  the fluid-particle Lagrangian model in turbulent (near wall) flow. 
In Section~\ref{sec:TKEmodeling}, 
we mathematically derive in several steps the 0D+time model for the instantaneous turbulent kinetic energy 
(TKE).  We analyse the wellposedness of the SDE describing the  instantaneous TKE process 
(existence and trajectorial uniqueness).  
We give semi-explicit form of the moments solution and analyse their long-time behaviour, 
in order to further simplify  the reduced model.  
We also present the wind speed observations  targeted by our instantaneous TKE model (see Subsection~\ref{sec:Data}) and  used to calibrate and to validate  the model in the next sections. 
We formulate the calibration procedure for the instantaneous TKE process in Section~\ref{sec:Calibration} in two main steps. Based on the mathematical analysis of the model, we first derive a   pseudo-likelihood maximisation procedure (Step zero in~\ref{sec:Step Zero}), we next improve the calibration process, introducing a Bayesian procedure to handle  the assumed  uncertainty  on the parameters (Step one in~\ref{sec:Step One}). Finally, in Section~\ref{sec:Results} we summarize the key findings and results of the calibration along with the validation of the model.

We end this introduction by motivating the calibration under uncertainty methodology that we promote in this paper. 

\paragraph{\bf Calibration with uncertainty for turbulent flow models}

As pointed out  in ~\cite{oden2010computer}, to produce accurate predictions of quantities of interest it is necessary to make a systematic treatment of the uncertainties within the models and observations, quantifying them along their propagation through a computational model. In particular, estimation of model parameters comes before assessing model performance. 

For instance in \cite{C14}, studying the uncertainty  of the RANS model parameters based on a Launder-Sharma 
turbulence closure relation, the authors used a Bayesian calibration method employing measured 
boundary-layer velocity profiles. 
By modelling  the spread of parameters within the flow-class,  
they show  the ability of the Bayesian calibration  to provide information about the 
values these parameters should take in each flow case. 
This  uncertainty  is  thus highlighted by its  quantified distribution,  
suggesting that the parameters must be seen as tuned to be associated with the $\tke$-$\pdissip$ 
turbulence closure, and in general,  parameters are not expected to be flow-independent.

\medskip

Given the stochastic nature of  Lagrangian modelling handled in this paper, 
the uncertainty quantification for the model parameters can be less costly to implement through the 
computation/approximation of densities. 
In particular the need for a surrogate model can be more easily mitigated. 
More precisely, considering  a (deterministic) model, a set of observations, 
and the set of  model parameters to estimate from this data, the  obvious idea is to minimize one of 
the performance measures with respect to the parameter values. Instead, with a stochastic model, 
a more statistically reliable method of parameter estimation can be used, since (in principle) 
it is possible to construct the likelihood function which is merely the probability that the model generates 
the observed data, given a parameter set.  
Once a model for parameters uncertainty is identified, 
by modifying/increasing the variables dependency of the  likelihood function, 
we may quantify the probability distribution of the parameters given the observational data, 
by the well-known method of maximum likelihood estimation. 
Note that the explicit computation of the likelihood requires to work with an explicit form for 
the  density function, which drastically restricts the class of possible stochastic models. 
Alternative approximation methods are available, based on discrete time  sampling  such as 
Markov Chain Monte Carlo (MCMC)  methods that we detail in Section \ref{sec:Calibration} 
and implement in Section \ref{sec:Results}.

\section{A short review of stochastic Lagrangian approach for turbulent flows}\label{sec:Models_and_Theory}

In this section we introduce the framework for the stochastic Lagrangian approach (also known as PDF approach) describing the dynamics of fluid particles within turbulent flows. 
Ideally, a fluid particle is a tracer that moves with the local flow velocity. Considering $X^+(t,Y)$ the position at time $t$ of a tracer initially located at the point $Y$ -called Lagrangian coordinate, or material coordinate- at a specified time $t_0$, the tracer evolves according to $\frac{\partial}{\partial t}X^+(t,Y)=U(t,X^+(t,Y))$ where $(t,x) \mapsto U(t,x)$ is here the Eulerian velocity field. The Lagrangian velocity  $U^+$ is then defined in terms of its Eulerian counterpart
\begin{align*}
U^+(t,Y)=U(t,X^+(t,Y)).
\end{align*}
In CFD, turbulence modelling gives  access only to  the averaged Eulerian velocity and other second moments according to the model. Stochastic Lagrangian approach focus on describing the dynamics of a fluid-particle -or virtual fluid parcel- and its characteristic position and instantaneous velocity $(X_t ,U_t)$, dynamics characterized by a SDE which suitably approximate the motion of $(X^+(t),U(t,X^+(t)))$. This SDE is constructed on the basis of a transport equation (or Fokker Planck equation) for the density function relative to the position and the velocity of the fluid particle. This joint probability density of the process $((X_t,U_t);0\leq t\leq T)$, denoted below by $\varrho$,  allows to interpret the Reynolds operator $\av{\cdot}$ as in \eqref{eq:Reynolds_average_interpretation}, the expectation symbol $\EE$ being notably associated to the probability measure $\PP$, under which the Brownian motion $(B_t)$ driving the SDE is defined.

A reference stochastic Lagrangian model is the \emph{Generalized Langevin Model} (GLM),
\begin{align}\label{eq:GLM}
\left\{
\begin{array}{l}
d X^{(i)}_t =U^{(i)}_t dt,\,1\leq i\leq 3, \\
d U^{(i)}_t  = -\frac{1}{\rho} \frac{\partial \av{P}}{\partial x_i} (t,X_t) dt +  \sum_{j=1}^3   G_{i j}(t,X_t) (U^{(j)}_t- \av{U^{(j)}}(t,X_t)) dt+ \sqrt{ C_0(t,X_t)\pdissip(t,X_t) } d B^{(i)}_t,\\
\text{$\av{P}$ is the mean pressure, $(B_t=(B^{(i)}_t),\,t\geq 0)$ is a standard 3D-Brownian motion,}
\end{array}
\right. 
\end{align}
designed to be consistent with the Navier-Stokes equations through formal developments on the Fokker-Planck equation derived from  \eqref{eq:GLM}. This reference model involves a  generalized return-to-equilibrium tensor $G_{i j}(U^{(j)}- \av{U^{(j)}})$ and the dissipation rate $\pdissip$ defined by
\[
\pdissip(t,x)=\tfrac{1}{2}\nu \sum_{i,j=1}^3 \av{ ({\partial_{x_j}} u'^{(i)} + {\partial_{x_i}} u'^{(j) })^2}(t,x),
\]
with $\nu$ the kinematic viscosity of the fluid. 

The probability measure  supporting the Brownian motion $B$ and the solution of the GLM equation \eqref{eq:GLM} 
allow us to define  the mean velocity field of the flow as the conditional expectation 
\[
\av{U^{(i)}}(t,x)= \Ee[U^{(i)}_{t}| X_{t}=x],\]
and to define the Lagrangian turbulent part of the instantaneous velocity:
\[
u'_t = U_t - \av{U}(t,X_t).
\]
Then the stochastic Lagrangian equation for the process $((X_t,u'_t);~t\geq 0)$ is 
\begin{equation}\label{eq:SDM}
\left\{
\begin{aligned}
&d X^{(i)}_t =(u'^{(i)}_t +\langle U^{(i)}\rangle(t,X_t)) dt,\,1\leq i\leq 3, \\
&d u'^{(i)}_t  = \sum_{j=1}^3   G_{i j}(t,X_t) u'^{(j)}_t dt+ \sqrt{ C_0(t,X_t)\pdissip(t,X_t) } d B^{(i)}_t,
\end{aligned}
\right.
\end{equation}
where the dimensionless coefficient $C_0(t,x)$ may be a constant or may depend on the local values of the Reynolds stress tensor, the dissipation rate and the drag force ${\partial_{x_j}}\av{U^{(i)}}$.

For simulation purposes (see for instance  \cite{BeBoChJaRo10} 
for a general presentation and application to downscaling methods,  
\cite{B16, B18} for wake turbine and wind farm simulation), 
a discrete time particle method can be applied 
to stochastic Lagrangian models to obtain an approximation for the mean flow components. 
The main difficulty of these approximation methods regards the non-linearity driven by the conditional expectation, and can be overcome using a kernel-regularization approach and particles in cell methods \cite{BeBoChJaRo10,B18}. 

In the literature, several models with distinct referential values for the parameters exist. 
Among them, we mention the Rotta model related to simplified Langevin model (SLM),  
the  isotropization-to-production model (IPM),  also the Shih-Lumley model (SL), and the Launder-Reece-Rodi Model (LRR). We further refer the interested reader to \cite{P93} 
to a detailed  presentation of these -and other- models.

\subsection{Models for the tensor $G$}\label{subsec:TensorG}

We present a selection of possible  models, chosen for their use in the ABL modelling, characterized by 
high-Reynolds number and anisotropic flows. 

\paragraph{The Simplified Langevin model (SLM) }  
Proposed by Pope  \cite{Pope85}, and later seen as a particular case of GLM \cite{P00},  the SLM assumes the tensor  $G$  as an isotropic (diagonal) tensor given by 
\begin{equation}\label{eq:GLMClosure}
G_{i j}(t,x) = -\left(\frac{1}{2}+\frac{3}{4}C_0\right)\frac{\pdissip}{\tke}(t,x) \ \delta_{ij},
\end{equation}
where $\delta_{ij}$ is the Kronecker delta\footnote{$\delta_{ij}= 1$ if $i=j$, $0$ otherwise.} and  $\tke(t,x)$ is the mean turbulent kinetic energy defined by 
\begin{equation}\label{eq:lagrang_tke}
\begin{aligned}
\tke(t,x)
&=\tfrac{1}{2}\sum_{i=1}^{3}\Ee[(u'^{(i)}_{t})^{2}|X_{t} =x].
\end{aligned}
\end{equation}
It has been shown (e.g. \cite{P00}) that the model tensor 
\eqref{eq:GLMClosure} is consistent with the simple Rotta's return-to-isotropy model for the dynamics of the Reynolds stresses $\av{u'^{(i)} u'^{(j)}}$ (so it is called simple). The condition for this consistency is to identify\footnote{ A more general relationship 
(see \cite[pp. 398]{DurbinSpeziale-94}) is given by $C_0=\frac23( \tfrac{\tke}{\tau\pdissip}C_R -1)$ for $\tau$ some characteristic time scale. In the case of Rotta's closure we have $\tau=\pdissip/\tke$.} the so-called Rotta's constant  $C_R$ with 
\begin{equation}\label{Relation C0 CR}
C_R= 1 + \tfrac32 C_0. 
\end{equation}

Notice that, the constant $C_0$ in the model \eqref{eq:SDM} must be non-negative in order to have a well-defined diffusion term. 
Thus, in view of \eqref{Relation C0 CR}, $C_R$ must satisfy  the Rotta condition, $C_R\geq1$~\cite{DurbinSpeziale-94}.

\paragraph{The isotropization-to-production (IP) model for homogeneous turbulence} 
This GLM version  is constructed to be consistent with the LRR-IP model \cite{P94,P00} (see also \cite{DurbinSpeziale-94,DrePop-98}): 
\begin{equation}\label{eq:IPmodel}
\begin{aligned}
G_{i j}(t,x) &= -\frac{C_{R}}{2}   \dfrac{\pdissip}{\tke}(t,x) \delta_{ij} + C_{2}   \partial_{x_{j}} \av{U^{(i)}}(t,x).
\end{aligned}
\end{equation}
The consistency with the LRR-IP model is ensured by the choice of a diffusion coefficient computed from the turbulent production tensor $\productke_{ij}$ (and coming with a realizability constraint \cite{DurbinSpeziale-94}) : 
\begin{equation}\label{eq:IPclosure}
\begin{aligned}
C_{0} \ \pdissip(t,x)  &= \tfrac{2}{3} \left(C_{R}\;\pdissip(t,x)  + C_2\;   \productke(t,x) -  \pdissip(t,x) \right), 
\end{aligned}
\end{equation}
where, adopting Einstein notation, $\productke(t,x)= \tfrac{1}{2}  \productke_{ii}(t,x)$ is the turbulent production derived from  
\begin{align*}
\productke_{ij} :=  - \sum_{l=1}^3 \av{u^{\prime(i)}u^{\prime(l)}}
\partial_{x_l} \av{U^{(i)}}
+ \av{u^{\prime(j)}u^{\prime(l)}}
{\partial}_{{x}_l} \av{U^{(j)}}.
\end{align*}
Notice that $C_0$ in the definition of the tensor $G$ for the IP model \eqref{eq:IPclosure} is no more considered as constant.  
Notice also that the realizability constraint requiring $C_0$ to be positive reads as $C_R\geq 1- C_2 \frac{\productke}{\pdissip}$. The SLM  is a particular case of the IP model considering $C_2=0$. 

\paragraph{Elliptic blending model} 
The purpose of this model is to add an anisotropic  effect  near the ground  (or wall effect, see e.g. \cite{Durbin93, durbin2002elliptic,manceau2002elliptic,Wacl}). It starts with the following form of $G$ and diffusion coefficient: 
\begin{equation}\label{eq:EBmClosure}
G_{i j}=  - \gamma_{ij} - \frac{1}{2} \frac{\pdissip}{\tke}\delta_{ij},\qquad
C_{0}~\pdissip  = \sum_{i,j=1}^3 \frac{2}{3}\gamma_{ij} \av{u^{\prime(i)}u^{\prime(j)}},
\end{equation}
where the tensor $\gamma$ is 'interpolated' between a near wall Reynolds stress model and the homogeneous 
model already described \cite{durbin2002elliptic}.   
In its simplified version  \cite{manceau2002elliptic}, 
$$\gamma_{ij}(t,x) = (1 - \alpha(t,x) \tke(t,x)) \gamma_{ij}^{\text{{\it w}}}(t,x)  
+ \alpha(t,x) \tke(t,x)  \gamma_{ij}^{\text{{\it h}}}(t,x),$$ 
where $\alpha(t,x)$ is the elliptic blending coefficient given as the solution of the elliptic partial 
differential equation:
\begin{align*}
L^2 \nabla^2 \alpha(t,x) - \alpha(t,x) &= -\frac{1}{\tke}(t,x),
\end{align*}
with $L$ being a characteristic length scale (possibly varying in time and space).  
We refer to \cite{Wacl} and the references therein for the modelling of $\gamma_{ij}^{\text{{\it w}}}$ and $\gamma_{ij}^{\text{{\it h}}}$. 
Notice  that, here again,  $C_0$  is no more considered as constant and depends in particular on $\alpha$.

\begin{rem}\label{rem:on the value of C0}
It should be noted that in both  Lagrangian and Eulerian approaches, the values of the coefficients $C_0$ and $C_R$, known as the Kolmogorov constant and the Rotta constant,  might vary according to the model and context. For instance, the values $C_0=2.1$ and $C_R=4.15$ are suggested in \cite{P93}. 
Nevertheless, the IP model (see Equation \eqref{eq:IPclosure}) may assume  $C_R=1.5$ or $C_R=1.8$, \cite{DurbinSpeziale-94}. 
Similarly for the LRR model, $C_R=1.5$, \cite{P93}. From \cite[Appendix]{P93}, we quote that $C_R = 1.0$ corresponds to no-return-to-isotropy, while values from $1.5$ to $5.0$ have been suggested by different authors. This implies that the values of $C_0$ might vary between $\tfrac13$ and $\tfrac83$ (using the identification \eqref{Relation C0 CR}).
\end{rem}

\subsection{Parameterization of the dissipation rate}\label{subsec:TurbulentClosure}
The GLM or its Fokker Planck equation contains no information on the turbulence timescale --unless thought the addition of a direct dissipation state variable \cite[12.5]{P00}-- 
 and needs to be supplemented  with a model for the dissipation rate of the kinetic energy. 
We present below only two simple models based on the relation $\tke-\pdissip$, but refinements are proposed in the literature. 

\paragraph{The mixing length parameterization} This relation is classically used in the ABL: 
\begin{equation}\label{eq:closure mixing length}
\pdissip(t,x)=\frac{C_\pdissip}{\lm}\tke^{\frac{3}{2}}(t,x),
\end{equation}
where $C_\pdissip$ is a  constant to be chosen, $\lm$ is   a characteristic length scale called {\emph mixing length}.  In general, near the ground,  the value of the mixing length $\lm$ does vary linearly with respect to $\kappa z$, where $z$ denotes the  distance to the wall from $x$, and  $\kappa$ is the Von K\'arm\'an constant (see \cite{BeBoChDrRoSa-09} and the references therein for further discussion). 

\paragraph{The turbulent viscosity parameterization} In the $\tke-\pdissip$ model framework \cite{P00}, the turbulence eddy viscosity is given by  ${\nut}=C_\mu\frac{\tke^2}{\pdissip}$ with $C_\mu=0.09$. In addition, near the wall (behind a height  of 150\xspace\si{\meter}, to give an order of magnitude in the ABL) $\nut$ is related with the velocity friction thought $\nut(z) = \ustar \kappa z$, leading to the following parameterization
\begin{equation}\label{eq:closure turbulent viscosity}
\pdissip(t,x)=\frac{C_\mu}{\ustar \kappa z}\tke^2(t,x).
\end{equation}
The velocity friction  $\ustar$ depends on the flow and on the ground  (the roughness length of the terrain among others factors).

\section{Reduced Lagrangian model for the instantaneous turbulent kinetic energy}\label{sec:TKEmodeling}

As already mentioned  in the introduction, modelling characteristics  of the wind at a fixed location  is of great interest  for many  applications. In this context, fluid-particle based turbulent flow models offer a simple way  to reduce from '3D+time' model to 'time' model. For a Lagrangian  stochastic model,  the '3D' field notion  is a conditional expectation that can be approximated at a fixed location by a calibration procedure. A step in this direction has been taken by  Baehr \cite{Baehr-10} which was interested in a filtering methodology for one  point (eventually mobile) of wind  observation and 
which proposed to use local random model such as Lagrangian models combined with nonlinear filtering techniques to clean wind measurements.
These ideas were applied later to Doppler wind LIDAR observations \cite{SuBaDa-11, baehr2012retrieval,rottner2017stochastic}.

In our case, we assume (filtered) data  available at a fixed location $x_\text{obs}$ (see Section \ref{sec:Data}), ready to be used to calibrate a time-model.  The idea developed in this paper is to somehow reduce the stochastic Lagrangian modelling  into instantaneous turbulent kinetic energy as a time-model,  while considering the uncertainty of the physical parameters of the model. 
Renewable energy development has raised  growing interest in energy production  forecasting and modelling. Several approaches are taking into account 
the uncertain nature of the forecast through stochastic modelling. 
For recent examples based on stochastic diffusion models, we refer for instance to \cite{BGGK17,murata2018modeling} for the construction of probabilistic forecast of solar irradiance, with simple linear drift form. In \cite{BGGK17},  a diffusion process is proposed, resulting from a deterministic forecast, and  the parameters involved are estimated by means of a variance-autocorrelation fitting. The diffusion coefficient used to model the solar irradiance has the power form $x^\alpha(1-x)^\alpha$, with  the constant $\alpha$ to be calibrate from data.
Concerning wind energy production, \cite{arenas2020stochastic} has proposed a data-driven Ornstein-Uhlenbeck model describing the wind speed on a scale of seconds. On the other hand, the Weibull distribution has been widely used in wind energy and other renewable energy sources \cite{carta2009review}, where the main issue has been the estimation of the distribution coefficients. Based on this, in \cite{bensoussan2016cox}, a stochastic model of the squared norm of the wind velocity as a CIR process have been proposed with coefficients to be calibrate.  These models have in common the fact that they are inspired from some a priori on their parametric forms.

In this section we derive a physical-based model describing the  instantaneous TKE localized at a fixed location~$\xobs$.  

\subsection{First step: localized dynamics of the norm of the turbulent  velocity} 
We start from the SLM equations (\ref{eq:SDM},\ref{eq:GLMClosure}), and applying the It\^o lemma, we obtain a first equation for the squared-norm of the turbulent velocity that we force to be localized at~$\xobs$
\begin{align}\label{eq:Ito_for_turbulent-velocity}
d \|u'_t\|^2 
&=- 2\left(\frac12+\frac34 C_0\right)\frac{\pdissip}{\tke}(t, \xobs) \|u'_t\|^2 dt+ 3 C_0 \pdissip(t,\xobs) dt+ 2\sqrt{C_0 \pdissip(t,\xobs) }\sum_i u^{\prime(i)}_t d B^{(i)}_t. 
\end{align}
Equation \eqref{eq:Ito_for_turbulent-velocity} is a version of the  fluid-particle model (\ref{eq:SDM},\ref{eq:GLMClosure}),  conditioned at each time $t$ by the event that the fluid-particle position $X_t$  is going through the point $\xobs$.
We proceed by defining the \emph{instantaneous turbulent kinetic energy}  at $\xobs$ as the stochastic process $(q_t;~ t\geq 0)$ given as the formal solution to  \eqref{eq:Ito_for_turbulent-velocity}.
In particular, we get the relation 
$$\tke(t,\xobs) = \frac12\EE[q_t].$$

In order to simplify notation, hereafter, $\tke_t$ (respectively $\pdissip_t$) will denote the mean turbulent kinetic energy at the fixed position $\tke(t,\xobs)$ (respectively $\pdissip(t,\xobs)$).  
According to the Levy's characterization Theorem of Brownian motion (see e.g. \cite{Protter-04}), the process $W_t  = \sum_i \int_0^t \frac{u_s^{\prime(i)}}{\|u^\prime_s\|} d B^{(i)}_s$ identifies as a one-dimensional Brownian motion. Further, assuming that the initial turbulent energy $q_0$ is known, we obtain the following SDE for the dynamics of the process $(q_t; t\geq 0)$:
\begin{align}\label{eq:kinetic_process}
d q_t =- C_R\frac{\pdissip_t}{\tke_t} q_t dt+ 3 C_0 \pdissip_t dt+ 2\sqrt{C_0 \pdissip_t }\sqrt{q_t}dW_t, \quad q_0 \text{ given.}
\end{align}

For this first SDE obtained from SLM, all that remains to be specified a priori is the dissipation $\pdissip_t$. However, in wind energy application context,  the fluid particle model is used in the vicinity of the ground,  and anisotropic effects cannot  be simply neglected, as well as buoyancy effects due to temperature variation.  In the one hand, these effects are complex to model. On the other hand, the mean long time estimate obtained in Lemma \ref{lem:longtime} below shows that the behaviour of the SLM alone is the systematic return to zero turbulence,  in contradiction with the  observations (see Section \ref{sec:Data}). This behaviour evidences the need of an additional  source term in the SDE accounting for the regime of turbulence production near the wall. In order to replicate this effect, we consider an extra drift term $\gamma$  accounting for the extra-diagonal (non-isotropic) contributions of the tensor $G_{ij}$ not retained in the SLM:
\begin{align}\label{eq:turbulentVelo-base-bis}
d q_t =\gamma \ dt - C_R\frac{\pdissip_t}{\tke_t} q_t \ dt+ 3 C_0 \pdissip_t dt+ 2\sqrt{C_0 \pdissip_t }\sqrt{q_t}dW_t, \quad q_0 \text{ given.}
\end{align}
In Section \ref{sec:Calibration}, we first propose a calibration procedure for $\gamma$, updated in time according to an intermediate time scale,  and next a higher one. We also propose to use the (renormalized)  wind turbulence intensity as a candidate to approximate $\gamma$. 
In order to simplify the discussion on the wellposedness analysis and the computation of the a priori estimators, in this first part we consider $\gamma$ as constant (in the sense that the variation in time of $\gamma$  is at a lower frequency than those of $q_t$).

\subsection{Second step: incorporating the dissipation parameterization} 

In order to close the term $\pdissip_t$ in the SDE \eqref{eq:turbulentVelo-base-bis}, we choose the mixing length  closure hypothesis \eqref{eq:closure mixing length} used in ABL  modelling.  
 For simplicity, we consider $\lm$ constant with $\lm=\kappa \zlm$, where $\kappa$ is the Von K\'arm\'an constant and $\zlm$ is the height at which the measurements were taken:
$$\pdissip_t = C_{\alpha}\tke_t^{3/2} = \dfrac{ C_\alpha}{2\sqrt{2}} \EE^{3/2}[q_t], $$
where $C_\alpha:=\frac{C_\pdissip}{\kappa z(\xobs)}$,  and where we use here and later the notation   $\EE^{\beta}[q_t]=(\EE[q_t])^{\beta}$ for any $\beta>0$.  Introducing this relation in Equation \eqref{eq:turbulentVelo-base-bis}, we obtain the following CIR-type {\bf stochastic mean-field  TKE model}:
\begin{equation}\label{eq:tke_mixinglength_bis2}
d q_t = \gamma dt - C_R\frac{C_\alpha}{\sqrt{2}}q_t{\EE^{1/2}[q_t]} dt+ 3 C_0\frac{C_\alpha}{2\sqrt{2}} \EE^{3/2}[q_t] dt+ \sqrt{\sqrt{2}C_0 C_\alpha}\ \EE^{3/4}[q_t]\sqrt{q_t}dW_t,\quad q_0 \text{ given.}
\end{equation}
\begin{rem}\label{rem:DeterministicTKE}
Formally, taking expectation on both sides of \eqref{eq:tke_mixinglength_bis2} and using \eqref{Relation C0 CR} leads to the ordinary differential equation for $\tke_t=\EE[q_t]/2$
\begin{align*}
\frac{d \tke_t }{dt}=
\left(\frac{\gamma}2-\pdissip_t\right)\ind_{\{\tke_t >0\}}, \quad \tke_0=\frac{q_0}{2}
\end{align*}
which corresponds to the classical equation  form for $\tke$, involving the turbulence production terms (here reduced to $\gamma$) deduced from RANS equations with a $\tke$-$\pdissip$ closure model \cite{DurbinSpeziale-94}. As the true production process is replaced by the model constant $\gamma$,  the indicator function in the right hand-side prevents $\tke$ to become non-positive. Coming back to the autonomous ODE form, 
\begin{equation}\label{eq:determinsticTKE}
\frac{d \tke_t }{dt}=
\left(\frac{\gamma}2-C_\alpha \tke_t^{3/2}\right)\ind_{\{\tke_t >0\}}, \quad \tke_0=\frac{q_0}{2},
\end{equation}
under the conditions that $q_0$ and $\gamma$ are non-negative,  Lemma \ref{lem:longtime} ensures that a unique non-negative solution to \eqref{eq:determinsticTKE} exists for all $t\geq 0$.
\end{rem}

The SDE  \eqref{eq:tke_mixinglength_bis2}  features two main characteristics. On the first hand, the SDE can be classified as of CIR type, since the dependence in $q$ of the diffusion coefficient is in a square root form, a non-Lipschitz dependence well studied in the literature (see \ref{app:wellposedness_MF_model} and \cite{IkedaWatanabe}). On the second hand, the SDE can be also classified as of mean-field type, since both drift and diffusion coefficients depend nonlinearly on the law's process through the expectation term $\EE[q_t] = 2 \tke_t$. The combination of these features categorizes the model as a  McKean-Vlasov SDE (we refer the interested reader to the surveys \cite{B05} and \cite{JW17} for some review on  theoretical aspects related to these models and their numerical approximation). At first sight, justifying the existence of a global-in-time solution to the equation \eqref{eq:tke_mixinglength_bis2} is not trivial.  While positiveness criteria for solution of SDEs with polynomial coefficients are well-understood, the dependency of the coefficients in fractional powers of $\EE[q_t]$ makes the application of such criteria not immediate. Nevertheless, an opportune aspect of the model lies in the dynamics  of $\EE[q_t]$, which as mentioned in Remark \ref{rem:DeterministicTKE} is driven by the autonomous ODE \eqref{eq:determinsticTKE}. We deduce the  existence and trajectorial uniqueness of  the positive solution to the SDE \eqref{eq:tke_mixinglength_bis2} from the a priori behaviour of $\tke$ stated in Lemma \ref{lem:longtime}. 

\begin{prop}\label{prop:SolNonLinear}
Consider the positive parameter set  $C_\alpha,C_0$, and let $\gamma \geq 0$. Then, there exists a unique strong positive solution $(q_t;~t\geq0)$ to the McKean-Vlasov SDE \eqref{eq:tke_mixinglength_bis2}. 
\end{prop}

The proof of Proposition \ref{prop:SolNonLinear} is postponed to \ref{app:wellposedness_MF_model}.

\begin{rem}
A negative production term $\gamma <0$ is not forbidden in the model, if it is designed to decrease to zero with $\tke$. For the sake of simplicity we assume $\gamma\geq 0$ as a constant updated on time sub-periods, and let the extension to possible function $q\mapsto \gamma(q)\in\mathbb{R}$ to future improvement. 
\end{rem}

\paragraph{On the dissipation of the energy} 

\begin{lem}\label{lem:longtime} Assume that there exists a solution $(q_t, t\geq 0)$ to \eqref{eq:tke_mixinglength_bis2}. Then, for any $\gamma\geq 0$ the $p$th-moment is bounded (uniformly in time), for all $p\geq1$, i.e. there exist some constants $\overline{C},\underline{C}>0$ independent on time such that
\begin{equation}\label{eq:p_moment_McKean}
\underline{C}~{\ind}_{\gamma>0} \leq \sup_{t\geq 0}\EE[q_t^p] \leq \overline{C}.
\end{equation}
In  particular for $p=1$, we have
\begin{equation}\label{eq:first_moment_McKean}
\min\big\{q_0,\big(\frac{\sqrt{2}\gamma\ind_{\{\gamma >0\}}}{C_\alpha}\big)^{2/3}\big\} \leq \sup_{t\geq 0}\EE[q_t]\leq \max\big\{q_0,\big(\frac{\sqrt{2}\gamma\ind_{\{\gamma >0\}}}{C_\alpha}\big)^{2/3}\big\},
\end{equation}
with long-time behaviour 
\begin{equation}\label{eq:mean_TKE_limit_gamma}
\lim_{t\rightarrow+\infty}\EE[q_t] = \left(\frac{\sqrt{2}\gamma \ind_{\{\gamma >0\}}}{C_\alpha}\right)^{2/3}.
\end{equation}
\end{lem} 
This lemma emphasises the role of  the (non-dissipative) source term $\gamma$ in the model. In particular when $\gamma>0$,  the instantaneous norm of the velocity fluctuation $q_t$ has the  non trivial limit \eqref{eq:mean_TKE_limit_gamma}. The proof is detailed in  \ref{app:proof_of_lemma}.

\subsection{Third step: deriving a CIR-like model for the instantaneous TKE}\label{sec:CalibrationNonErg}

In this last step, we move from the McKean nonlinear model  \eqref{eq:tke_mixinglength_bis2} to a  linear one.  The reason for this is that calibration methods for McKean processes are still little developed and limited to particular cases.  Instead, we make use of the long-time convergence of the moment $\EE[q_t]$ to simplify the dynamics \eqref{eq:tke_mixinglength_bis2}. More precisely, from the nature of our observations (see Section \ref{sec:Data} below), we assume that the production term $\gamma$ is always positive in order to get a non zero long-time limit. Then, taking one by one the coefficients of \eqref{eq:tke_mixinglength_bis2}
\begin{align*}
& \lim_{t\rightarrow +\infty}  C_R \frac{C_\alpha}{\sqrt{2}} (\EE[q_t])^{\tfrac{1}{2}} = C_R 
(\frac{C_\alpha^2\gamma}{2})^{1/3} := \Theta(C_\alpha,\gamma) \\
& \lim_{t\rightarrow +\infty} 
\gamma + 3 C_0 \frac{C_\alpha}{2\sqrt{2}}  (\EE[q_t]) ^{\frac{3}{2}} =  \gamma +\frac{3}{2} C_0 \gamma := \mu(C_\alpha,\gamma)\Theta(C_\alpha,\gamma)\\
& \lim_{t\rightarrow +\infty}  \sqrt{\sqrt{2}C_0 C_\alpha}\ (\EE[q_t])^{\tfrac34}= \sqrt{ 2 C_0 \gamma} := \sigma(\gamma).  
\end{align*}
By doing so, we 
erase a part of the complexity of the transition to equilibrium, obtaining  the following  {\bf CIR model for the instantaneous TKE}:
\begin{equation}\label{eq:tke_equilibrated_CIR}
dq_t = \Theta(C_\alpha,\gamma)\left(\mu(C_\alpha,\gamma)-q_t\right)dt + \sigma(\gamma)\sqrt{q_t}dW_t, \quad q_0\text{ given, and }\gamma>0,
\end{equation}
where
\begin{align*}
\Theta(C_\alpha,\gamma) =C_R\big(\frac{C_\alpha^2\gamma}{2}\big)^{1/3},
&\qquad \mu(C_\alpha,\gamma) = \big(\sqrt{2}\frac{\gamma}{C_\alpha}\big)^{2/3} ,\qquad \sigma(\gamma) = \sqrt{2{C_0}{\gamma}}.
\end{align*}

Remarkably, starting from a well established  turbulence model, and following the path of the reduction to time-model, we recover the class of diffusion models  suggested  in \cite{bensoussan2016cox}. 
Our major contribution in this context is thus to provide  an intrinsic physical meaning on the model  parameters. Further, a more descriptive dynamics for the turbulent kinetic energy can be obtained from the non-linear model \eqref{eq:tke_mixinglength_bis2}, where additional efforts are needed to propose an adequate calibration procedure. This last point will be the subject of future works.

A strong advantage of the model \eqref{eq:tke_equilibrated_CIR} is precisely that its solution, the CIR process \cite{CIR}, have been widely studied,  although mainly in mathematical finance. For CIR processes, the parameter  $\Theta>0$ is the speed of adjustment to the mean $\mu$, and the well-known assumption  on the parameters
\[2\Theta(C_\alpha,\gamma)\mu(C_\alpha,\gamma)\geq \sigma^2(\gamma),\]
(from Feller's criteria \cite[Theorem 5.5.29]{KarShr-91}) excludes the trajectories to visit zero, through a pushing upward effect of the drift against the diffusion when the process gets close to zero. In the particular case of Equation \eqref{eq:tke_equilibrated_CIR}, this condition reads as 
\[2 C_R  \geq 2 C_0,\]
which, by the Rotta's relation \eqref{Relation C0 CR}, is always satisfied. Then, starting from a positive $q_0$, the uniqueness of a (strictly) positive trajectory satisfying Equation \eqref{eq:tke_equilibrated_CIR} is always  guaranteed.  

The CIR process is also an ergodic process with known stationary density. The CIR process  has a well-known explicit relation with chi-squared random variables, which provides the transition density associated to the solution of Equation \eqref{eq:tke_equilibrated_CIR} in terms of Bessel functions. From this, the moments of the process $(q_t; t\geq 0)$ can be explicitly computed (see, e.g., \cite{DNS11}) with 
\begin{equation}\label{eq:cir-moments}
\EE[q_t^p] = \left(\dfrac{2\Theta}{\sigma^2\left(1-\exp\{-\Theta t\}\right)}\right)^{-p} \dfrac{\Gamma\left({2\mu\Theta}/{\sigma^2}+p\right)}{\Gamma\left({2\mu\Theta}/{\sigma^2}\right)} \ {}_1F_1\left(-p,\dfrac{2\mu\Theta}{\sigma^2};-\dfrac{2\Theta~q_0}{\sigma^2(\exp\{\Theta t\}-1)}\right),
\end{equation}
where $\Gamma$ is the gamma function and ${}_1 F_1$ is the confluent hypergeometric function\footnote{also known as Kummer's function: ${}_1 F_1 (a,b,z) = \sum_{m=0}^{+\infty} \frac{a^{(m)} z^m}{b^{(m)} m!}$, 
with $a^{(m)}= a (a+1)\ldots (a+m-1)$, the rising factorials.}  (here, we have been omitted the dependence on $(C_\alpha,\gamma)$ in the coefficients in order to simplify notations).  Then, $\EE[q_t^q]$ is finite for all $p>-\tfrac{2\mu\Theta}{\sigma^2}$. 

Notice that the computation of the moments in  \eqref{eq:cir-moments} allows to recover the identity \eqref{eq:mean_TKE_limit_gamma}, underlying the consistency between the McKean nonlinear model \eqref{eq:tke_mixinglength_bis2} with its  linear version  \eqref{eq:tke_equilibrated_CIR}.

\subsection{Observational data}\label{sec:Data}

The data used in our study was obtained from the open observation platform of SIRTA\footnote{Site Instrumental de Recherche par T\'el\'ed\'etection Atmosph\'erique.} (Haeffelin et al.~\cite{haeffelin2005sirta}). 
We used wind measurements taken at a mast of 30 meters height  with a sonic anemometer. This instrument measures the wind components at a  fixed single point, considered as an observer particle located at $z(x_{\text{obs}})=$30\xspace\si{\meter}.  
Precisely, the observational data is a family of time series containing, among other measures, the three components of the wind, registered with a frequency of 10\xspace\si{\hertz} during the year 2017. In order to take advantage of the great variability of wind situations represented in a so huge data set, without the inconvenience of processing and analysing such a large series for a summarized presentation, we have cut out the series, retaining only one day per week chosen arbitrarily and favouring daytime periods from 4 a.m. to 8 p.m. (measured in local time). 
We thus have  extracted  a time series composed of $46$ periods of 16 hours spanned for all Wednesdays of the year 2017. Our data set is then representative of the range of possible values for the temperature, degree of humidity, direction of the wind, intensity, and therefore a wide variety of wind profiles during the cycle of a year.

\begin{figure}[htbp]
\centering
{\includegraphics[width=0.5\linewidth, height=0.24\linewidth]{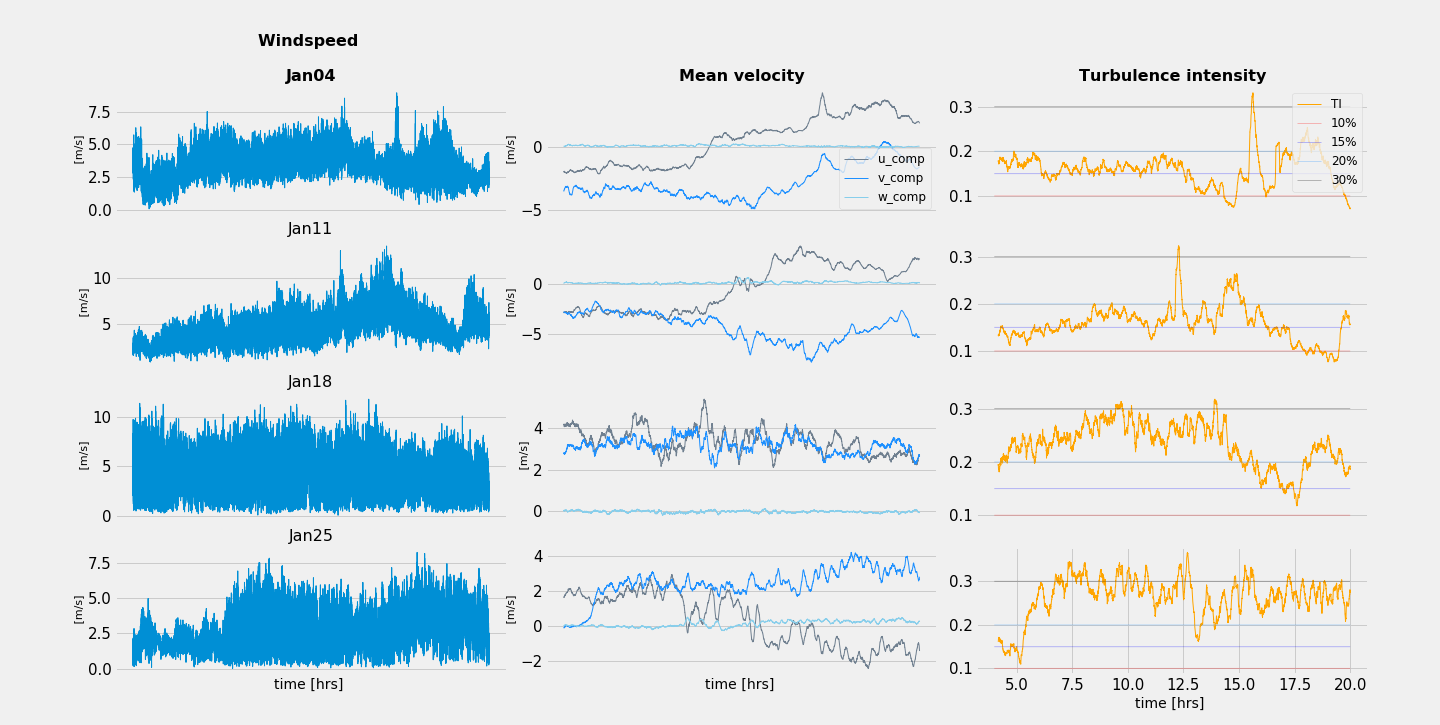}\label{january}}
{\includegraphics[width=0.48\linewidth, height=0.24\linewidth]{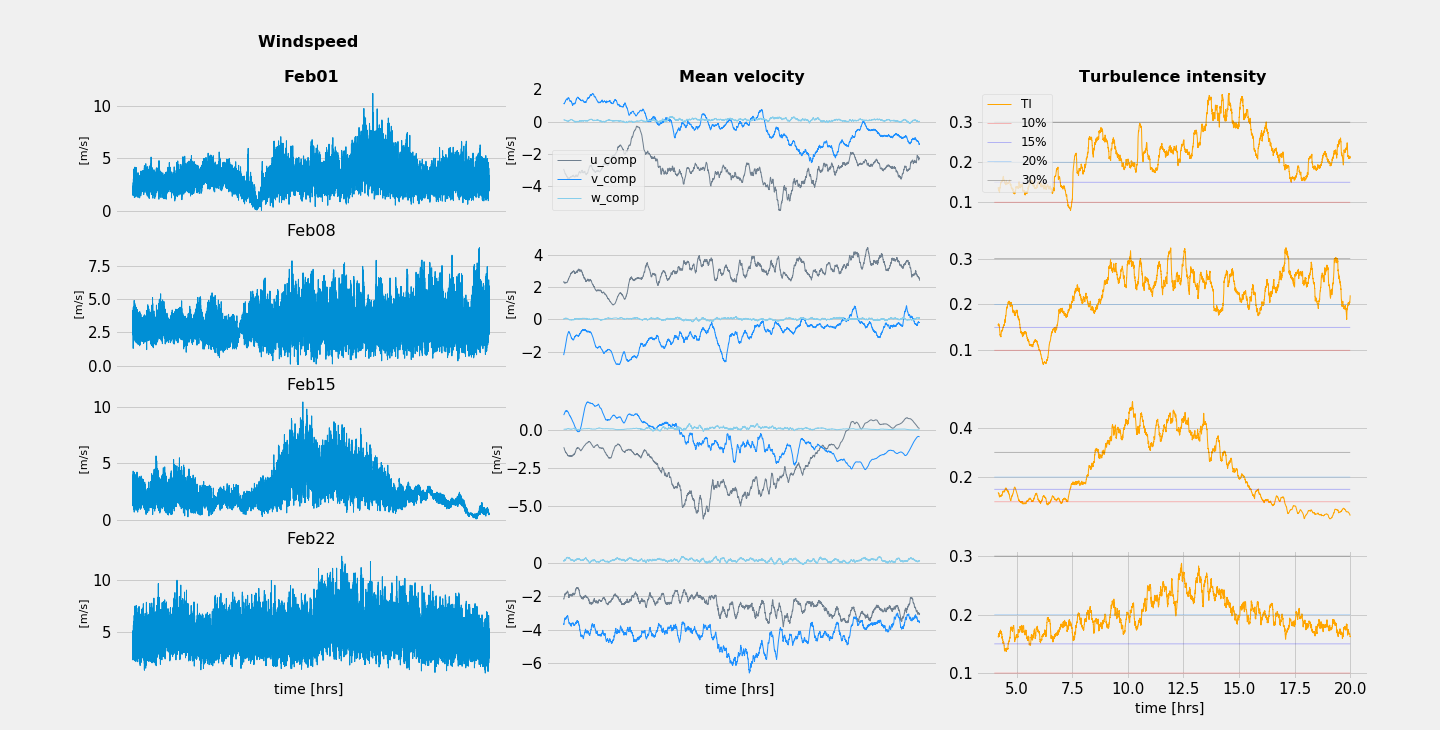}\label{february}}\\
{\includegraphics[width=0.5\linewidth, height=0.24\linewidth]{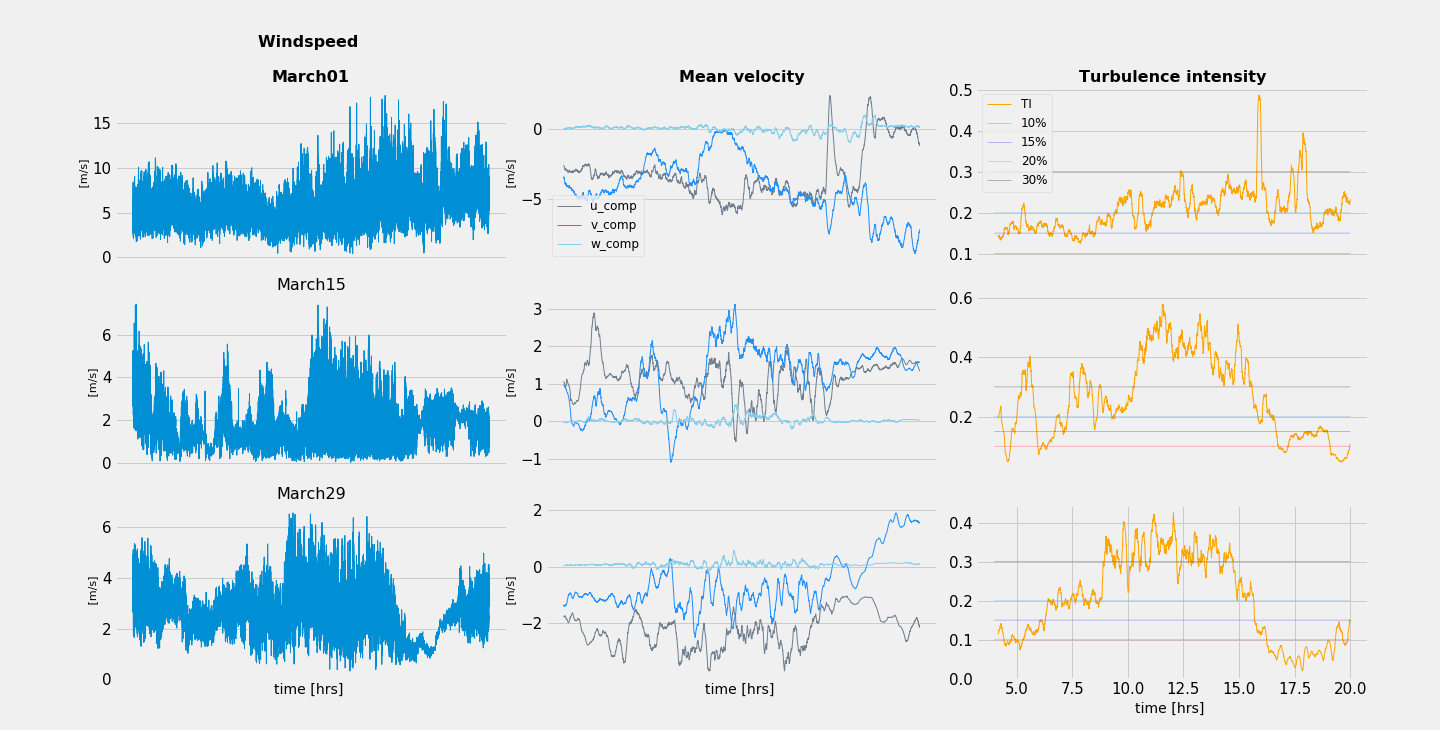}\label{march}}
{\includegraphics[width=0.48\linewidth, height=0.24\linewidth]{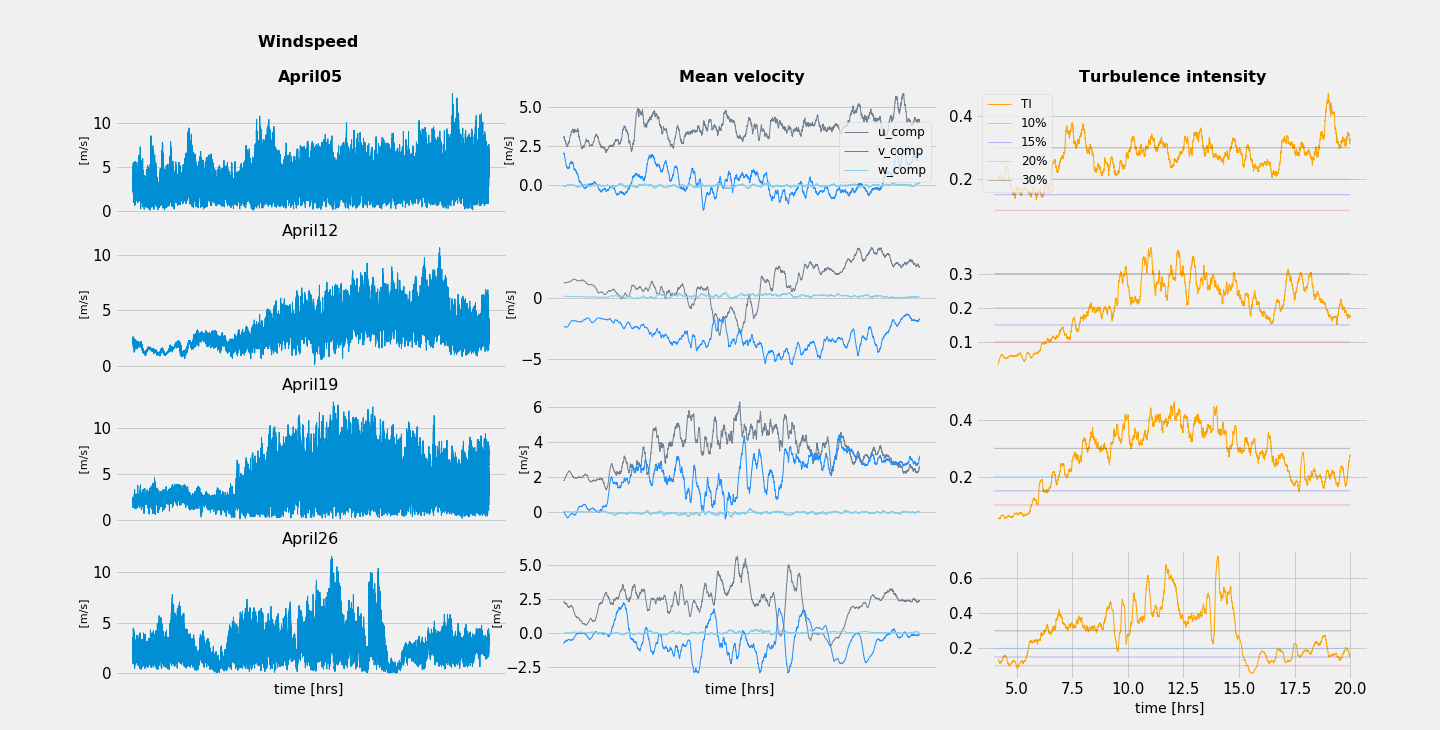}\label{april}}\\
{\includegraphics[width=0.5\linewidth, height=0.24\linewidth]{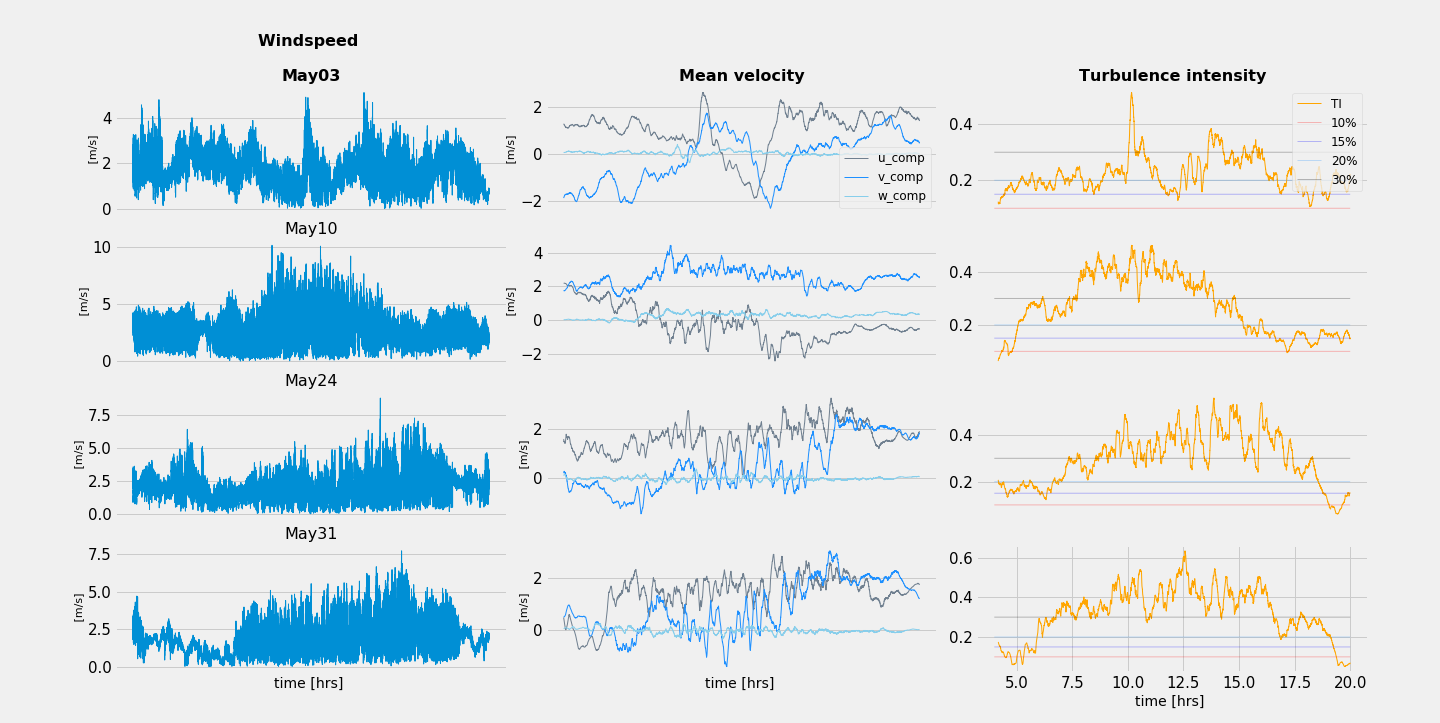}\label{may}}
{\includegraphics[width=0.48\linewidth, height=0.24\linewidth]{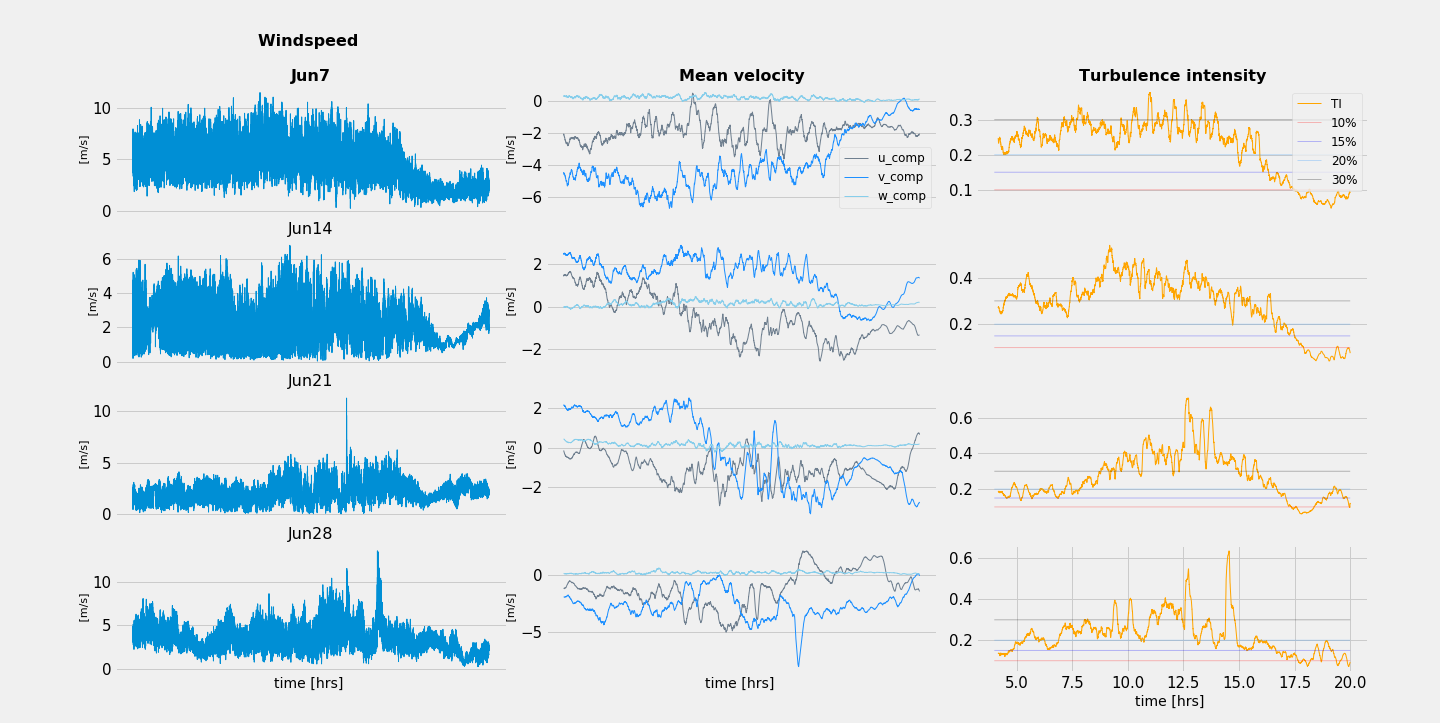}\label{june}}\\
{\includegraphics[width=0.5\linewidth, height=0.24\linewidth]{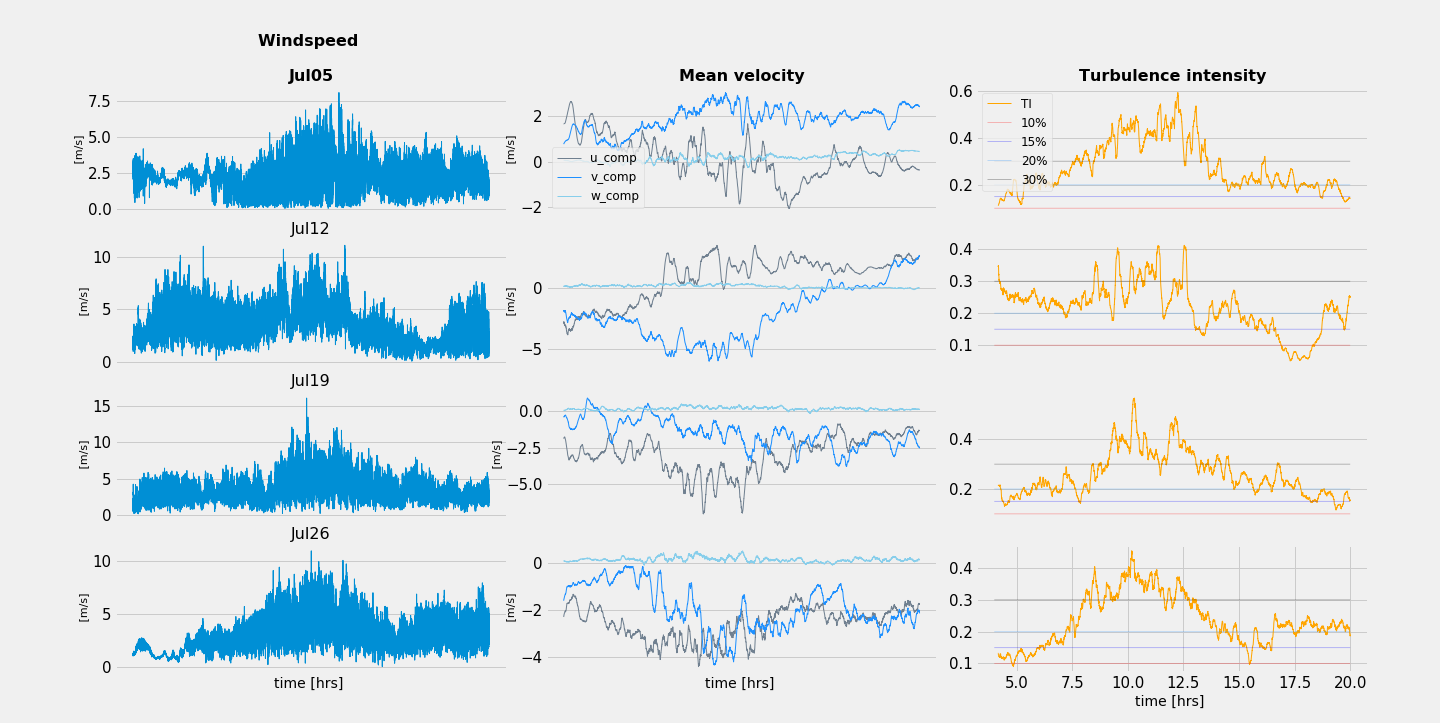}\label{july}}
{\includegraphics[width=0.48\linewidth, height=0.24\linewidth]{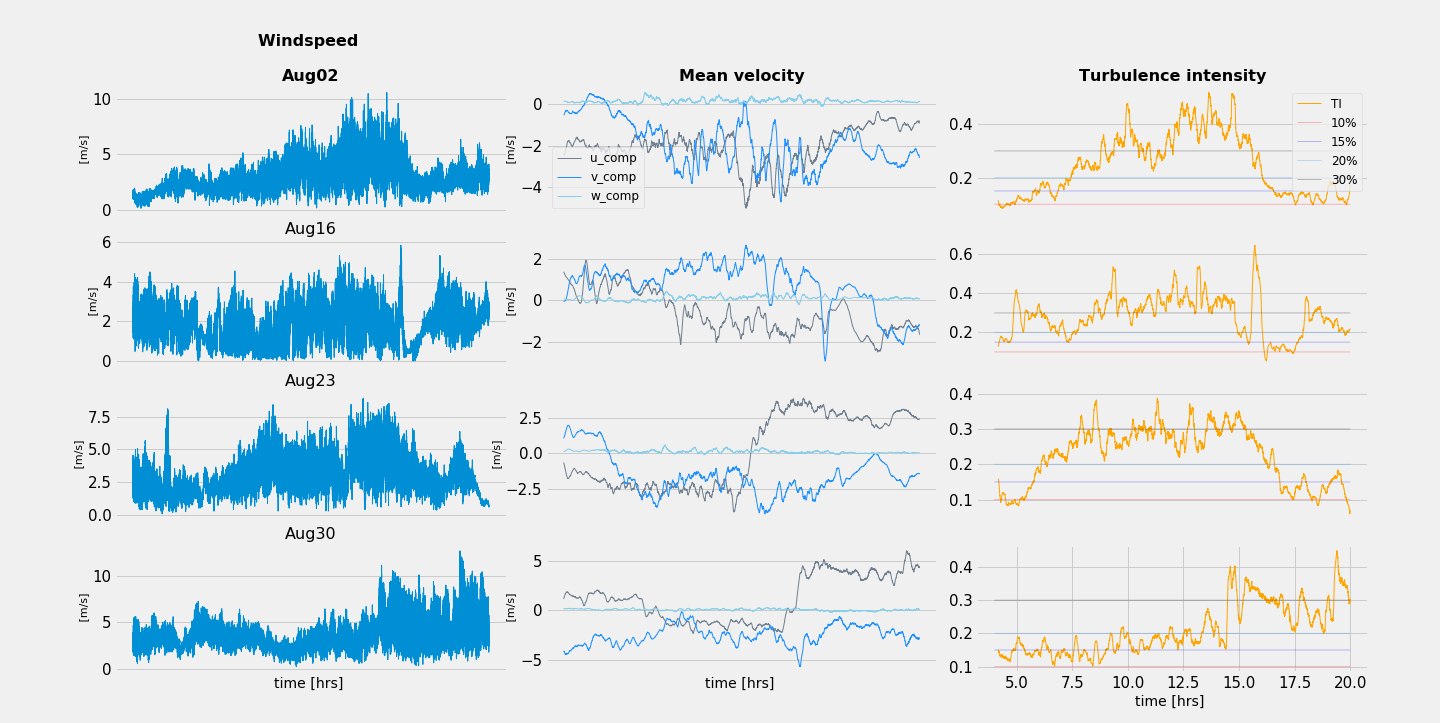}\label{august}}\\
{\includegraphics[width=0.5\linewidth, height=0.24\linewidth]{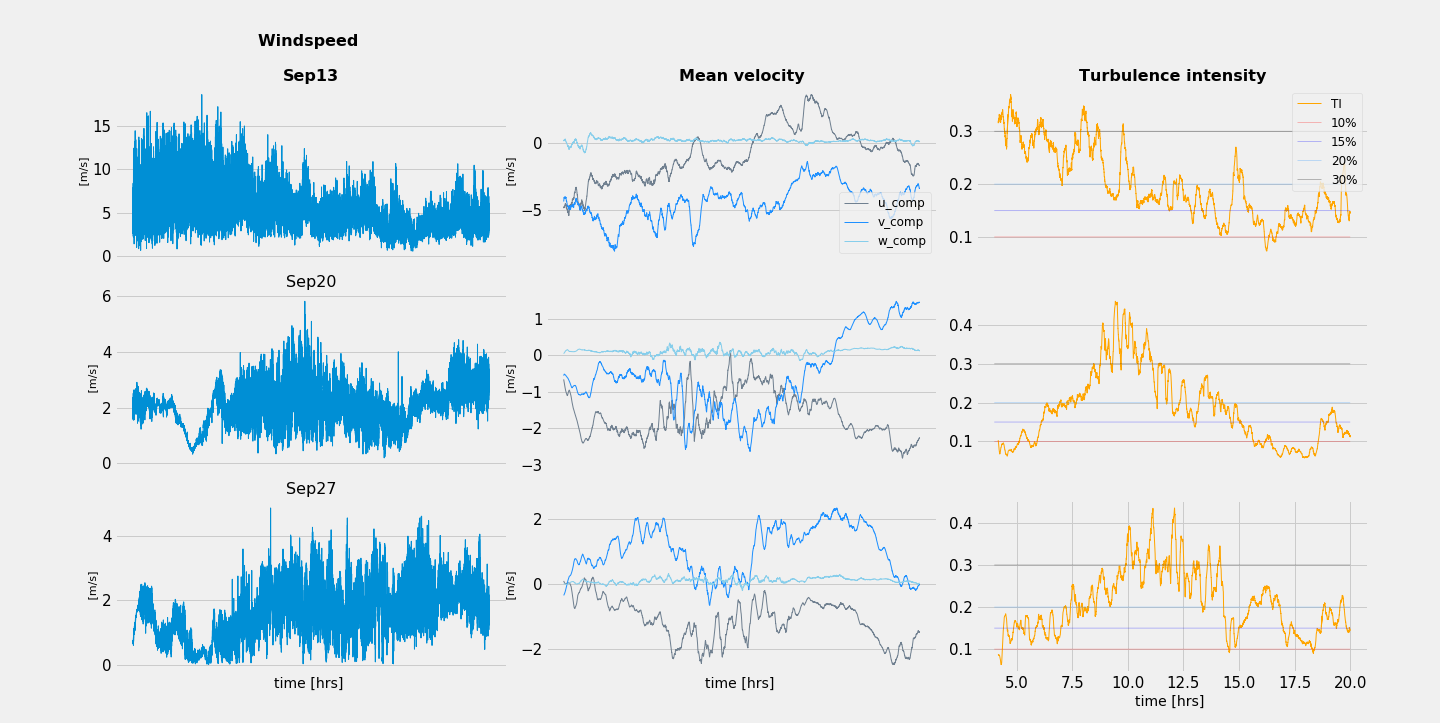}\label{september}}
{\includegraphics[width=0.48\linewidth, height=0.24\linewidth]{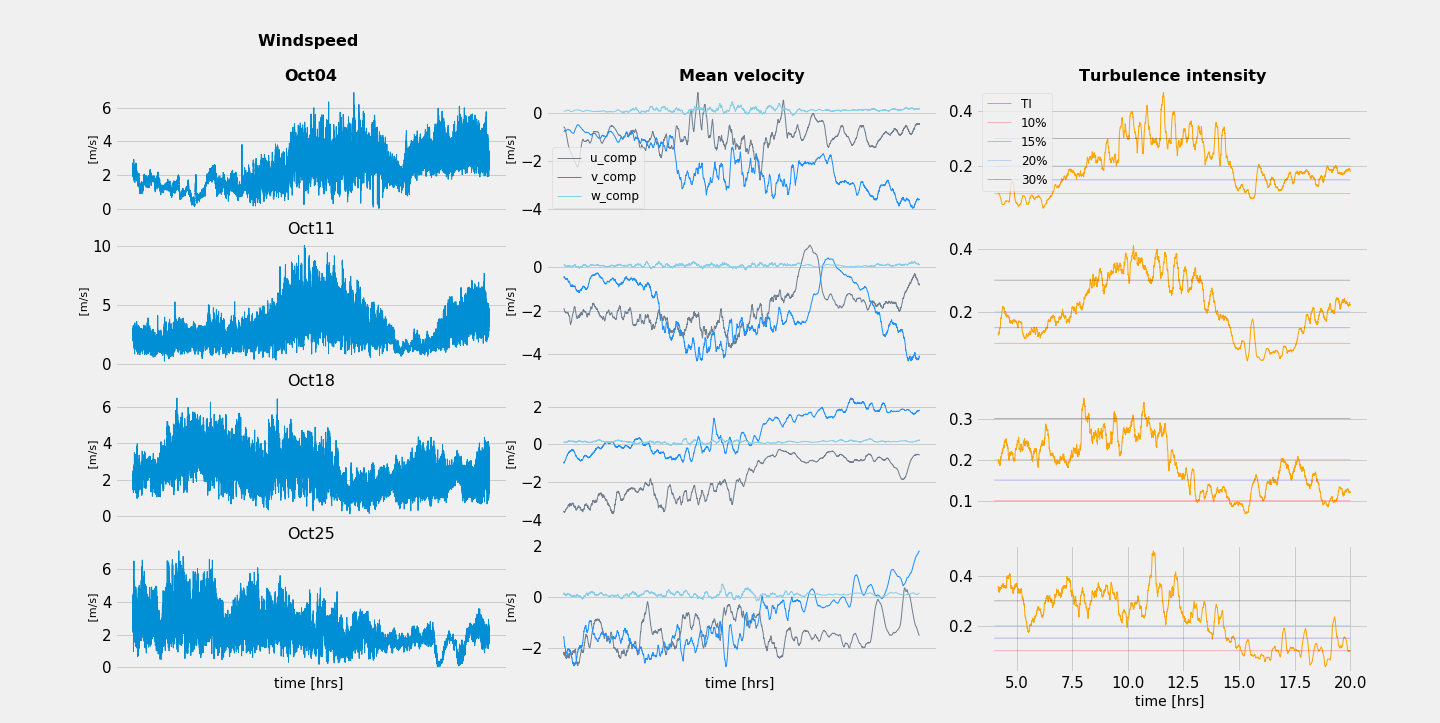}\label{october}}\\
{\includegraphics[width=0.5\linewidth, height=0.24\linewidth]{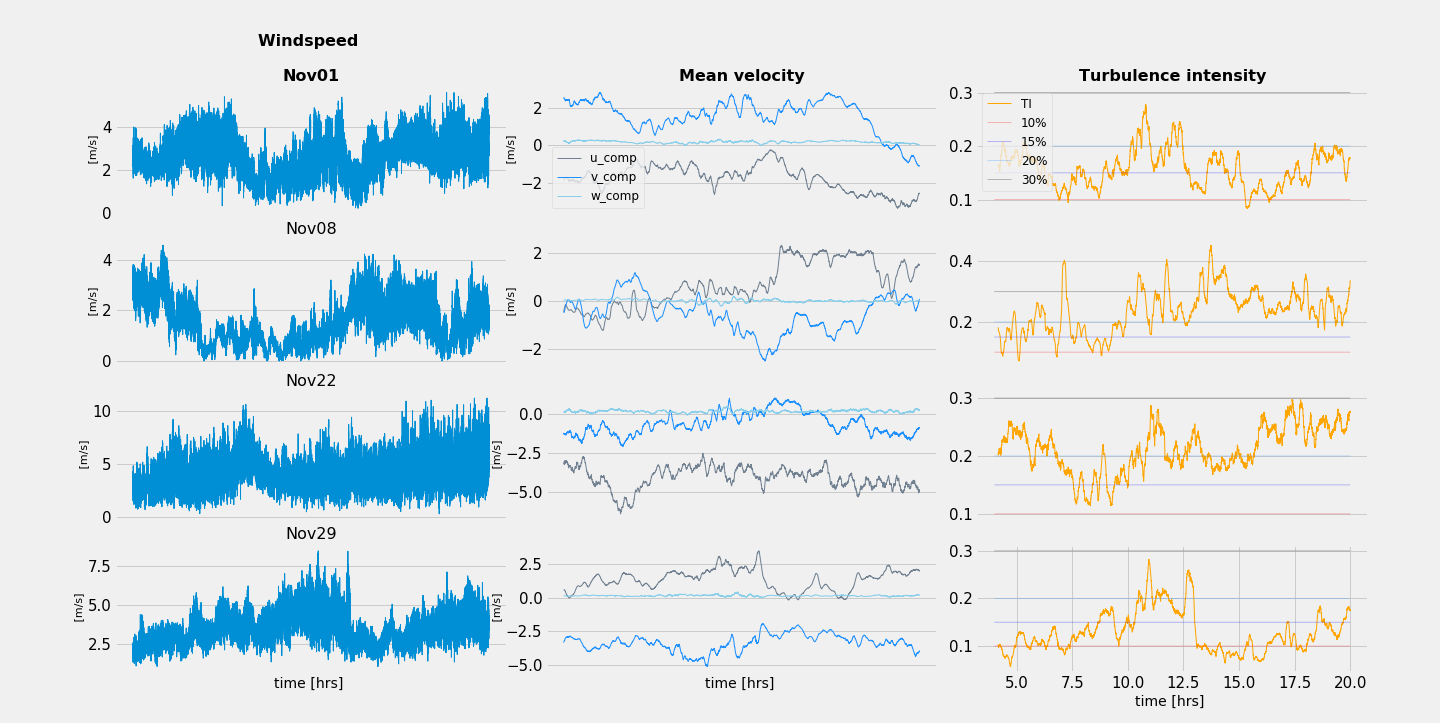}\label{november}}
{\includegraphics[width=0.48\linewidth, height=0.24\linewidth]{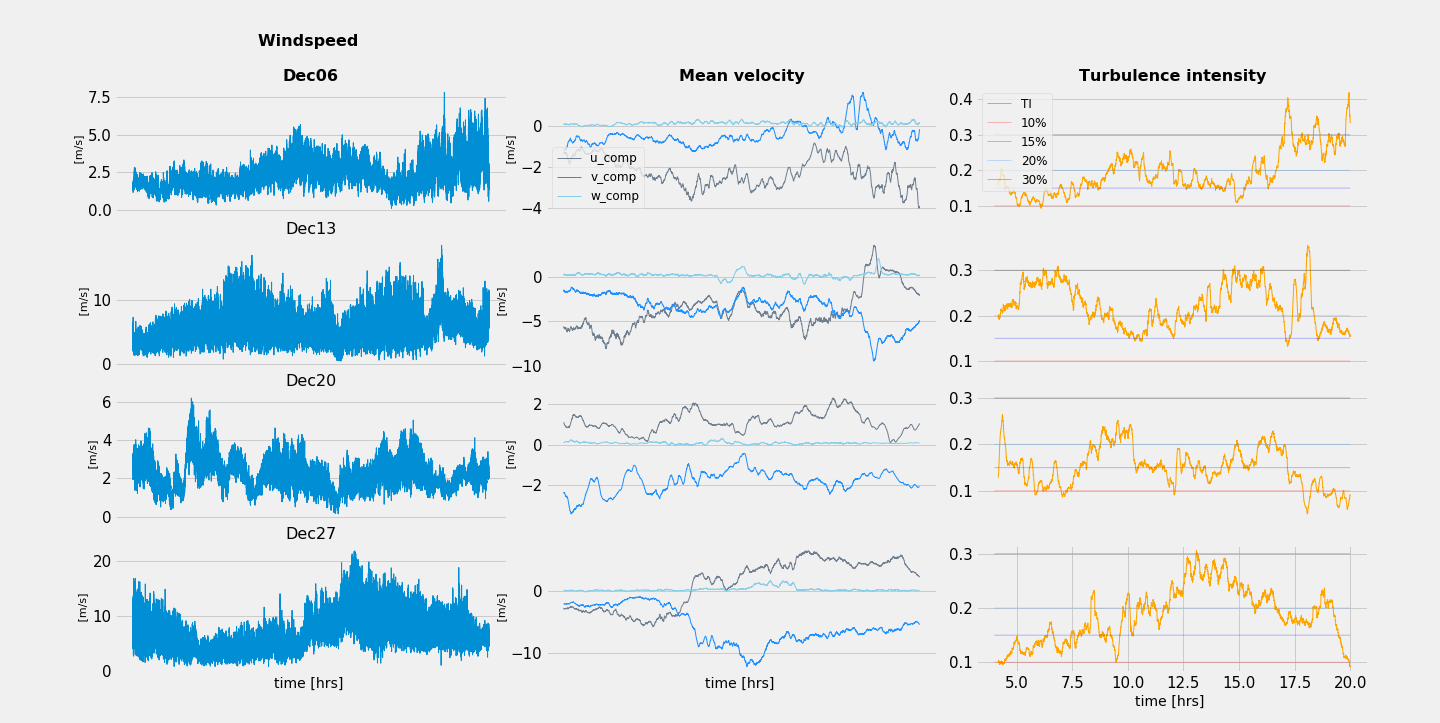}\label{december}}
\caption{For each Wednesday in 2017, during the period of  4 a.m. to 8 p.m.  (local time), we plot the observed wind speed (in blue), the corresponding mean velocity (in blue scale) and turbulence intensity (in orange). \label{fig:data_all_year}}
\end{figure}

Given this time series $(U_t^{(i),\text{obs}}, i=1,2,3,t\geq 0)$,  where $t$ is incremented  each $\frac1{10}$ seconds, we first extract from it the observed  instantaneous TKE process  ${\bm q}^{\text{obs}} = (q_t^{\text{obs}}; ~t\geq 0)$ through the instantaneous turbulent velocity $(u_t';~ t\geq0)$. In practice, it is very common in wind energy industry to approximate  the mean velocity $\av{U}$  by an average in time over an interval of  10 minutes to 60 minutes, corresponding to a minimum in the wind power spectral density. Hence, we compute:
\begin{equation}\label{eq:qt conversion}
q_{t}^{\text{obs}}\approx \big\| U_{t}^{\text{obs}}-\frac1\zeta\sum_{t-\zeta\leq s <t}U_{s}^{\text{obs}}\big\|^2, \qquad \mbox{for $t$ in the selected signal interval, $\Delta t=\frac{1}{10}$},
\end{equation}
with the time-window $\zeta=40$ minutes. While seeking wind homogeneity periods, we may be interested in classify the observations in wind condition regimes. A way to do so is provided by 
 the turbulence intensity  (TI) measure of the wind, defined as (see e.g.  \cite{HoViGo} and the reference therein) the quotient between the standard deviation of wind speed series and a representative mean velocity:
\begin{equation}\label{eq:turbulent_intensity}
I_t := \frac{\sqrt{\av{q_t^{\text{obs}}}}}{\sqrt{3}\|\av{ U^{\text{obs}}_{(d)}}\|}.
\end{equation}
For low wind speeds, it is observed that high turbulence will enhance wind power performance. At the contrary when the wind speed is high and stable, high turbulence will lower the wind power  production. 
In many references (e.g.  \cite{HoViGo,goccmen2016estimation}) the classification of the turbulence intensity of the wind is given by the following five thresholds: 10\%, 15\%, 20\% and 30\%.  Figure \ref{fig:data_all_year} illustrates the wind observation  for each  Wednesday of the year 2017,  during  the 4 a.m. to 8 p.m. period, plotting the (10\si{\hertz} frequency) wind speed, and in the same line the 40-minutes mean velocity components  and the TI  in \eqref{eq:turbulent_intensity}. The TI is estimated  using the  mean $\av{q_t^{\text{obs}}}$ computed over the time-scale $\zeta_{\text{TI}} = 10$-minutes and the  reference mean velocity $\av{U_{(d)}^{\text{obs}}}$ is computed on the entire period from 4 a.m. to 8 p.m. and updated for each considered days. As shown in Figure\xspace\ref{fig:data_all_year}, the (arbitrary) selection of 46 Wednesday signals reveals a wide variety of behaviour. The lowest wind speed can be observed during November 8th, in contrast with December 27th having the highest wind speed during the afternoon. Regarding the values of the TI, the highest values were captured during June 21th and August 16th. We can observe also that, since the measurements were taken  near the ground, the vertical component of the wind velocity stays  closed to zero. Nevertheless, it adds a smoothing effect on the reconstruction of $q_t^{\text{obs}}$ in \eqref{eq:qt conversion}.

\subsection{Time depend regimes in the reduced model} \label{sec:time}

The variety  of magnitudes, and of jumpy  behaviours illustrated in Figure \ref{fig:data_all_year}  suggests    some regimes dependency in the signal,  easily explained by the dynamics of meteorological situations   (wind direction changing the wall turbulence with the terrain, temperature, pressure, relative humidity conditions, modifying  the stability of the ABL).  
In order to represent this regime dynamics in the uncertainty modelling, we allow the parameters to be period-depend, with periods limited in the presentation of this study to diurnal period (precisely from 4 a.m. to 8 p.m.  in local time) that we call 'day-period' arbitrary  chosen along the annual cycle of the year 2017.  

Formally, we are defining sub-periods, numbering  from $1$ to $N_p$ and we allow the parameter values to be period dependent. This assumption is first made possible by the abundance of observations in each sub-period. Moreover Bayesian inference and the evaluation of the level of uncertainty in the calibration process allows us to assess it a posteriori (see Figure \ref{fig:calib_calpha}).

With this in mind, we can go further in the temporal dependence, in order to better capture the variation in the turbulent  regimes represented by the parameter $\gamma$ (see the orange curves of turbulence intensity in Figure \ref{fig:data_all_year}).  
For the parameter $\gamma$ only, and only during the posterior calibration  Step one, we will  refine each day-period, partitioning  them in sub-signals of 20-minutes long, for a total of 48 sub-signals per day. Analysing the results of the time-dependence reduced model in Section \ref{sec:Results}, we will be able to show the relation between the $\gamma$-posterior calibration and  the observed turbulent intensity  statistic, and finally propose this last  quantity as a predictive value for $\gamma$ (see Relation  \ref{eq:gamma_stationary} in Section \ref{sec:Results}).  

\section{Calibration and analysis of the reduced model}\label{sec:Calibration}

We now move forward to the next step: infer on the possible values of the parameters of the model \eqref{eq:tke_equilibrated_CIR}, considering relevant observational dataset for wind energy application (see Section \ref{sec:Data}). 
To this aim, one could  follow a frequentist approach 
by considering a single-value estimation for each parameters. Or one could follow a Bayesian approach, by assigning a probability to each parameter possible values. 
The Bayesian approach is often combined with  Markov Chain Monte Carlo (MCMC) techniques: a Markov chain is constructed,  sampling a distribution that converges with time to the stationary distribution of the parameters.  
Nevertheless, within Bayesian methods it is necessary to set a prior distribution for the parameters to then update this information using the observations. 

\begin{figure}
\centering
\includegraphics[scale=0.22]{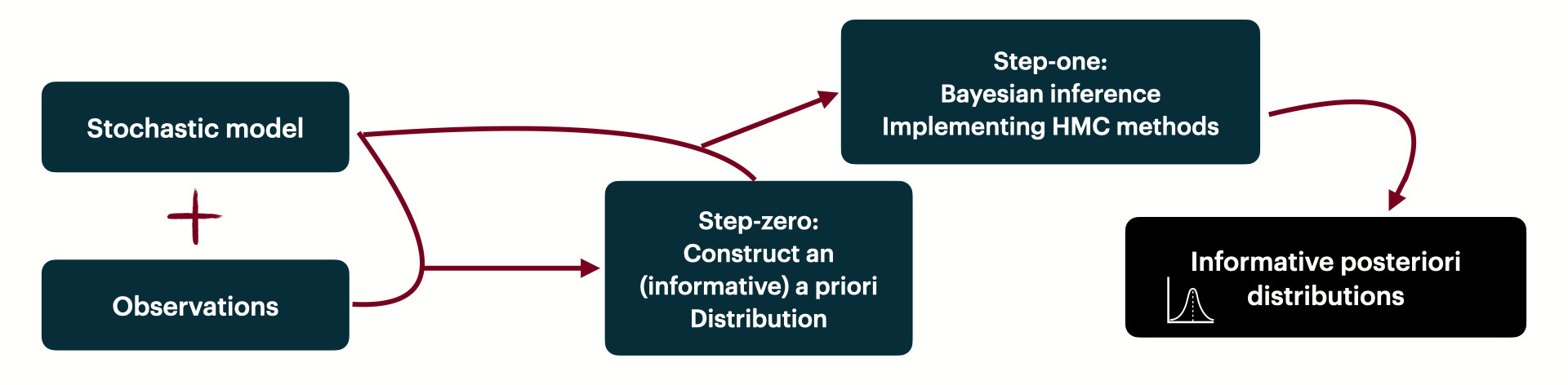}
\caption{the proposed two-steps calibration procedure.}\label{fig:calib_procedure}
\end{figure}

In this section we  construct  a calibration procedure to infer the values of the parameters 
$$C_\alpha=\frac{C_\pdissip}{\kappa z(x_{\text{obs}})},$$ 
and $\gamma$. In order to simplify the analysis  of the results, the Kolmogorov constant $C_0$ is  considered as a prescribed constant equal to 1.9.   Nevertheless, the Bayesian method used here can be extended in future work to include the calibration of $C_0$, which in  view of Remark \ref{rem:on the value of C0} may vary with cases. 
For the parameter $C_\alpha $,  a physically admissible prior distribution support is set using the typical values in the literature (see Remark \ref{rem:on the value of Calpha} below).
For the parameter $\gamma$ however, we do not have a priori information besides an estimation of the  support of its distribution.  In order to provide a reliable calibration without additional parameters, we propose a two steps method summarized in Figure \ref{fig:calib_procedure}: in a Step zero, we construct the a priori distribution of the parameters through a learning step. In a Step one,  we quantify the uncertainty of the parameters with Bayesian inference.

Before detailing the two steps of the calibration procedure, it is important to recall  the nonlinear dependence of the coefficients of the CIR process \eqref{eq:tke_equilibrated_CIR} $\Theta$, $\mu$ and $\sigma$  in terms of the vector parameter $(C_\alpha,\gamma)$. Although the moments \eqref{eq:cir-moments} and the transition density of the CIR process \eqref{eq:tke_equilibrated_CIR} are explicit in terms of Bessel functions, the tricky dependence of the parameters in these formulas  complicates the computation of stable estimators by an optimisation problem. Then, for both steps  of the calibration we introduce a discrete time approximation for the solution of the model \eqref{eq:tke_equilibrated_CIR}. To this aim, we consider an  homogeneous time step $\Delta t>0$, some discrete times $t_n = n\Delta t $, and define the symmetrized Euler scheme \cite{BBD08} associated to the SDE \eqref{eq:tke_equilibrated_CIR} as:
\begin{equation}\label{eq:tke_timediscretization}
\qn1= |p_{t_{n+1}}|,\quad p_{t_{n+1}} =\qnn + \Theta(C_\alpha,\gamma)\left(\mu(C_\alpha,\gamma)-\qnn\right)\Delta t + \sigma(\gamma)\sqrt{\qnn}\left(W_{t_{n+1}}-W_{t_n}\right), \mbox{ for }n\leq N-1,
\end{equation}
with initial condition $\widehat{q}_0 = q_0$, and $N$ such that $N\Delta t$ equals the duration of the signal (scaled in seconds). Under the assumption $C_0<2$, the scheme \eqref{eq:tke_timediscretization} converges in law with a rate of order  one~\cite[Theorem 2.3]{BD15}, i.e. for $f\in \mathcal{C}^4$ a bounded function with bounded derivatives, and sufficiently small $\Delta t $, there exists a constant $C$ (independent on $\Delta t$) such that:
\[\sup_{n \leq N}\left|\EE[f(q_{t_n}) - f(\widehat{q}_{t_n})]\right|\leq C(t_N)~\Delta t.\]

\begin{rem}\label{rem:on the value of Calpha}
In order to set an interval of referential values for $C_\alpha=\frac{C_\pdissip}{\kappa z(x_{\text{obs}})}$, we fix $z(x_{\text{obs}})= 30$\xspace\si\meter. We make varying the Von K\'arm\'an constant $\kappa\in [0.287, 0.615]$~\cite{C14}. We make also varying $C_\pdissip = C_\mu^{3/4},$ with $C_\mu\in[0.054,0.135]$~\cite{C14}. 
Therefore, we expect the values of $C_\alpha$ to be within the interval $[0.0061,~0.0259]$. 

The Kolmogorov constant is fixed to $C_0=1.9$, in  accordance with the values proposed  within the literature pointed in Remark \ref{rem:on the value of C0}.  With $C_0<2$,  the condition ensuring the first order convergence of the time discretization scheme \eqref{eq:tke_timediscretization} is fulfilled.
\end{rem}

We now detail the calibration steps.
\subsection{Step zero: Prior calibration}\label{sec:Step Zero}

The maximum likelihood estimator (MLE) is a classical parameters calibration method for statistical models having an explicit density~\cite{CasellaB}. As already  noticed in our case, due to the complexity of the CIR density in terms of the  parameters,  it is preferable (and common in such situation) to have the help of a discrete time  approximation leading to a pseudo-likelihood computation.  
The symmetrized Euler scheme is far to be the only  possible scheme for power-form coefficients SDEs --see e.g. \cite{bossy2018strong} and references therein, for other schemes and order rates, depending on $C_0$ and $\gamma$-- but it has the advantage to be time-explicit, having the folded Gaussian law as transition density from $\widehat{q}_{t_k}$ to $\widehat{q}_{t_{k+1}}$. Moreover, knowing $\widehat{q}_{t_k}$, the probability $\PP(\widehat{p}_{t_{n+1}}<0)$ decreases with $\Delta t$~\cite[Lemma 3.7]{BD15},  allowing to approximate the folded Gaussian density on $[0,+\infty)$ for $\widehat{q}_{t_{k+1}}$, by the Gaussian one:  for each $n=0,\ldots N-1$, we then assume in this a priori step the random variable $\qn1$ to follow the distribution:
\begin{equation}\label{eq:gaussian_model}
\qn1\;\sim\;\mathcal{N}\big(\qnn + \Theta(C_\alpha,\gamma)(\mu(C_\alpha,\gamma)-\qnn)\Delta t,\;\sigma^2(\gamma)\qnn \Delta t\big).
\end{equation}

From this,  considering the compact set $D\subset\RR^+\times\RR^+$ supporting the admissible values of the  vector $\theta = (C_\alpha,\gamma)$, choosing $\Delta t$ according to some data frequency,  we compute the pseudo-maximum likelihood estimator as 
\begin{align*}
\widehat{\theta}&=\argmax{\theta \in D}\;\log p_{\Delta t}^\theta
({q}^{\text{obs}}_0,\ldots,q^{\text{obs}}_{t_N}), 
\end{align*}
where, using the Markov property of solution of \eqref{eq:tke_timediscretization} and assuming   ${q}^{\text{obs}}_{t_{n+1}}$ knowing ${q}^{\text{obs}}_{t_{n}}$
is distributed according to \eqref{eq:gaussian_model}, the model density $p_{\Delta t}^\theta$  is expressed  in terms of a product of Gaussian densities as  follows
\begin{align}\label{eq:MLE}
\widehat{\theta}
&=\argmax{\theta \in D}\;\sum_{n=0}^{N-1}\log\Big(\frac1{\sqrt{2\pi\sigma^2(\gamma)\qbn \Delta t}}\exp\Big\{-\;\frac{\big|\qb1-\qbn - \Theta(C_\alpha,\gamma)\left(\mu(C_\alpha,\gamma)-\qbn\right)\Delta t \big|^2}{2\sigma^2(\gamma)\qbn \Delta t}\Big\}\Big) \nonumber
\\
&=\argmax{\theta \in D}\;  N\Big\{-\frac{1}2\log\gamma - \frac{\widehat{M}_{2,-1}}{4C_0\Delta t \gamma} -\frac{C_R \widehat{M}_{1,0}}{2^{4/3}C_0}\left(\frac{C_\alpha}{\gamma}\right)^{2/3}- \frac{C_R^2 \Delta t \widehat{M}_{0,1}}{C_0 2^{8/3}}\Big(\frac{C_\alpha}{\gamma}\Big)^{1/3}C_\alpha\nonumber
\\
&\qquad\qquad\qquad\qquad\qquad\qquad\qquad\qquad\qquad\quad\qquad\qquad\qquad+ \frac{C_R^2 \Delta t}{C_0 2^{4/3}}\left(\frac{C_\alpha}{\gamma}\right)^{2/3}\gamma - \frac{C_R^2 \Delta t\widehat{M}_{0,-1}}{4C_0 }\gamma\Big\},
\end{align}
and where we use the notation: for any integers $m_1$ and $m_2$ equal to 0, 1 or 2.
\begin{equation}\label{eq:notation_for_moments}
\widehat{M}_{m_1,m_2} = \frac1{N}\sum_{n=0}^{N-1}{(\qb1-|\qbn|)^{m_1}}{|\qbn|^{m_2}}.
\end{equation}
But here again -- and contrary to the direct $(\Theta, \mu, \sigma)$--parameters  pseudo-MLE  calibration of CIR process  that  can be solved explicitly~\cite{TANG200965} -- 
the optimal pair $(\widehat{C}_\alpha, \widehat{\gamma})$ is not easy to identify as our numerical tests  leads to rather  unstable approximation of an optimum.  To avoid this computational issue, we appeal to the construction of an estimator for $\gamma$ by using the formal convergence of the quadratic variation of a diffusion process as it was proposed in \cite{Mi04}. The main advantage in using quadratic variation estimator (QVE) is its simplicity from a computational point of view. In our case, the QVE for $\gamma$ reads as:
\begin{equation}\label{eq:gamma_estimator}
\widehat{\gamma} = \dfrac{\widehat{M}_{2,0}}{2C_0\Delta t\widehat{M}_{0,1}},
\end{equation}
which is always positive and converges in probability, as $N$ tends to infinity, to the parameter $\gamma$~\cite[Corollary 3.3]{Mi04}.

Considering then the estimator $\widehat{\gamma}$ in \eqref{eq:gamma_estimator}, the optimisation problem \eqref{eq:MLE} reduces to 
\begin{align*}
\widehat{C}_\alpha=\argmax{c > c_*}\varphi(c),
\end{align*}
with $\varphi(c)= ({\widehat{\gamma} C_R \Delta t}-{\widehat{M}_{1,0}})~c^{2/3}
- \frac{\widehat{\gamma}^{1/3} C_R \Delta t \widehat{M}_{0,1}}{ 2^{4/3}}~c^{4/3}$, and $c_*>0$ is a given lower bound including the values of $C_\alpha$ from expert knowledge (see Remark \ref{rem:on the value of Calpha}). Then, applying the first order optimality condition, 
\begin{equation}\label{eq:MLE_Calpha}
\widehat{C}_\alpha = \frac{\sqrt{2}}{\sqrt{\widehat{\gamma}}} 
\max\Big\{c_*\frac{\sqrt{\widehat{\gamma}}}{\sqrt{2}} 
, \ 
\Big(\frac{ \max\{\widehat{\gamma}\Delta t C_R - \widehat{M}_{1,0} \ ,0\}}{\widehat{M}_{0,1}\Delta t C_R}\Big)^{3/2} \Big\}.
\end{equation}
Furthermore, since $\widehat{\gamma}\Delta t C_R - \widehat{M}_{1,0}>0$, we have $\varphi''(c) < 0$, for all $c>0$. Then, the second order optimality condition allows to conclude that the maximum of $\varphi$  is achieved with  \eqref{eq:MLE_Calpha}. 

Finally,  we get our Step zero estimators \eqref{eq:MLE_Calpha} and \eqref{eq:gamma_estimator}. From Identity~\eqref{eq:gamma_estimator}, the condition $\widehat{\gamma}\Delta t C_R - \widehat{M}_{1,0}>0$ becomes  $C_R\widehat{M}_{2,0} - 2C_0\widehat{M}_{1,0}\widehat{M}_{0,1} >0$. This condition has been checked to be always satisfied for our wind time series described in Sections \ref{sec:Data}. A summary of the results produced with this first calibration step is presented in Section \ref{sec:Results}. 

\subsubsection{Time dependent regimes and prior calibration}

To go further on the quantification of uncertainty for the model parameters, one has to propose an a priori distribution which is both sufficiently informative and possibly admits values coming from expert knowledge. 
To this aim, for each day-period of interest $d$ in a selection $\mathcal{S}$,  we first compute the prior estimators $\widehat{\gamma}(d)$ and $\widehat{C}_\alpha(d)$  given, respectively, by Equations \eqref{eq:gamma_estimator} and \eqref{eq:MLE_Calpha} from the time series corresponding to segment $d$. From this, we obtain a family of estimators 
\[\Sigma: = \{(\widehat{C}_\alpha(d),~\widehat{\gamma}(d)),~ \mbox{for $d$ in the selection of day-periods $\mathcal{S}$ in the year 2017}\},\]
from which we define the truncated Gaussian a priori distributions:
\begin{align}\label{eq:aprioris}
\gamma \sim \mathcal{N}^+\left(\overline{\Gamma}(\mathcal{S}) , \mathbb{V}_{\Gamma}(\mathcal{S}) \right), \qquad C_\alpha \sim \mathcal{N}^+\left(\overline{C}(\mathcal{S}) , \mathbb{V}_{C}(\mathcal{S}) \right),
\end{align}
where $\overline{\Gamma}(\mathcal{S})  = \frac1{\#\mathcal{S}}\sum_{d\in\mathcal{S}} \widehat{\gamma}(d)$ and $\overline{C}(\mathcal{S})  = \frac1{\#\mathcal{S}}\sum_{d\in \mathcal{S}} \widehat{C}_\alpha(d)$ denote the empirical means of the $\widehat{\gamma}$ and $\widehat{C}_\alpha$ estimates,  and $\mathbb{V}_{\Gamma}$ and $\mathbb{V}_{C}$ their corresponding empirical variances over the day-periods in $\mathcal{S}$.

\subsection{Step one: Posterior calibration}\label{sec:Step One}

Inspired by the work of Edeling et al. \cite{C14}, we improve the estimations computed in Step zero by quantifying the uncertainty on the two  parameters distribution of the reduced model \eqref{eq:tke_equilibrated_CIR}. With the help of MCMC method, we apply a Bayesian inference to update our initial guess in  \eqref{eq:aprioris} on the distribution  of the parameters in the prior calibration step. 

\subsubsection{Statistical model for the uncertain parameters}\label{sec:statistical_model}

Assume that the observed TKE on the interval $[0,T]$ are independent and identically distributed observable vector random variables ${\bm q}^{\text{obs}}\in (0,+\infty)^{N+1}$, indexed with $\{0=t_0,t_1,\ldots,t_N=T\}$. As the name suggests, Bayesian inference uses Bayes' Theorem  in order to extract information from the observed values and infer a more realistic distribution for the parameters. Denoting by $\pi(\theta|{\bm q}^{\text{obs}})$ the probability density of the parameter vector $\theta=(\gamma,C_\alpha)$ given the realisation ${\bm q}^{\text{obs}}$, by  $p(\cdot|\theta)$ the probability density of the model given the parameters (likelihood function), and by $p_\theta$ the prior distribution of 
$\theta$ in \eqref{eq:aprioris},  we get from Bayes' Theorem that
\begin{equation}\label{eq:Bayes}
\pi(\theta|{\bm q}^{\text{obs}})=\frac{p({\bm q}^{\text{obs}}|\theta) \ p_\theta(\theta)}{p({\bm q}^{\text{obs}})},
\end{equation}
where $p({\bm q}^{\text{obs}})$ is the distribution of the observed data. For this latter distribution, we set a statistical model through our model for instantaneous TKE   which comes with observational error $\bm{  \mathcal{E}}$ as:
\begin{equation}\label{eq:statistical model}
{\bm q}^{\text{obs}}(\theta)= \;\widehat{\bm q}(\theta)+{\bm{ \mathcal{E}}},
\end{equation}
where the independent random vector variable  $\widehat{\bm q}(\theta)$ is identically distributed with the equilibrium law (for a given $\theta$) of   the discrete-time process  ~\eqref{eq:tke_timediscretization}; the random vector $\bm{\mathcal{E}}$ is assumed following the logistic distribution with zero mean and scale parameter $\sigma$ to be estimated from the data.  The choice of the logistic distribution is  made from  a preliminary analysis comparing the histogram of errors with a set of explicit mean-variance parametrized  probability densities (illustrated in Figure \ref{fig:observation_error} of Section \ref{sec:Results}).

\paragraph{Metropolis-Hastings (MH) Algorithm}\label{subsec:Metropolis}

A widely used method to generate samples from the posterior distribution is the Metropolis-Hasting algorithm. In our case, this iterative algorithm, sampling for the Markov chain $(\theta_n, n)$, takes the following form: start from an initial value $\theta_0$. At the $n$th-iteration, from $\theta_n$ to $\theta_{{n+1}}$, proceed as follows:
\begin{enumerate}
\item Simulate $\widetilde{\theta}\sim \rho({\widetilde{\theta}|\theta_n}),$ and $u\sim\textit{Uniform}(0,1),$ where $\rho$ is a proposed transition density.
\item Compute
\begin{align}\label{eq:Metr parameter}
a:=\min\Big\{1,\frac{p_{\theta}(\widetilde{\theta})p({\bm q}^{\text{obs}}|\widetilde{\theta})\rho({{\theta_n}|\widetilde{\theta}})}{p_{\theta}({\theta_n})p({\bm q}^{\text{obs}}|{\theta_n})\rho({\widetilde{\theta}|\theta_n})}\Big\}.
\end{align}
\item If $u< a$, the simulated state is accepted and $\theta_{n+1}=\widetilde{\theta}$. Else, the state $\widetilde{\theta}$ is rejected and we keep the previous state, i.e. $\theta_{n+1}=\theta_n.$
\end{enumerate}
Notably, the second step of the algorithm maximizes the posterior of the parameters with their transition, given the observed data.  The choice of a prior distribution close to the posterior one improves the convergence of the chain to its stationary distribution $\pi(\cdot|{\bm q}^{\text{obs}})$.    Nevertheless, a common issue in the implementation of this method is the computational time needed to explore properly the state space in high dimension. Through the sampling process, we not just infer the values of the parameters $\theta$ but also  the statistical model \eqref{eq:statistical model}. 

Hamiltonian Monte Carlo (HMC) is an alternative method to efficiently explore the state space of the MH framework. HMC uses the well-known Hamiltonian system to mimic the dynamics of a particle  having the logarithm of the target posterior distribution (which must be smooth enough)  as   potential energy. 
To this aim, the HMC method introduces a synthetic variable $\psi$, (usually sampled as a Gaussian random variable), describing the momentum. 
Assuming the conservation of the total energy, we consider the Hamiltonian function as the sum of kinetic energy $K$ (fixed to $\tfrac{1}{2} \|\psi\|^2$ for instance), and potential energy (as the log posterior density)  
\begin{align}\label{eq:Henergy}
H(\theta,\psi) = K(\psi) 
- \sum_{q\in\bm{q}^{\text{obs}}} \log p(q|\theta) -  \log p_\theta(\theta),
\end{align} 
and the parameter vector $\theta$ represents now the position of a  particle following the Hamiltonian equations:
\begin{align}\label{eq:Hamiltonian_system}
\left\{
\begin{aligned} 
\frac{d\theta}{dt} &= \frac{\partial H}{\partial \phi},\\
\frac{d \psi }{dt} &= - \frac{\partial H}{\partial \theta}.
\end{aligned}\right.
\end{align}
The exchange between kinetic and potential energies  during the discrete  time simulation of the Hamiltonian dynamics ensures that the evolution of the particle generates the contours of the target distribution $\pi$, where the latter is obtained by computing the marginal distribution of the joint position-momentum posterior. 
Among the main advantages within HMC methods,  we  underline  the preservation of the volume, the efficient exploration of the space, the reversibility of the dynamics (which is crucial to ensure that the MCMC updates preserves the  target distribution). These features allow HMC algorithms to converge to high-dimensional target distributions much more quickly than simpler methods such as random walk Metropolis or Gibbs sampling. Nonetheless, the introduction of this auxiliary momentum duplicates the number of variables, and therefore the computational cost.

Sampling the momentum from the conditional distribution (assuming that the current point $\theta_n$ is in the contour of the target distribution) we generate trajectories moving -in time- through the entire phase space while being constrained to the typical set. Thus, to propose a new state $\theta_{n+1}$ for the Markov chain, the trajectory is projected back to the parameter space, and finally the state is accepted/rejected  following a decision criterion  similar to the one implemented in the Metropolis-Hasting algorithm. For a more detailed introduction to HMC methods we refer the interested reader to the work of Neal~\cite{neal2011mcmc}.

Since usually the likelihood function involved in \eqref{eq:Henergy} is not explicit, and even so, the solution to Hamiltonian system 
is not  explicit as well, the HMC method requires a discrete-time approximation  to Equations \eqref{eq:Hamiltonian_system}. The quality  of numerical approximation  relies on  the ability of the simulated  trajectories  to not drift away from the exact energy level set. One common approach is the leapfrog scheme  which start with half step for the momentum variables, then do a full step for the position using the updated momentum, and finally complete the remaining half step for the momentum. The main drawback of solving (by approximation) the dynamics  \eqref{eq:Hamiltonian_system} is the need to specify  the terminal time and the size of the steps in the trajectory. Among the methods implemented for the sampling  of Monte Carlo Markov Chains with Hamiltonian method, we use the No U-Turn sampler~\cite{HoGe} that  allows automatic tuning of the step size and number of simulation steps. 

\subsubsection{Algorithm for the Bayesian calibration}
We now detail the entire Bayesian calibration algorithm  and discuss  its  specificities for the  model \eqref{eq:tke_equilibrated_CIR}. 

Consider an arbitrary selection of  day-periods $\mathcal{S}$. For $d$ in $\mathcal{S}$, we proceed as follows for the  Bayesian calibration of $\widehat{\theta}(d) = (\widehat{C}_\alpha(d), \widehat{\gamma}(d))$.

\paragraph{Selection of frequencies} We recall that for the Bayesian algorithm, we need independent realisations of variables $\bm{q}^{\text{obs}}$ selected from the observed time series. To do so,  we compute the autocorrelation of the signal and select a compromise between independence and the length of the sample: we therefore  choose a frequency $\xi_{C_\alpha}=1/30\ \si\second^{-1}$ for the Bayesian calibration of  $C_\alpha$ on the 16 hours duration of the day-period $d$, and a frequency of  $\xi_\gamma=1/5\ \si\second^{-1}$ for the Bayesian calibration of $\gamma$ for each sub periods of 20 minutes of the day-period. 
 
For the calibration coherency between Step zero and Step one,
in Step zero, we stay with the frequency $\xi_{C_\alpha}= \xi_\gamma=1/30\ \si\second^{-1}$,  calibrating with QVE and QMLE 
  the parameter pair $(\widehat{C}_\alpha(d), \widehat{\gamma}(d))$, assumed  constant on the entire day-period.  These chosen frequencies set a unique  time step $\Delta t$ to use in \eqref{eq:gaussian_model}. 
Moreover, due to the high frequency of jumps in the observations, this  rather low frequency helps to bring out the diffusive behaviour in the observations.

For convenience we denote $[0,T]$ the day-period $d$, with dates $\{0=t_0,t_1,\ldots,t_N=T\}$, and proceed as follows: 
\begin{itemize}
\item[0.~~] \emph{Compute the prior distribution \eqref{eq:aprioris}.}   

\item[1.1]\emph{Estimate the observational error. }  We estimate the scale parameter $\widehat{\sigma}(d)$ of the logistic distribution appearing in the statistical model  retained for   the observational error ${\bm{\mathcal{E}}}$ in \eqref{eq:statistical model}. 
For this, we first compute the empirical mean 
\[\EE[{\bm{\mathcal{E}}}(d)] 
\simeq \frac{1}{N}\sum_{n=1}^N q^{\text{obs}}_{t_n} - \widehat{q}_{t_n} (\widehat{\theta}(d)),\]
simulating $\widehat{q}$ using \eqref{eq:tke_timediscretization} with $\Delta t= \frac{1}{\xi_{C_\alpha}}=30\ \si\second$ (the lower frequency is used here as we need an independent sample to approximate the mean)   and  with $\widehat{\theta}(d) = (\widehat{C}_\alpha(d), \widehat{\gamma}(d))$ computed in Step zero. Computing the  empirical variance  with the same sample, we obtain the approximate scale $\widehat{\sigma}^2 (d) = \tfrac{3}{\pi^2} \mathbb{V}[\bm{\mathcal{E}}(d)]$. 

\item[1.2] \emph{Bayesian calibration of $\gamma$. } As announced in Section \ref{sec:time}, $\gamma$ is allowed to be estimated on  refined sub-signals  along a day-period $[0,T] = {\bigcup_{0\leq i\leq S}[T_i, T_{i+1}]}$, with $T_{i+1}-T_i=$ 20-minutes long. In this step, the parameter  $\widehat{C}_\alpha(d)$ is fixed, given by the Step zero estimator.  
For each $i\in\{1,\ldots, S=48\}$, we estimate the density of $\gamma(d,i)$  considering the sub-signal ${\bm q}^{\text{obs}}\big|_{[T_i,T_{i+1}]}$ with the higher frequency $\xi_\gamma$ and the time step $\Delta t = \frac{1}{\xi_\gamma} = 5\ \si\second$ for the simulation of $\widehat{\bm q}_t$. 
Starting from the prior distribution $\gamma(d,i)\sim\mathcal{N}^+(\overline{\Gamma}, \mathbb{V}_{\Gamma})$ in \eqref{eq:aprioris}, and observation error $\mathcal{E}\sim\textit{Logistic}(0,\sigma)$ with  scale $\sigma\sim\textit{Lognormal}(\widehat{\sigma}(d),1)$, we apply the HMC method in order to sample the posterior distribution of $\gamma(d,i)$  with the statistical model
\[{\bm q}^{\text{obs}}\big|_{[T_i, T_{i+1}]} =  \widehat{\bm q}(\gamma(d,i),\widehat{C}_\alpha(d)) + {\bm{\mathcal{E}}}.\]
In particular, by computing $\EE[\gamma(d,i)]$ from the empirical mean of the HMC sampling of size 2000,  we obtain a piecewise constant mean production term  process
\begin{align}\label{eq:definition_gamma_bar}
\bar\gamma_t = \sum_{i=0}^S \EE[\gamma(d,i)] {\ind}_{[T_i, T_{i+1}]}(t).
\end{align}

\item[1.3] \emph{Bayesian calibration of $C_\alpha$. }  Starting from the prior distribution $\mathcal{N}^+\left(\overline{C}(\mathcal{S}), \mathbb{V}_{C}(\mathcal{S})\right)$ in \eqref{eq:aprioris}, and observation error ${\bm{\mathcal{E}}}$ as in~ 1.2, we apply a HMC method in order to sample the posterior distribution of $C_\alpha(d)$ 
with the statistical model
\[{\bm q}^{\text{obs}}_t|_{[0, T]} =  \widehat{\bm q}_t(\bar\gamma_t, C_\alpha(d)) + \mathcal{E}_t.\]
In this step we consider ${\bm q}^{\text{obs}}$ with frequency $\xi_{C_\alpha}$ and $\Delta t = \frac{1}{\xi_{C_\alpha}}$.  
\end{itemize}
For the HMC implementation,  we have used the Python package {PyMC3}~\cite{salvatier2016probabilistic} with NUTS step method.

\paragraph{Convergence Diagnostic Tests} 
To check the convergence of the HMC method  to the posterior distribution, several convergence diagnostics can be assessed. For instance, we can verify if the Markov chain explores the state space thoroughly, check the convergence of the empirical mean or analyse if the simulated values are uncorrelated. Along with ad hoc convergence diagnostics,  formal statistical diagnostics have been implemented. The convergence of each of the Markov chains constructed in steps 1.i  has been tested through  the following:
\begin{itemize}
\item[--] \emph{Geweke diagnostic} that analyses the similarity (mean and variance) of segments from the beginning and the end of a single chain, and check if the samples are drawn from the stationary distribution.

\item[--] \emph{Gelman-Rubin diagnostic} (also called $\widehat{R}$-hat test) that uses multiple chains to check the lack of convergence by computing the ratio between the within-chain variance and the posterior variance estimate for the pooled traces. This quotient converges to 1 when each of the traces is a sample from the target posterior.
\end{itemize}

\section{Calibration results and model evaluation against data}\label{sec:Results}

First, we recall that for the presentation of the results summarized in this section, we have chosen to show a selection set $\mathcal{S}$ arbitrary day-periods corresponding to wind measurements observed every Wednesday (46 day-periods) of the year 2017,  starting from 4 a.m. to 8 p.m. (local time), maintaining the representativeness of  the different regimes and seasonality (see Figure \ref{fig:data_all_year}).

In the following, we illustrate the results of the calibration, describing the key findings of each step. 
Before calibrating with the observations, we have tested each step using synthetic data generated from the time-discretisation of the SDE \eqref{eq:tke_equilibrated_CIR} used with fixed parameters, from which we recovered the input values (results not shown here).

\begin{figure}
\centering
\includegraphics[scale=0.35, trim={0 0 0 2cm},clip]{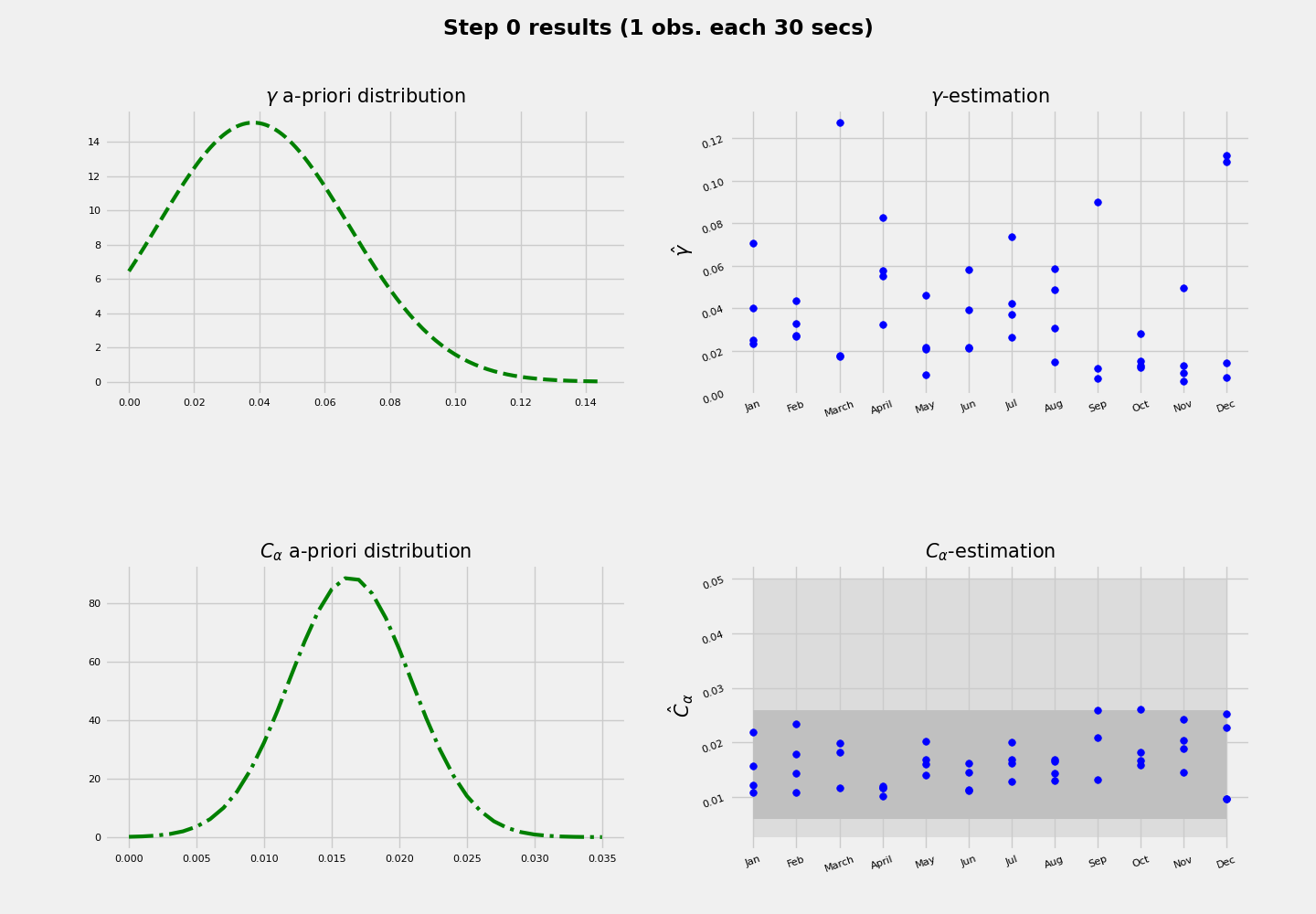}
\caption{Prior calibration results: point-estimations (right figures) for each Wednesday of 2017, and resulting prior distribution (left figures)  from  \eqref{eq:aprioris}. The dark grey area in the representation of the $C_\alpha$ calibration corresponds to the reference interval given in Remark \ref{rem:on the value of Calpha}.\label{fig:Step0_CIR}}
\end{figure}

In Step zero, after computing the point-estimates $(\widehat{C}_
\alpha(d),\widehat{\gamma}(d), d\in \mathcal{S})$  with \eqref{eq:gamma_estimator} and \eqref{eq:MLE_Calpha}, we construct the prior distribution of the parameter vector as described in \eqref{eq:aprioris}. The results of this step are illustrated in Figure~\ref{fig:Step0_CIR}, with the collection of  point-estimations (right) and the informative a priori distribution (left)  for each parameter. 
Concerning  $\widehat{C}_\alpha$ point-estimations, it should be stressed that all  the obtained estimates  fall within the reference range of ${C}_\alpha$ values given in the literature and reported in Remark~\ref{rem:on the value of Calpha}, without the need to impose the minimum value $c_*$ in \eqref{eq:MLE_Calpha}, which glimpses the accuracy of the model.  Concerning the prior calibration of $\gamma$, the $\widehat{\gamma}(d)$ estimates are all positive, spread in a relatively larger interval (and with a large variance), which is coherent with the idea of decomposing the calibration of $\gamma$ on sub-signals.  

In order to validate the estimates obtained from Step zero, we consider the theoretical  moment $\EE[q_\infty]$ given in  \eqref{eq:mean_TKE_limit_gamma} and, along each day, we compare the quotient $(\tfrac{\sqrt{2}\widehat{\gamma}(d)}{\widehat{C}_\alpha(d)})^{2/3}$ against the time-averaging $\frac1{\# \mbox{obs}}\sum_{t}{q}_t^{\text{obs}}$. Table~\ref{tab:validation_step_zero} summarizes these values showing the accuracy of the prior step of the calibration procedure. 

\begin{table}
\centering
{\footnotesize{
\begin{tabular}{l l l l}
\toprule
\tabhead{Month}
& \tabhead{Quotient $(\tfrac{\sqrt{2}\widehat{\gamma}(d)}{\widehat{C}_\alpha(d)})^{2/3}$ }& \tabhead{Time-average $\frac1{\# \mbox{obs}}\sum_{t}{\bm q}_t^{\text{obs}}$} & \tabhead{Absolute error}\\
\midrule
  & 1.32326987 & 1.32352805 & 1.95e-04 \\%
January  & 2.34663793 & 2.34799496 & 5.78e-04\\
 & 4.0458171&	4.047094	& 3.16e-04\\
 & 2.19174731& 2.1923935 	& 2.95e-04 \\
 \midrule
   & 1.64077689 & 1.64175207	&5.94e-04 \\%
February  & 1.92520322 & 1.92621081		& 5.23e-04\\
 & 1.5771361 &	1.57724267	& 6.76e-05\\
 & 3.18971747 & 	3.18980458	& 2.73e-05\\
  \midrule
   & 6.14997814 &	6.15392597	& 6.42e-04\\%
March  & 1.17901587 &		1.17918501	& 1.43e-04\\
 & 1.22671331 &	1.22699872	& 2.33e-04\\
 \midrule
   & 3.99699439 &	3.99936725	& 5.93e-04\\%
April  & 2.43719514 &	2.43752582	& 1.36e-04\\
 & 4.6508576 &	4.65245935	& 3.44e-04\\
 & 3.5330159 & 	3.53318812	& 4.87e-05\\
  \midrule
   & 0.80623555 &	0.80670673	& 5.84e-04\\%
May  & 2.18575829 &	2.18624388	& 2.22e-04 \\
 & 1.53782389 &	1.53811257	& 1.88e-04\\
 & 1.63677343 & 1.63688688	& 6.93e-05\\
  \midrule
   & 3.76661667 &	3.76674479	& 3.40e-05\\%
June  & 1.51005644 &	1.51051385	& 3.04e-04\\
 &1.65372866 &	1.65381007	& 4.92e-05\\
 & 2.88200534 &	2.88234525	 & 1.18e-04\\
  \midrule
   & 1.51360975 &	1.51388343	& 1.81e-04\\%
July  & 2.55681663 &	2.55779937	& 3.84e-04\\
 & 3.36248203 &	3.3629028 & 1.25e-04\\
 & 2.39722227 & 2.39762117 & 1.66e-04 \\
  \midrule
   & 2.59128586 &	2.59280967	& 5.88e-04\\%
August  & 1.29836408 &	1.29888482	& 4.01e-04\\
 & 2.22652455 &2.22708633 &2.52e-04\\
 & 2.88813526 & 2.88946423	 & 4.60e-04\\
  \midrule
   & 4.53006814 &	4.53197261	& 4.20e-04\\%
September  & 0.74429092 &	0.74442679	& 1.83e-04 \\
 & 0.62351139 &	0.62381067	& 4.80e-04\\
 \midrule
   & 1.12645897 &	1.12662812	& 1.50e-04\\%
October  & 1.78036412 &1.78109514	& 4.10e-04\\
 & 0.77009635 &	0.77025073	& 2.00e-04\\
 & 1.11176981 & 1.11209757	& 2.95e-04\\
 \midrule
   & 0.68282708 &	0.68306123 & 3.43e-04\\%
November  & 0.53527354 &	0.53539273 & 2.23e-04\\
 & 2.85532726 &	2.8558594	& 1.86e-04\\
 & 0.98429323 & 	0.98442425	& 1.33e-04\\
  \midrule
   & 0.9223277 &	0.92314383	& 8.84e-04\\%
December  & 6.28067272 &	6.28366521	& 4.76e-04\\
 & 0.55240658 &	0.55471823 & 4.17e-03\\
 & 6.44366922 & 6.44440406 & 1.14e-04\\
\bottomrule
\end{tabular}
}}
\caption{ Validation of the prior calibration through Identity \eqref{eq:mean_TKE_limit_gamma}.\label{tab:validation_step_zero}}
\end{table}

\begin{figure}
\centering
\subfigure[February 1st]{\includegraphics[trim={0 24cm 0  0.1cm}, clip, width=0.48\textwidth]{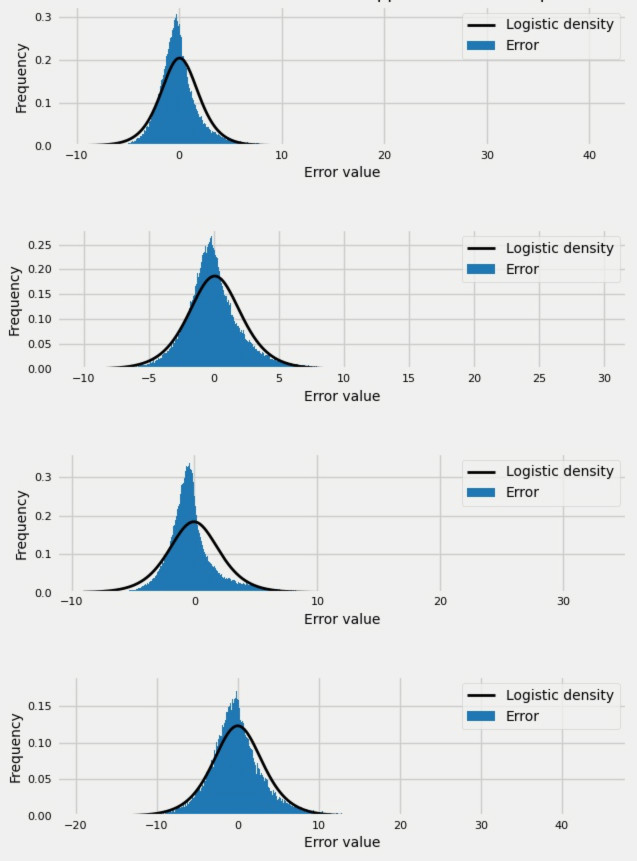}\label{im:Feb1}}
\subfigure[February 8th]{\includegraphics[trim={0 16cm 0  8cm}, clip, width=0.48\textwidth]{./Observed_priori_error_distribution_Feb.jpg}\label{im:Feb8}}\\
\subfigure[February 15th]{\includegraphics[trim={0 8cm 0  16cm}, clip, width=0.48\textwidth]{./Observed_priori_error_distribution_Feb.jpg}\label{im:Feb15}}
\subfigure[February 22th]{\includegraphics[trim={0 0cm 0  24cm}, clip, width=0.48\textwidth]{./Observed_priori_error_distribution_Feb.jpg}\label{im:Feb22}}
\caption{Observation error distribution, introduced in Subsection \ref{sec:statistical_model}, approximated with Monte Carlo method. Histogram (blue bars) are compared with adjusted logistic density (in black).}\label{fig:observation_error}
\end{figure}

Concerning the calibration of the distribution of the observation error $\bm{\mathcal{E}}$ in \eqref{eq:statistical model}, we mention that after several numerical tests, we have chosen a logistic distribution with scale parameter $\widehat{\sigma}^2(d) = \tfrac{3}{\pi^2}\mathbb{V}[{\bm {\mathcal{E}}}(d)]$ approximated with a Monte Carlo method. In Figure \ref{fig:observation_error}, we plot the empirical distribution of ${\bm {\mathcal{E}}}(d)$ resulting from \eqref{eq:statistical model}, for $d = \text{February}$ 1st, 8th, 15th, 22th. The black curves plotted in the figure represent the logistic densities adjusted with the moments of the sample. Here, we can see that some days adapt better to the density (for example February 22th in contrast to February 15th). However, given the thinness of the empirical density tails, in general the logistic distribution is the simplest distribution that best fits.

\begin{figure}
\centering
\subfigure[Box plots within month of posterior distribution of $\widehat{C}_\alpha$.]{\includegraphics[trim={0 0cm 0  8.5cm}, clip, width=0.85\textwidth]{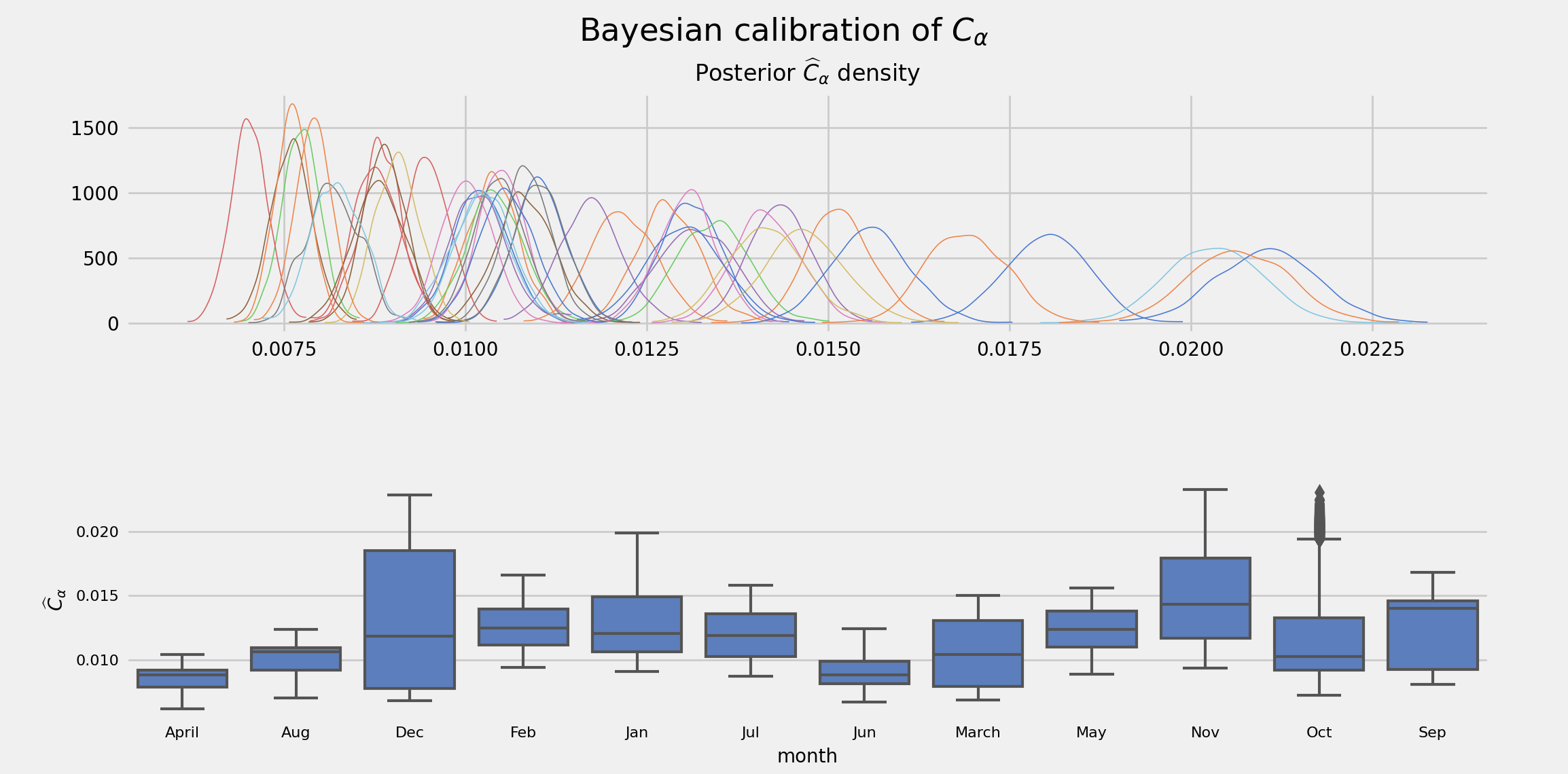}\label{im:BoxPlotCalpha}}\\
\subfigure[Posterior mean estimations of $\widehat{C}_\alpha(d)$.]{\includegraphics[width=0.85\textwidth]{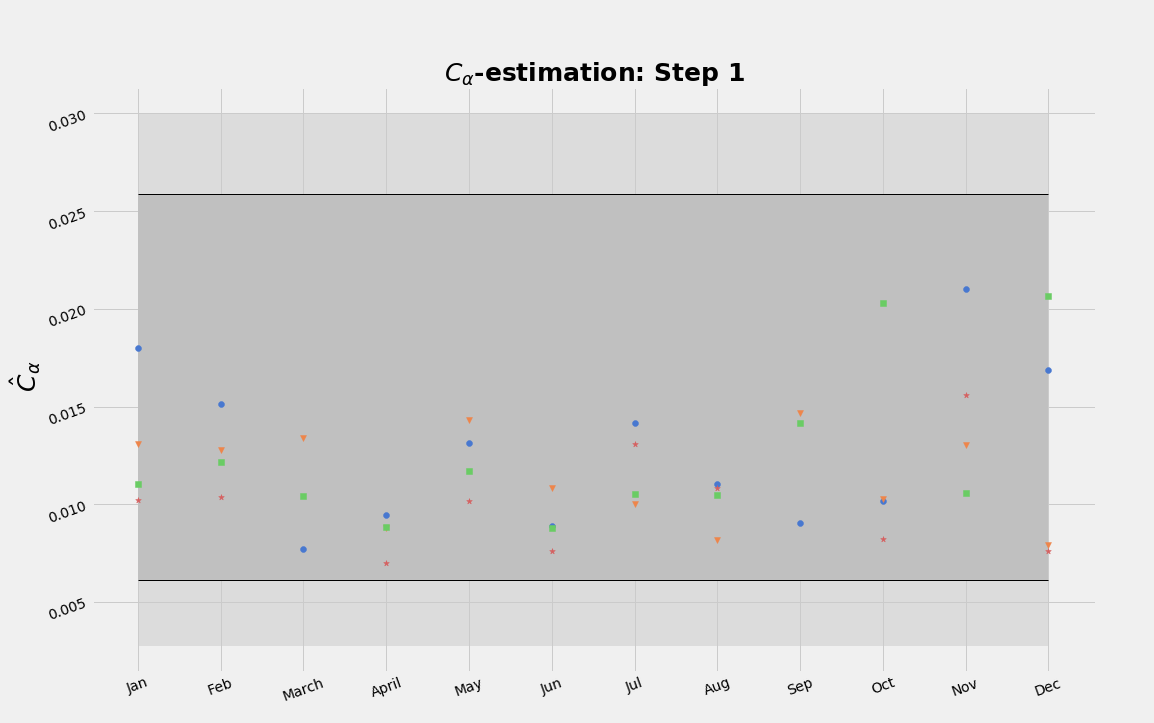}\label{im:Step1Calpha}}\\
\subfigure[Exploration of the state space and posterior distribution.]{\includegraphics[width=0.85\textwidth]{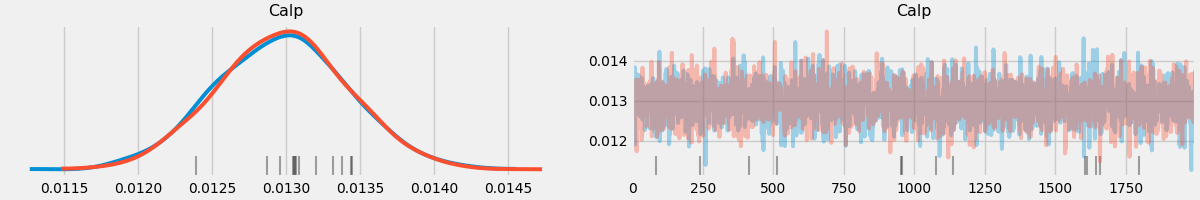}\label{im:traceplot_calpha}}
\caption{Bayesian calibration of $C_\alpha$ for all Wednesdays of 2017. In (a) the box plots constructed from the Markov chains  samples in a given month. In (b) the  obtained posterior mean estimations for each day. The filled symbols circle (blue), triangle (orange), square (green), star (red) correspond respectively to occurrence order of the Wednesdays in each month.   In (c), two examples of  exploration of the state space with the Markov chain are illustrated (right), with the corresponding posterior distribution of $C_\alpha$ (left). \label{fig:calib_calpha}}
\end{figure}

In Figure \ref{fig:calib_calpha}, we summarize the results obtained for the Bayesian calibration of $C_\alpha$, with the  posterior mean estimations reported in Figure \ref{im:Step1Calpha} for each selected day. The posterior mean estimates have been computed as the average on the Markov chain sample $\{C_{\alpha,n}(d), n\geq1\}$ that passes convergence diagnostic tests according to the chosen   length of the chain. So we can consider these empirical means converging  towards the true parameter value. 
The box plots reported in Figure \ref{im:BoxPlotCalpha} have been constructed from the combination of  all the Markov chain samples in a given month, by computing the quartiles of each set. Then, for each month, the range of the box comprises values between 25\% and 75\% of the inferred values for $C_\alpha$ during that month. Additionally, Figure \ref{im:traceplot_calpha} illustrates the exploration (right) of the state space for two samples (red and blue) along the Markov chain,  for the particular day of February 15th, with their associated posterior distribution (left), having convergence $\widehat{R}$-hat statistic equals 1.

We emphasize that the estimated $C_\alpha$ values fall again within the reference interval $[0.0061, 0.0259]$, this time the average values being closer to 0.01, except during October, November and December, months for which we can observe more variability and lower turbulence intensity.

\begin{figure}
\centering
\subfigure[Box plot of the $\gamma(d,i)$ for $i=1,\ldots,48$.]{\includegraphics[width=0.85\textwidth]{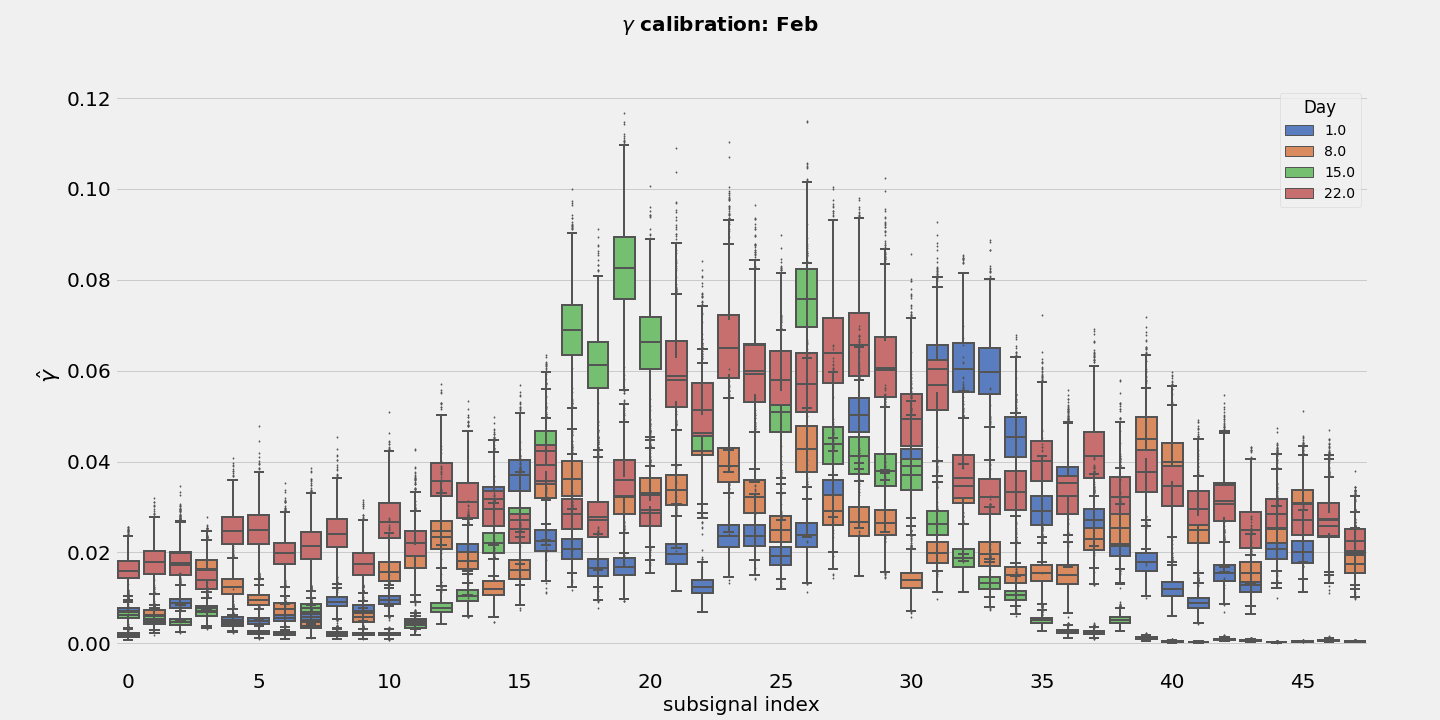}\label{im:Step1_gamma_Feb}}\\
\subfigure[The dynamics $t\mapsto\bar\gamma_t$ obtained from Step one  for the same four days; the horizontal lines are the level of the means over the period (the black line is the Step zero estimator).]{\includegraphics[trim={0 0.6cm 0  1.3cm}, clip, width=0.85\textwidth]{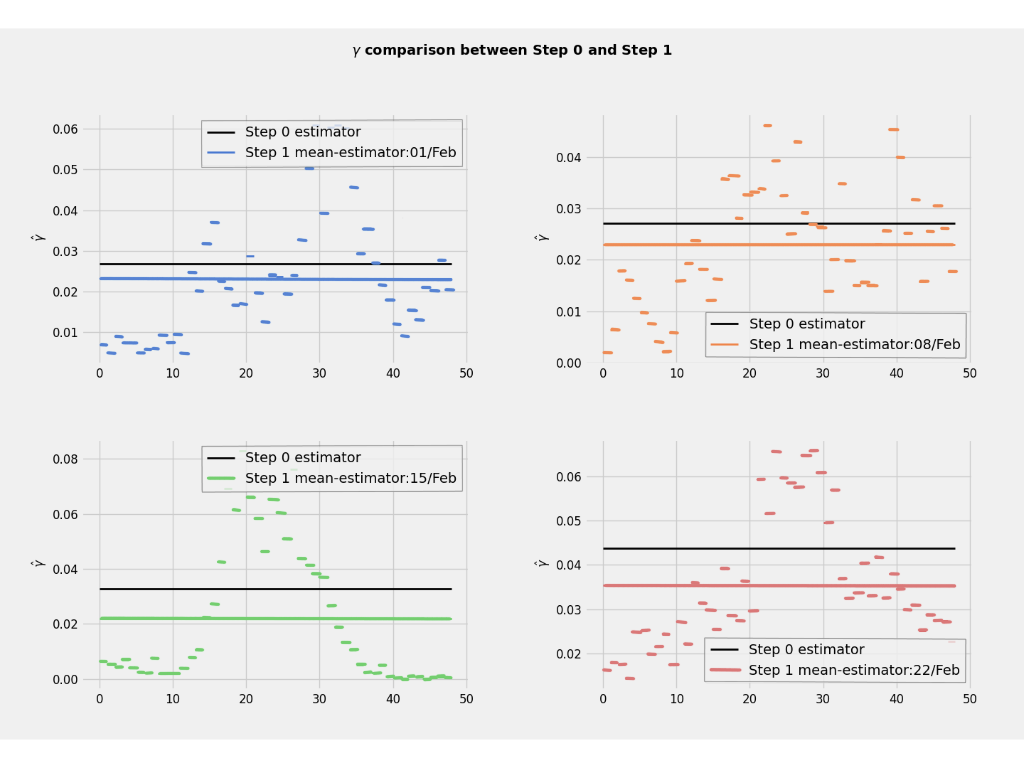}\label{im:Gamma_comp_Feb}}
\caption{Bayesian calibration of $\gamma$ during February, 2017: box plot for each 20 minutes-length sub-signal \eqref{im:Step1_gamma_Feb}  and comparison between Step zero and Step one with mean posterior estimations \eqref{im:Gamma_comp_Feb}  (small line-segments) day-mean posterior estimator (solid colour line) and prior estimator (black line). The colours blue(top left), orange (top right), green (bottom left), red (bottom right) correspond respectively to occurrence order of the Wednesdays  in the month.}\label{fig:calib_gamma}
\end{figure}

Regarding the calibration of the parameter $\gamma$, in Figure \ref{fig:calib_gamma} we summarize the results of the calibration of each sub-signal during February, 2017. More precisely, in Figure \ref{im:Step1_gamma_Feb} we show the box-plot for each 20 minutes-length sub-signal, where we notice that the variance of those parameters that reach low values is quite small, suggesting that the state 0 is a kind of attractor point. On the other hand, Figure \ref{im:Gamma_comp_Feb} shows the comparison between the mean posterior estimations (associated with each sub-signal), the mean $\gamma$ during the 16 hours period for each day (coloured solid lines) and the prior estimator computed with quadratic variation estimator. From the results of the calibration of $\gamma$ we highlight that, despite the fact that the training step estimator and the (daily) average estimator are quite close, the variation of $\gamma$ is significant. The calibration by segments represents an undoubted improvement taking into account the time dependence of the production coefficient and the subsequent ability of the model to adapt to the data. 

\begin{figure}
\centering
\subfigure[${\bm q}^{\text{obs}}$ for $d=$\text{ Feb} 1st]{\includegraphics[trim={2.1cm 17cm 18.5cm  3cm}, clip, width=0.48\textwidth]{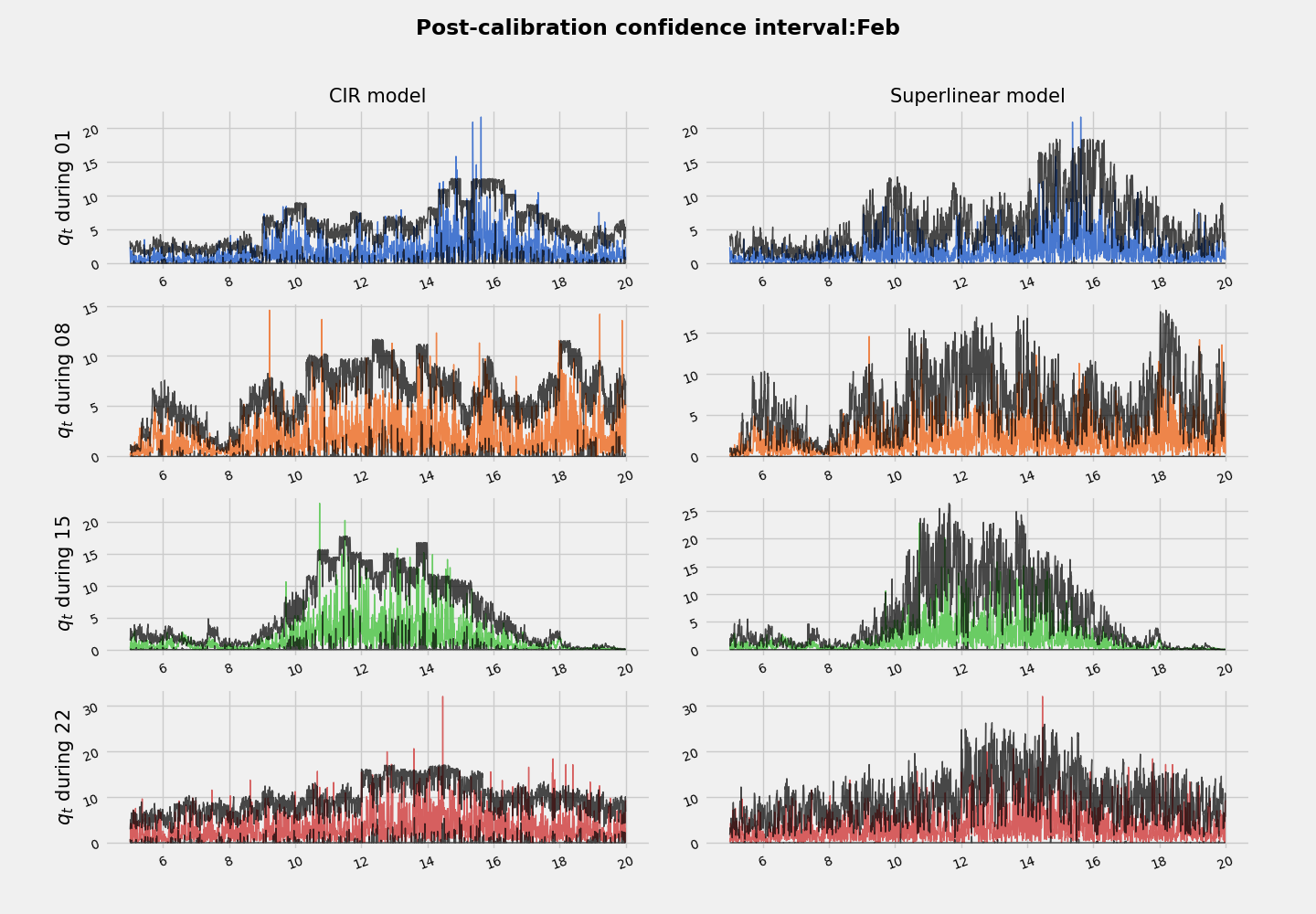}\label{im:February1}}
\subfigure[${\bm q}^{\text{obs}}$ for $d=$\text{ Feb} 8th]{\includegraphics[trim={2.1cm 11.8cm 18.5cm  8.2cm}, clip, width=0.48\textwidth]{./Repl_CIRvsSL_step1_Feb.png}\label{im:February8}}
\subfigure[${\bm q}^{\text{obs}}$ for $d=$\text{ Feb} 15th]{\includegraphics[trim={2.1cm 6.4cm 18.5cm  13.5cm}, clip, width=0.48\textwidth]{./Repl_CIRvsSL_step1_Feb.png}\label{im:February15}}
\subfigure[${\bm q}^{\text{obs}}$ for $d=$\text{ Feb} 22th]{\includegraphics[trim={2.1cm 1.1cm 18.5cm  18.8cm}, clip, width=0.48\textwidth]{./Repl_CIRvsSL_step1_Feb.png}\label{im:February22}}
\caption{Validation of the calibration procedure: instantaneous turbulent kinetic energy observed during February, 2017, between 5 a.m. and 8 p.m. (colour plots) using the frequency of 1/30\xspace$\si\second^{-1}$, and confidence interval (plotted in black) of $(\qnn,n)$ in \eqref{eq:ses_bis} using the posterior mean values  $\widehat{C}_\alpha(d)$  and the time dependent mean $\bar\gamma(t) = \sum_{i=0}^S \EE[\gamma(d,i)] \ind_{[T_i, T_{i+1}]}(t)$, $\Delta t =30$\xspace$\si\second$. \label{fig:replicating_Feb}}
\end{figure}

Once the calibrations in Steps zero and one are performed, we would like to verify the ability of the model in replicating the observations. To do so, we compute the 95\% confidence interval for the trajectories $(\qnn,n)$ and compare this confidence interval against the observations. Precisely, we use the inferred  values  $\widehat{C}_\alpha(d)$ illustrated in Figure \ref{im:Step1Calpha} and the time dependent mean $\bar\gamma(t) = \sum_{i=0}^S \EE[\gamma(d,i)] \ind_{[T_i, T_{i+1}]}(t)$, for each calibrated day $d$, to construct the associated discrete time instantaneous turbulent kinetic energy with the symmetrized Euler scheme:
\begin{align}\label{eq:ses_bis}
&\qn1 =|p_{t_{n+1}}|, \nonumber\\
&p_{t_{n+1}}= \qnn + C_R \bar\gamma_t\Delta t - C_R\left(\frac{\widehat{C}_\alpha^2(d)\bar\gamma_t}2\right)^{1/3} \qnn \Delta t + \sqrt{2C_0\bar\gamma_t}\sqrt{\qnn}(W_{t_{t+1}}-W_{t_n}),~\widehat{q}_0 = {\bm q}_0^{\text{obs}},
\end{align}
where $(W_{t_{t+1}}-W_{t_n})\sim\mathcal{N}(0,\Delta t)$, with the time step $\Delta t$ chosen according to the desired data frequency of $1/30\ \si\second^{-1}$. 
Time by time, we estimate the 95\% confidence interval of the trajectories $(\qnn,n)$. In Figure~\ref{fig:replicating_Feb}, we observe that the confidence interval (in black) tightly envelops the observed trajectory (coloured) for $d =$\ February 1st, 8th, 15th, 22th, validating the calibration process and the probabilistic model \eqref{eq:tke_equilibrated_CIR}. Similar results were obtained for the whole data set.

\subsection{The 10-minutes turbulence intensity as a substitute to the calibration of $(\bar \gamma_t)$}

When comparing the curves of the time evolution $t \mapsto \bar\gamma_t$ in Figure \ref{im:Gamma_comp_Feb} with those of the turbulent intensity $t\mapsto I_t$, defined in Equation \eqref{eq:turbulent_intensity} and illustrated in Figure \ref{fig:data_all_year}, for the same days we can observe a good similarity of their behaviours  even if the magnitudes of the two curves differ. 

Let us going further into the connection between the turbulence intensity $I_t$  and the value of the production coefficient $\gamma_t$. 
Assuming the stationary regime, and considering Equation \eqref{eq:tke_equilibrated_CIR},  we would expect the instantaneous kinetic energy to oscillate around the mean.  Then, if we apply the  expectation operator on both sides of  \eqref{eq:tke_equilibrated_CIR}, and consider --by argument of stationarity --  the zero value for the left-hand side, we get that for all $t\geq0$
\[\gamma_t - \left(\dfrac{C_\alpha^2 \gamma_t}2\right)^{1/3}\EE[q_t]= 0,\]
from which we deduce with \eqref{eq:turbulent_intensity}  the formal relation:
\begin{equation}\label{eq:gamma_stationary}
\gamma_t
= \frac{C_\alpha}{\sqrt{2}}\left( \sqrt{3} \|\av{ U^{\text{obs}}_{(d)}}\| I_t \right)^{3}.
\end{equation}

\begin{figure}
\centering
\subfigure[Turbulence intensity $I_t$.]{\includegraphics[width=0.85\textwidth]{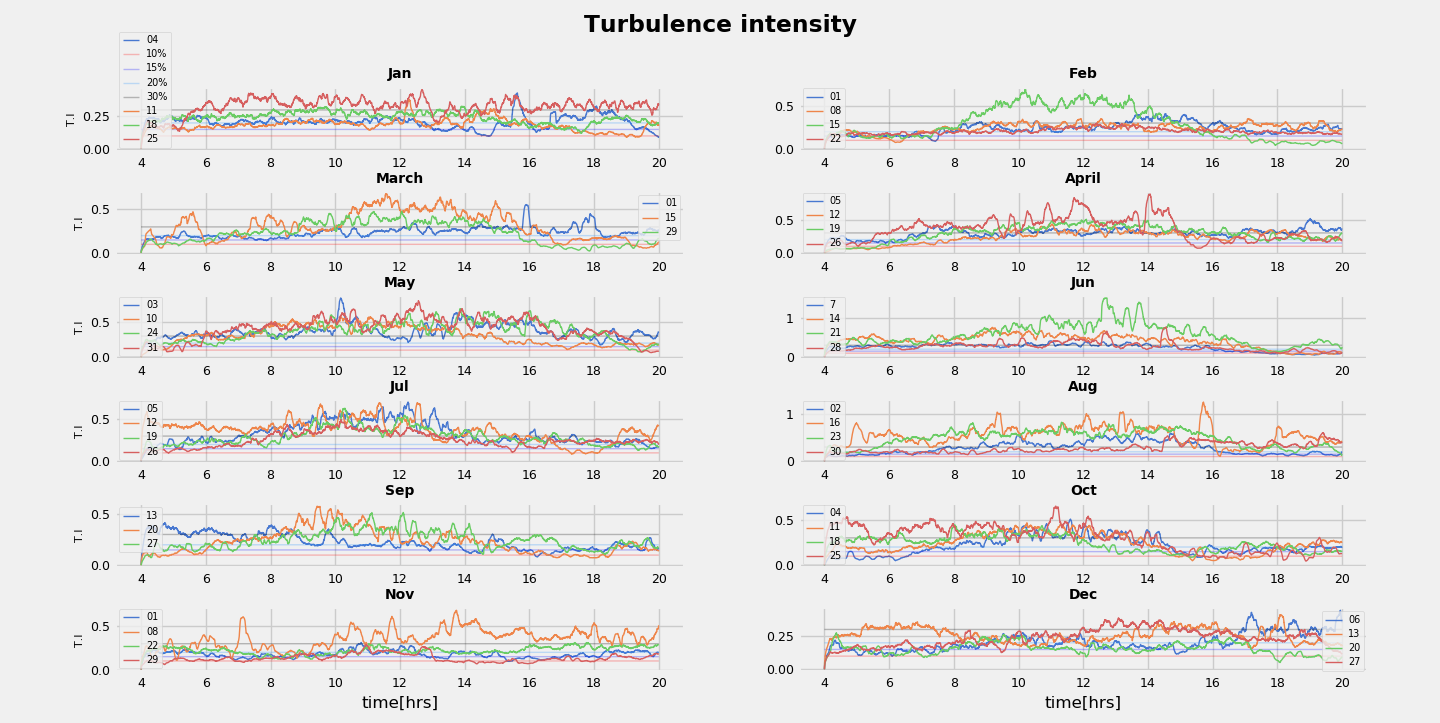}\label{im:turbulence_intensity}}\\
\subfigure[Transformation $\frac{(\sqrt{2}\bar\gamma_t)^{1/3}(d)}{\sqrt{3}\widehat{C}_\alpha^{1/3}(d)\|\av{ U^{\text{obs}}_{(d)}}\|}$.]{\includegraphics[width=0.85\textwidth]{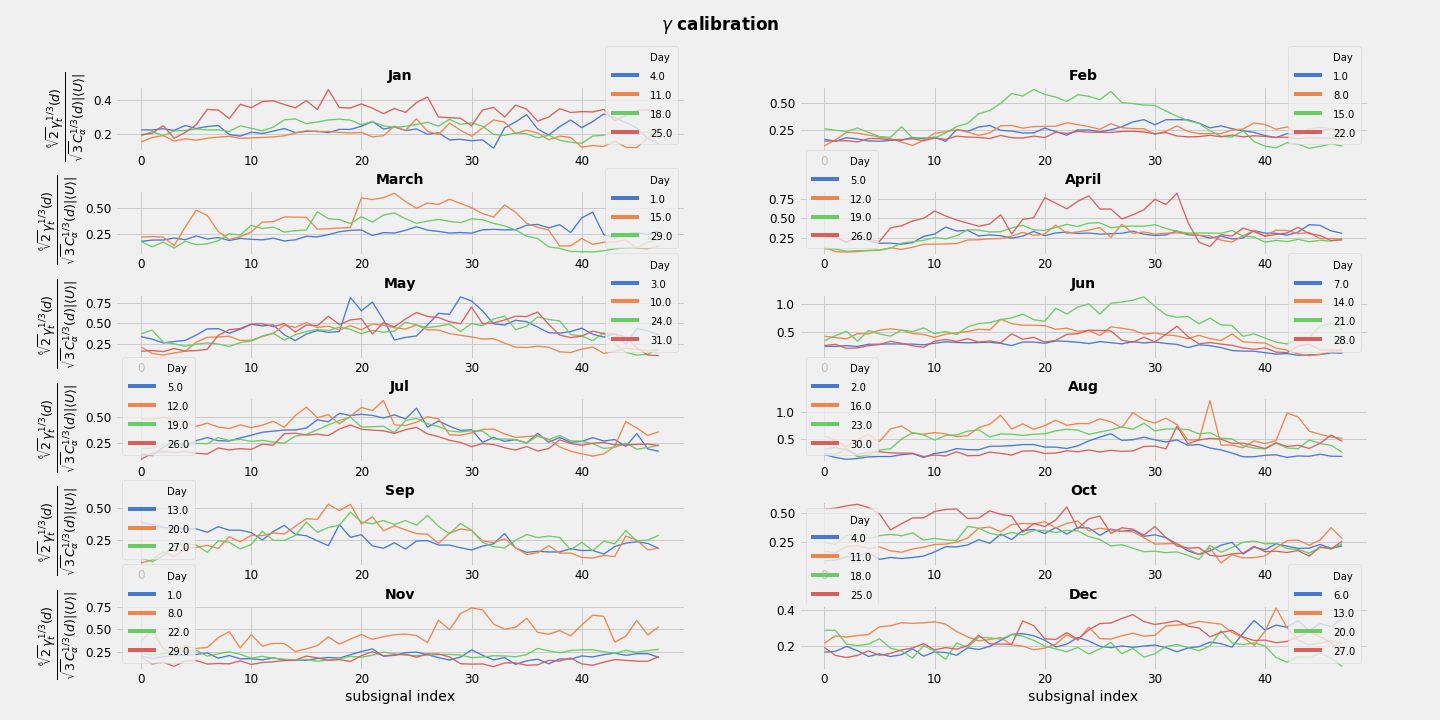}\label{im:gamma_transformation}}
\caption{Validation of Identification  \eqref{eq:gamma_stationary}.}\label{fig:intensity_vs_gamma}
\end{figure}

In Figure \ref{fig:intensity_vs_gamma} we verify the validity of the  relation \eqref{eq:gamma_stationary} by comparing the observed turbulence intensity $I_t$ against the transformation $\frac{(\sqrt{2}\bar \gamma_t)^{1/3}}{\sqrt{3}\widehat{C}_\alpha^{1/3}(d)\|\av{ U^{\text{obs}}_{(d)}}\|}$ for each calibrated day.  Then, assuming the stationary regime, the unquestionable similarity between TI and the transformation of  $\bar\gamma_t$ from \eqref{eq:gamma_stationary} suggests to consider directly $\bar\gamma_t$ in the model to be given dynamically by the 10-minutes turbulence intensity  \eqref{eq:turbulent_intensity}. This observation opens to a simpler and much more efficient calibration procedure for wind forecasting or near-casting  applications. 

\begin{figure}
\centering
\subfigure[CI of ${\bm q}^{\text{obs}}$ obtained with mean posterior estimators.]{\includegraphics[width=0.7\textwidth]{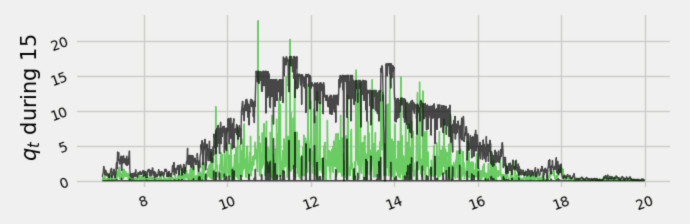}\label{im:Predict_Feb_Baye}}\\
\subfigure[CI of ${\bm q}^{\text{obs}}$ obtained by sampling posterior distribution of $C_\alpha$ and   $\bar\gamma(t)$ from TI.]{\includegraphics[width=0.7\textwidth, ]{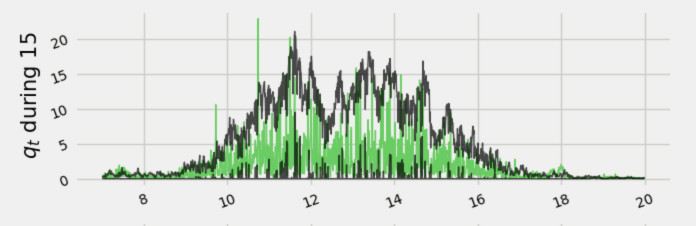}\label{im:Predict_Feb_TI}}
\caption{Comparison of the accuracy of the model to predict the instantaneous turbulent kinetic energy against the observation: the green  plot is the instantaneous turbulent kinetic energy observed during February 15th, 2017, between 7 a.m. and 8 p.m.  using the frequency of 1/30\xspace$\si\second^{-1}$. In Figure \ref{im:Predict_Feb_Baye} (left), the black plots correspond to the 
confidence interval of $(\qnn,n)$ in \eqref{eq:ses_bis} using the posterior mean values  $\widehat{C}_\alpha(d)$  and the time dependent mean $\bar\gamma(t) = \sum_{i=0}^S \EE[\gamma(d,i)] \ind_{[T_i, T_{i+1}]}(t)$, $\Delta t =30$\xspace$\si\second$. In Figure \ref{im:Predict_Feb_TI} (right), the black plots correspond to the confidence interval of $(\qnn,n)$ in \eqref{eq:ses_bis} using the within-year posterior distribution of  ${C}_\alpha$  and the time dependent mean $\bar\gamma(t)$  replaced  by the turbulent intensity statistic through Equation \eqref{eq:gamma_estimator}, $\Delta t =30$\xspace$\si\second$.
\label{fig:predicting_feb_15} }
\end{figure}

\begin{figure}
\centering
\subfigure[${\bm q}^{\text{obs}}$ for $d=\texttt{November}$ 1st and its predicted 95\% CI]{\includegraphics[trim={0cm 0 0 2.2cm}, clip, width=0.7\textwidth]
{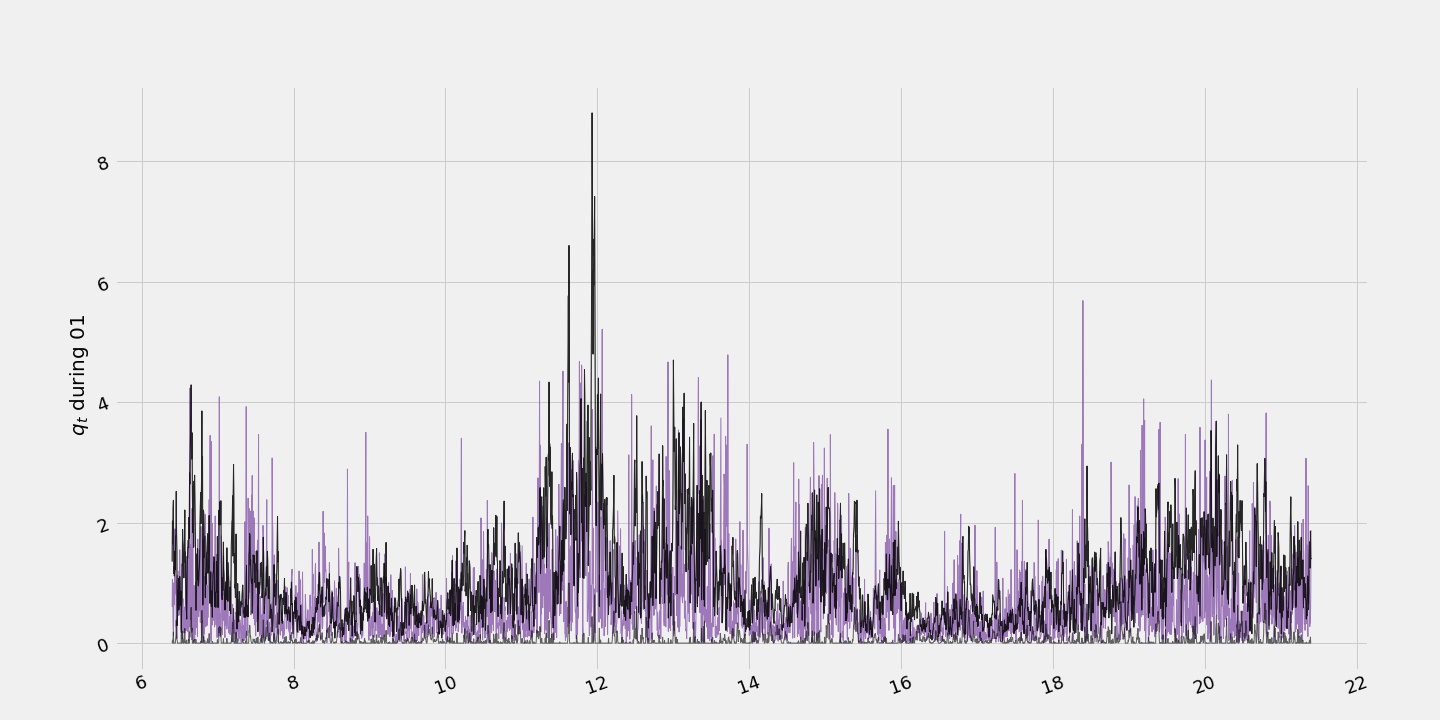}\label{im:Predict_Nov1}}
\subfigure[${\bm q}^{\text{obs}}$ for $d=\texttt{November}$ 10th and its predicted 95\% CI]{\includegraphics[width=0.7\textwidth]
{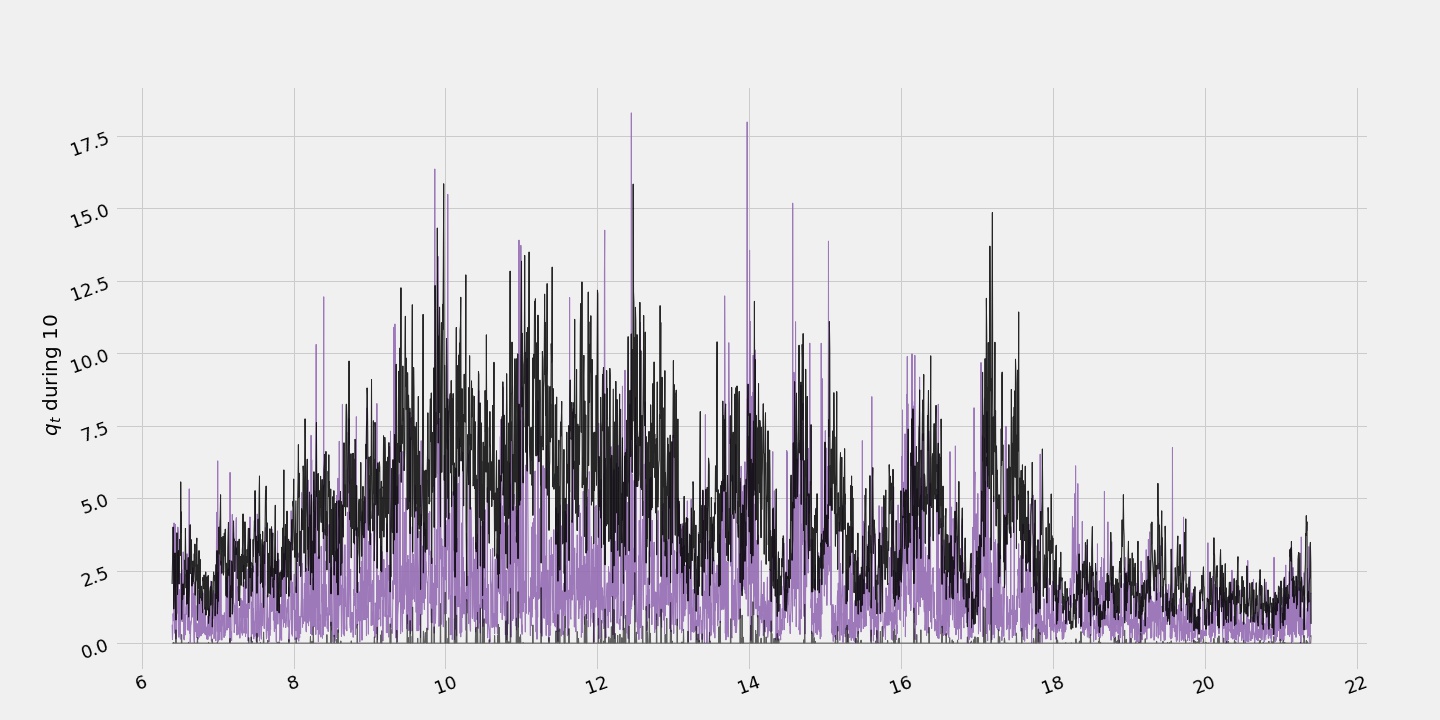}\label{im:Predict_Nov10}}
\caption{Prediction of the instantaneous turbulent kinetic energy: construction of 95\% confidence intervals (in black) from  $(\qnn,n)$ in \eqref{eq:ses_bis} using the within-year posterior distribution of  ${C}_\alpha$  and the time dependent mean $\bar\gamma(t)$  replaced  by the turbulent intensity statistic through Equation \eqref{eq:gamma_estimator}. Observations were taken during  November 1st and November 10th between 7 a.m. and 10 p.m. (colour plots).\label{fig:predicting}}
\end{figure}

Following the previous procedure  leading to Figure \ref{fig:replicating_Feb} for the validation of the calibration method, we generate confidence intervals (with 95\% confidence level) for the solution of the model \eqref{eq:tke_equilibrated_CIR} using the numerical scheme \eqref{eq:ses_bis}, but with input $\bar\gamma_t$ given by relation \eqref{eq:gamma_stationary}. An updated information on the turbulence intensity is needed at each time sub-interval. The parameter ${C}_\alpha$ was sampled from the distribution $\mathcal{N}(1.18\times 10^{-2}, 1.21\times10^{-5})$ obtained from its within-year posterior distribution. In Figures \ref{fig:predicting_feb_15} and \ref{fig:predicting} we illustrate the ability of this calibration method to predict the instantaneous turbulent kinetic energy. Figure \ref{fig:predicting_feb_15} shows the impact of using the turbulent intensity statistic to predict $\bar\gamma_t$ combined with the sampling of the posterior $C_\alpha$  distribution, comparing with the Bayesian calibration procedure result in Figure \ref{im:February15}. Figure \ref{fig:predicting} shows the predicted 95\% confidence interval for two days chosen randomly in the year 2017. 

We end this section with a more global in time evaluation of the model, by comparing the distribution of the turbulent wind speed $\|u'\|$  obtained over a year from the TKE model with the one from the observations. 
More precisely, we consider now the observed 10-minutes-averaged turbulent wind speed $\|u^{\prime\text{\,obs}}\|= \tfrac{1}{\xi}\sum_{t-\xi \leq s\leq t}\sqrt{q^{\text{obs}}_s}$ during all Wednesdays of 2017, with $\xi =$10 minutes, and we compute its empirical distribution. Similarly, we consider the modelled 10-minutes-averaged turbulent wind speed  $\tfrac{1}{\xi}\sum_{t-\xi \leq s\leq t}\sqrt{\widehat{q}_s}$ using the numerical scheme \eqref{eq:ses_bis} with $C_\alpha$ sampled from the posteriori distribution and $\gamma_t$ given by the observed TI and Equation \eqref{eq:gamma_stationary}, and we compute its empirical distribution.

\begin{figure}
\centering
\subfigure[Histograms for the average turbulent wind speed.]{\includegraphics[width=0.85\textwidth]{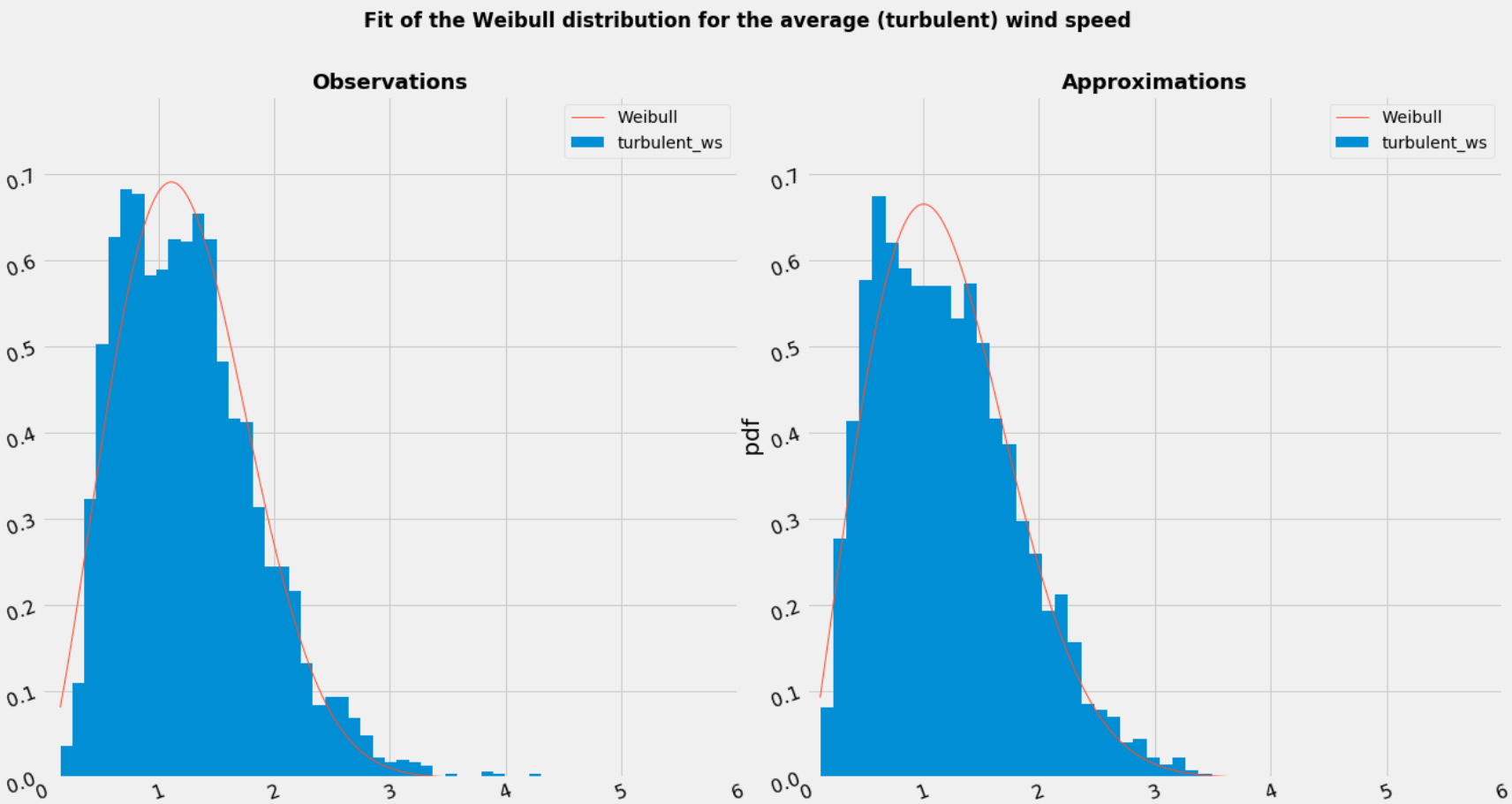}\label{im:weibull_obs}}
\subfigure[Comparison of the fitted densities and QQ-plot for the approximated turbulent wind speed against the observed turbulent wind speed.]{\includegraphics[width=0.85\textwidth]{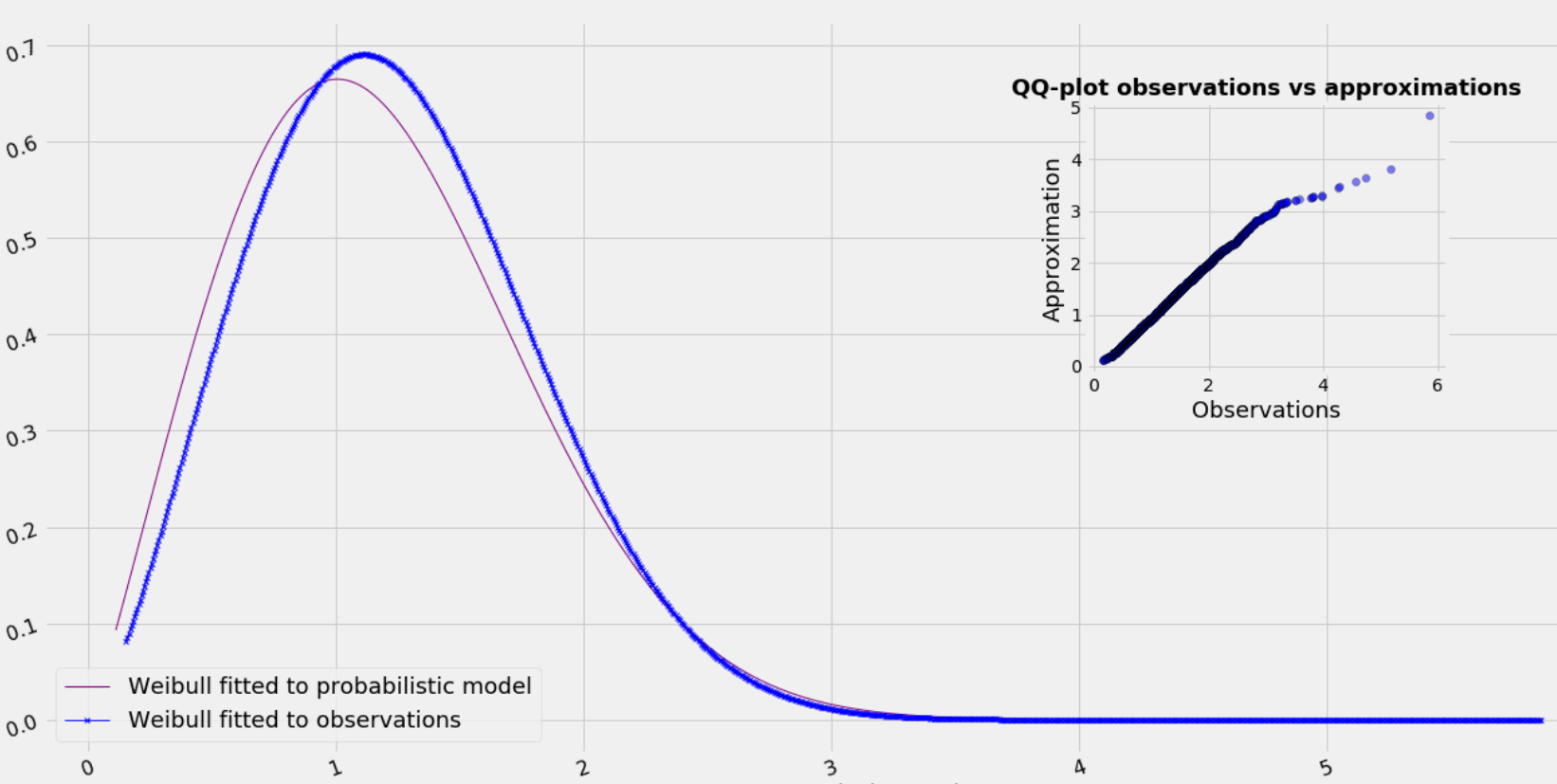}\label{im:weibull_dens}}
\caption{Empirical density for the average turbulent wind speed computed with 10 minutes-period for all Wednesdays (2017) between 5 a.m. and 8 p.m. Fig. \ref{im:weibull_obs} shows the empirical density corresponding to the observations $\tfrac{1}{\xi}\sum_{t-\xi \leq s\leq t}\sqrt{q_s^{\text{obs}}}$ (left) and the probabilistic model $\tfrac{1}{\xi}\sum_{t-\xi \leq s\leq t}\sqrt{\widehat{q}_s}$ (right). Figure \ref{im:weibull_dens} refines the comparison of the associated Weibull densities in Fig.\ref{im:weibull_obs}. Additionally, the graphical method of the quantile-quantile plot for observations against approximations is shown.\label{fig:weibull}}
\end{figure}

For a better comparison of the obtained distributions in Figure 
\ref{im:weibull_obs}, we also adjust a model density to the empirical distributions. As mentioned in the introduction, within local forecasting, the Weibull probability distribution has been fitted to the average wind speed measurements. More precisely, the average wind speed over a 10 minutes-period is commonly considered as a Weibull random variable with shape parameter $k>0$ and scale $\lambda>0$ ~\cite{bensoussan2014generalized,drobinski2015surface}. The parameters $k$ and $\lambda$ can be estimated with different statistical methods, among which we can mention maximum likelihood methods, Anderson–Darling estimation of tails methods and with the Cram\'er–von Mises statistic~ \cite{drobinski2015surface}.

 The Weibull densities in Figure \ref{fig:weibull} have been fitted with simple estimators computed from the mode and median of the sample in order to illustrate density distributions.   From this fit, we have obtained a Weibull distribution with shape parameter of $k_{\text{obs}} = 2.37$ and scale parameter of $\lambda_{\text{obs}} = 1.40$ for the observed turbulent wind speed, and shape parameter of $k_{\text{mod}}  = 2.0$ and scale parameter of $\lambda_{\text{mod}}  = 1.32$ for the approximation of the turbulent wind speed. Additionally, in Figure \ref{im:weibull_dens} we include a Q-Q plot comparing the approximated turbulent wind speed obtained from the CIR model against the observed one.  Considering the comparison of the percentiles, we can conclude that the turbulent wind speed approximated from the proposed model fits very well to the Weibull distribution, showing a slight discrepancy only in the tails.

\section{Conclusions}

Starting from stochastic Lagrangian models for turbulent flows, we have modelled the turbulent kinetic energy filtered at a given point,  introducing a production term denoted by $\gamma$. 
Remarkably, from the derived model, we recover the CIR-like process previously proposed in the literature, with our approach also providing a direct connection from the 3D+time physical model and parameters to the 0D+time model. 
The main feature of the production term $\gamma$ (which has been proved analytically and numerically) is indeed to replicate the non-dissipative nature of the real wind data. 
We have proposed a complete calibration procedure (without external parameters) composed of a preliminary and a Bayesian inference stages. In the prior calibration (Step zero) we implemented the quadratic variation and maximum pseudo-likelihood estimators for $\gamma$ and $C_\alpha$, respectively. From these estimators we propose an a priori distribution for the vector parameter associated to the kinetic energy model \eqref{eq:tke_equilibrated_CIR} and estimate the observational error. Later, in the Bayesian step we quantify the uncertainty of the parameters by implementing Hamiltonian Monte Carlo methods within the construction of Markov chains converging to the stationary distribution of the model. This step provides a distribution for each parameter and allows variation in time for the parameter $\gamma$. The numerical study yields accurate results for both steps of the calibration, recovering the values of the physical parameter $C_\alpha$, validating the model, quantifying the uncertainty and suggesting a model for the production term directly connected to the turbulence intensity statistic. 
Hence, if we have an initial guess on the turbulence intensity, (or if we separate the time interval into regimes with constant intensity), the model \eqref{eq:tke_equilibrated_CIR} with the closure \eqref{eq:gamma_stationary} can be successfully used in the prediction of the turbulent kinetic energy. 

In order to improve the replication of the fast oscillation in the observed process, a superlinear drift term can be considered in the model equation \eqref{eq:tke_equilibrated_CIR} with an exponential Euler scheme~\cite{BoJaMa2021} for its discrete time version. 
For such drift with superlinear growth,  the mean reverting term can act faster and the model can adapt better to the jumps present in the real data. This last point will be consider in a future work.

\subsection*{Acknowledgement}
The authors would like to acknowledge SIRTA for providing high frequency data used in this study. 

The second author acknowledges the support of the Russian Academic Excellence Project 5--100.

The third author acknowledges the support of ANID FONDECYT/POSTDOCTORADO N${}^\circ$321011.

\appendix
\section{On the mean-field turbulent kinetic energy model}
\subsection{Proof of Proposition \ref{prop:SolNonLinear}}\label{app:wellposedness_MF_model}

By means of Lemma \ref{lem:longtime}, the wellposedness of the mean-field model \eqref{eq:tke_mixinglength_bis2} can be easily deduced from the wellposedness of CIR processes with time-dependent coefficients. The latter can be proved with a time-change technique (see e.g. \cite{maghsoodi1996}). Indeed, let us consider the map $t\mapsto \ttke(t)$ as the solution of the ordinary differential equation
\begin{equation}\label{eq:ODE_tke}
\frac{d \ttke(t)}{dt} = \gamma - \frac{C_\alpha}{\sqrt{2}}\ttke^{3/2}(t),\qquad\ttke_0= \EE[q_0],
\end{equation}
and the extended CIR model:
\begin{equation}\label{eq:time_dependent_CIR}
dq_t=\gamma dt - C_R\frac{C_\alpha}{\sqrt{2}}q_t{\ttke(t)^{1/2}} dt+ 3 C_0\frac{C_\alpha}{2\sqrt{2}} \ttke(t)^{3/2} dt+ \sqrt{\sqrt{2}C_0 C_\alpha}\ \ttke(t)^{3/4}\sqrt{q_t}dW_t,\quad q_0>0 \mbox{ given}. 
\end{equation}
Assuming $\gamma >0$, by  Lemma \ref{lem:longtime},  we have that, for all $t\geq0$:
\begin{align}\label{eq:control}
\min\big\{q_0,\big(\tfrac{\sqrt{2}\gamma}{C_\alpha}\big)^{2/3}\big\} \leq \ttke(t)\leq \max\big\{q_0,\big(\tfrac{\sqrt{2}\gamma}{C_\alpha}\big)^{2/3}\big\},
\end{align}
and consequently, for all $t\geq0$, we compute 
\begin{equation}\label{eq:explosion_test_coef}
\delta(t) := \frac{4\left(\gamma + 3 C_0\frac{C_\alpha}{2\sqrt{2}} \ttke(t)^{3/2}\right)}{\sqrt{2}C_0 C_\alpha\ \ttke(t)^{3/2}} = \frac{2\sqrt{2}\gamma}{C_0C_\alpha\ttke(t)^{3/2}} + 3~ >2.
\end{equation}
Then, from Theorem 2.5 in \cite{maghsoodi1996} there exists a unique strictly positive solution $(q_t;t\geq0)$ to Equation \eqref{eq:time_dependent_CIR}. Now, considering this solution, we notice that $\EE[q_t]$ is a non-negative solution to
\[
\frac{d\EE[q_t]}{dt}=F(t,\EE[q_t]),\quad\mbox{with}\quad F(t,x)=\gamma-C_R\frac{C_\alpha}{\sqrt{2}}(\ttke(t))^{1/2}x+3 C_0\frac{C_\alpha}{2\sqrt{2}} \ttke(t)^{3/2}.
\]
Estimate \eqref{eq:control} ensures that $x\mapsto F(t,x)$ is uniformly Lipschitz continuous, and, since $\ttke(t)$ in \eqref{eq:ODE_tke} satisfies the same equation as $\EE[q_t]$, we have $\EE[q_t]=\ttke(t)$. This immediately guarantees that $(q_t ;t\geq 0)$ is a solution to \eqref{eq:tke_mixinglength_bis2}. 

\subsection{Proof of Lemma \ref{lem:longtime}}
\label{app:proof_of_lemma}

For any $p\geq 1$, from  It\^o formula, applied on $q_t^p$ with \eqref{eq:tke_mixinglength_bis2}, we have
\[\EE[q_t^p] = q_0^p +p \gamma\int_0^t\EE[q_s^{p-1}]ds -pC_R\frac{C_\alpha}{\sqrt{2}}\int_0^t\EE[q_s^p]\sqrt{\EE[q_s]}ds + p(p+\tfrac12)C_0\frac{C_\alpha}{\sqrt{2}}\int_0^t\EE^{3/2}[q_s]\EE[q_s^{p-1}]ds.\]
By identifying $\ttke_p(t) : = \EE[q_t^p]$, we deduce the following linear ordinary differential equation
\begin{align}\label{eq:ode_ttke}
\frac{d \ttke_p(t)}{dt}&= -pC_R\frac{C_\alpha}{\sqrt{2}}\ttke_1^{1/2}(t)~\ttke_p(t) + p\big((p+\tfrac12)C_0\frac{C_\alpha}{\sqrt{2}}\ttke_1^{3/2}(t)+\gamma\big)\ttke_{p-1}(t)
\end{align}
with $\ttke_1(t)$ the non-negative solution to
\[
\frac{d \ttke_1(t)}{dt}= (\gamma - \frac{C_\alpha}{\sqrt{2}} \ttke_1^{3/2}(t))\ind_{\{\ttke_1(t)>0\}},
\]
such that $\ttke_1(t)=0$ for all $t\geq \tau_1:=\inf\{t>0\,:\,\ttke_1(t)=0\}$.
\paragraph{In the case $\gamma = 0$} From Equation \eqref{eq:ode_ttke}, we have $\frac{d \ttke_1(t)}{dt}\leq 0$. Thus, the map $t\in[0,+\infty)\mapsto \ttke_1(t)$ is non-increasing, leading immediately to the bound  \eqref{eq:first_moment_McKean}. We prove the bound \eqref{eq:p_moment_McKean}, by applying a comparison principle on the ODE \eqref{eq:ode_ttke}. Indeed, from H\"older inequality we know $\ttke_1(t)\ttke_{p-1}(t)\leq \ttke_p(t)$, and then 
\begin{align*}
-pC_R\frac{C_\alpha}{\sqrt{2}}\ttke_1^{1/2}(t)~\ttke_p(t) + p(p+\tfrac12)C_0\frac{C_\alpha}{\sqrt{2}}\ttke_1^{3/2}(t)\ttke_{p-1}(t) &\leq p \frac{C_0(p-1)C_\alpha}{\sqrt{2}}\ttke_1^{3/2}(t)\ttke_{p-1}(t)\\
&\leq - p C_0(p-1) \frac{d\ttke_1}{dt}(t)\ttke_{p-1}(t).
\end{align*}
Hence, for any $p\geq 1$, by means of a comparison principle we obtain
\begin{align*}
&\ttke_p(t) \leq q_0^p - p C_0(p-1)\int_0^t\dfrac{d\ttke_1(s)}{dt}\ttke_{p-1}(s)ds.
\end{align*}
Iterating the previous inequality, and noticing that,
\[0\leq-\int_0^t\dfrac{d\ttke_1(s)}{dt}ds\leq q_0,\qquad \mbox{and} \qquad 0\leq-\int_0^t\dfrac{d\ttke_1(s)}{dt}\ttke_1(s)ds\leq q_0^2/2,\] 
we deduce  \eqref{eq:p_moment_McKean} from 
\[0\leq \ttke_p(t) \leq q_0^p\left(1+ \frac1{p(p-n)}\sum_{n=1}^{p-1}C_0^n \prod_{i=0}^n(p-i)^2\right), \quad\mbox{ for all }t\geq0.\]

When $\gamma=0$, Equation \eqref{eq:ode_ttke} can be explicitly  solved for $p=1$ with
\begin{equation}\label{eq:mean_TKE_dissipative}
\widetilde{\tke}_1(t) = \EE[q_t] = \left(q_0^{-1/2} + \frac{C_\alpha t}{2\sqrt{2}}\right)^{-2}.
\end{equation}
Therefore $\tau_1 = +\infty$ and  we conclude on \eqref{eq:mean_TKE_limit_gamma} with the  dissipation of the moment of the TKE at large times.

\paragraph{In  the case $\gamma>0$} Equation \eqref{eq:ode_ttke} can be written for $p=1$ as:
\begin{equation}\label{eq:ode_p1_gamma_positive}
C_{q_0} - \frac{C_\alpha}{\sqrt{2}}t = \frac{1}{\gamma' }\left\{\frac13 \log\Big(\frac{(\gamma'-\sqrt{\ttke_1(t)})^2}{\ttke_1(t) + \gamma' \sqrt{\ttke_1(t)} + \gamma^{\prime 2}}\Big)+\frac{2}{\sqrt{3}}\arctan\Big(\frac{2\sqrt{\ttke_1(t)} +\gamma'}{\gamma'\sqrt{3}}\Big)\right\},
\end{equation}
with $\gamma' = (\frac{\sqrt{2}\gamma}{C_\alpha})^{1/3}$ and $C_{q_0}$ a constant depending on the initial condition.  Then, we can analyse the limit behaviour of $\ttke_1$ by means of the isocline method applied to the ODE \eqref{eq:determinsticTKE} and the Equation \eqref{eq:ode_p1_gamma_positive}. Indeed, from \eqref{eq:ode_p1_gamma_positive} we have $\ttke_1(t) \neq  (\tfrac{\sqrt{2}\gamma}{C_\alpha})^{2/3}$ for all $t\geq0$, and then $\tfrac{d\ttke_1(t)}{dt}$ is either strictly negative or strictly positive. Further, by computing the second derivative of $\ttke_1(t)$:
\begin{equation*}
\frac{d^2 \ttke_1}{dt^2}(t)=-\frac{3C_\alpha}{2\sqrt{2}}(\ttke_1)^{1/2}(t)\frac{d \ttke_1}{dt}(t),
\end{equation*}
we can check that $\ttke_1$ does not change its curvature sign, and
\[\ttke_1(t) > \Big(\frac{\sqrt{2}\gamma}{C_\alpha}\Big)^{2/3}~\mbox{ implies }\quad\frac{d \ttke_1}{dt}(t)<0~\mbox{ and }~\frac{d^2 \ttke_1}{dt^2}(t)>0.\]
Thus, in the set $\{t:~\ttke_1(t)>(\tfrac{\sqrt{2}\gamma}{C_\alpha})^{2/3}\}$, the map $t\mapsto\ttke_1(t)$ decreases with time and is convex. Likewise, when $\ttke_1(t) < (\tfrac{\sqrt{2}\gamma}{C_\alpha})^{2/3}$, the map $t\mapsto\ttke_1(t)$ increases and is concave. We conclude on  the boundedness of $\ttke(t)$ and on  its behaviour at large times:
\begin{align*}
\lim_{t\rightarrow+\infty}\ttke_1(t)= \left(\frac{\sqrt{2}\gamma}{C_\alpha}\right)^{2/3}.
\end{align*}

Similarly to the case $\gamma=0$, for any $p>1$ we can bound the right hand-side term in Equation \eqref{eq:ode_ttke} in order to show that the $p$th-moment is uniformly bounded. This time, we use the non-null lower bound of $\ttke_1$: $\min\{q_0,(\tfrac{\sqrt{2}\gamma}{C_\alpha})^{2/3}\}$, obtaining:
\begin{align*}
-\underline{\xi}_{1,p}\ttke_p(t) + \underline{\xi}_{2,p}~\ttke_{p-1}(t) &\leq -pC_R\frac{C_\alpha}{\sqrt{2}}\ttke_1^{1/2}(t)~\ttke_p(t) + p\big((p+\tfrac12)C_0\frac{C_\alpha}{\sqrt{2}}\ttke_1^{3/2}(t)+\gamma\big)\ttke_{p-1}(t) \\
&\leq -\overline{\xi}_{1,p}\ttke_p(t) + \overline{\xi}_{2,p}~\ttke_{p-1}(t),
\end{align*}
for some positive constants $\overline{\xi}_{1,p}, \overline{\xi}_{2,p},\underline{\xi}_{1,p}$ and $\underline{\xi}_{2,p}$ depending on $p,C_\alpha,C_0, \gamma $ and $C_R$. Then, from an induction argument on $p$ and a comparison principle we deduce that
\begin{align*}
\frac{\underline{\xi}_{2,p}}{\underline{\xi}_{1,p}}\inf_{s\geq0}\ttke_{p-1}(s)&-\left(\frac{\underline{\xi}_{2,p}}{\overline{\xi}_{1,p}}\inf_{s\geq0}\ttke_{p-1}(s)~-q_0^p\right)\exp\left\{-\underline{\xi}_{1,p}~t\right\}\\
&\qquad\qquad\leq \ttke_p(t)\leq \frac{\overline{\xi}_{2,p}}{\overline{\xi}_{1,p}}\sup_{s\geq0}\ttke_{p-1}(s)-\left(\frac{\overline{\xi}_{2,p}}{\overline{\xi}_{1,p}}\sup_{s\geq0}\ttke_{p-1}(s)~-q_0^p\right)\exp\left\{-\overline{\xi}_{1,p}~t\right\},
\end{align*}
for all $t\geq0,$ i.e. $\ttke_{p}$ is bounded uniformly in time, which ends the proof.


\end{document}